\documentclass[a4paper,10pt,twoside,openright]{report}
\usepackage{geometry}
\usepackage{multirow}
\usepackage {longtable}
\usepackage[english,french]{babel}
\usepackage[T1]{fontenc}
\usepackage[utf8]{inputenc}

\usepackage{epstopdf}

\usepackage{pst-all}
\usepackage{pst-node}
\usepackage{pst-tree}
\usepackage{graphicx}

\graphicspath{{Figures/}}

\usepackage{amsfonts}
\usepackage{amsmath}
\usepackage{amssymb}
\usepackage[ruled,vlined,french,titlenumbered]{algorithm2e}
\usepackage{fancyhdr}

\newtheorem{theorem}{Théorème}
\newtheorem{definition}{Définition}
\newtheorem{lemma}{Lemme}
\newtheorem{proposition}{Proposition}
\newtheorem{exemple}{Exemple}

\fancypagestyle{plain}{ 
\fancyhead{} 
}

\begin{document}
\thispagestyle{empty}
\newcommand{\member}[3]{{\small #1, #2,\dotfill #3\\}}
\newcommand{\Gras}[1]{\normalfont\bfseries\scshape #1}

\newcommand{\spec}{Informatique}
\newcommand{\fd}{Informatique}
\newcommand{\ed}{Information, Structure, Syst\`emes}
\thispagestyle{empty}

  \begin{center}
      
        \vfill\mbox{}\par\vfill
        \vskip-8ex
        {\Large\textsc{{U}niversit\'e~~{M}ontpellier~~{II}}}\par
        --- {\large\Gras{S}ciences et \Gras{T}echniques du \Gras{L}anguedoc} ---\par\vfill
        {\LARGE{\textsc{Habilitation à Diriger des Recherches}}}\par\vskip2ex\vfill
        \parbox{11cm}{\center%
            \begin{tabular}{lll}
                \textsc{Discipline} &:~&\textsc{~\spec}\\
                \textit{Sp\'ecialit\'e Doctorale} &:~&\textit{~\fd}\\
                \textit{École Doctorale} &:~&\textit{~\ed}
            \end{tabular}
        }\par\vskip2ex\vfill
        pr\'esent\'ee et soutenue publiquement par\\
        \par\vskip4ex
        {\Large Sadok \textsc{Ben Yahia}}\\
          \par\vskip2ex
        le 3 avril 2009 \vfill
        {\LARGE{\bf \textsc{\textbf{Contributions à la formalisation et à
l'extraction de bases génériques de règles associatives}}}}\par\vfill
        \par\vskip6ex
        \parbox{1cm}{\center {\sc Jury}}\par\vskip4ex
        \member{Ramon Lopez \textsc{De Mantaras}, Professeur}{Spanish National Research Council, Espagne}{Rapporteur}
        \member{Lhaouari \textsc{Nourine}, Professeur}{Université Blaise Pascal}{Rapporteur}
        \member{Mohamed \textsc{J}. \textsc{Zaki}, Professeur}{Rensselaer Polytechnic Institute, USA}{Rapporteur}
        \member{Marianne \textsc{Huchard}, Professeur}{Universit\'e Montpellier II}{Examinateur}
        \member{Jean-\textsc{M}arc \textsc{Petit}, Professeur}{Université de Lyon}{Examinateur}
        \member{Pascal \textsc{Poncelet}, Professeur}{Universit\'e Montpellier II}{Examinateur}

 \end{center}
       

\newpage

\chapter*{Remerciements}
\markboth{Remerciements}{Remerciements}


Je suis très reconnaissant à Monsieur Ramon Lopez \textsc{de Mantaras},
Professeur au Spanish National Research Council, à Monsieur
Lhaouri  \textsc{Nourine}, Professeur à l'Université Blaise-Pascal et à Monsieur Mohamed Jaweed \textsc{Zaki}, Professeur à l'Université de Rossynaeler
 qui ont bien voulu prendre de leur temps pour évaluer et
rapporter ce travail.
Je tiens également à remercier Madame Marianne \textsc{Huchard},
Professeur à l'Université de Montpellier II, Monsieur Jean-Marc \textsc{Petit}, Professeur à l'Université de Lyon  ainsi qu'à
Monsieur Pascal \textsc{Poncelet}, Professeur à l'Université de Montpellier II, qui se sont intéressés à
mes travaux de recherche et qui ont accepté d'examiner cette
habilitation.

\bigskip
Je ne saurais jamais trouver les mots pour remercier Monsieur
Yahya \textsc{Slimani}, Professeur à la Faculté des Sciences de
Tunis, pour son soutien, ses remarques et ses conseils. Qu'il en
trouve en ce modeste travail les prémisses de la passion de la
recherche, qu'il a incrusté en moi.

\bigskip
Je demeure reconnaissant à Monsieur Ali Jaoua, Professeur à la
Faculté des Sciences de Tunis et Directeur de thèse de doctorat
pour son soutien moral et pour m'avoir intégré dans son unité de
recherche.

\bigskip
\`{A} mon ami et collègue Engelbert Mephu \textsc{Nguifo}, Professeur à l'Université Blaise Pascal pour ses
encouragements et ses suggestions constructives au cours de
discussions intenses, à Tunis et à Lens, que nous avons eues ces
dernières années.

\bigskip
\`{A} tous mes collègues du Département des Sciences de
l'Informatique qui m'ont aidé durant la préparation de mon
habilitation, et en particulier à Monsieur Samir \textsc{Elloumi}
pour tous les instants que nous avons pu partager.

\bigskip
\`{A} tous mes étudiant(e)s, spécialement Tarek, Ghada, Sami,
Leila, Imen, et Slim, qui se sont investis avec moi dans ces
travaux, j'ai partagé avec vous beaucoup de moments de bonheur.

\bigskip
\`{A} toute ma famille et particulièrement à mes "\textsc{3r}".
Aucun mot ne peut faire oublier mes moments d'absence (parfois
même en étant physiquement près de vous). L'aboutissement de ce
travail est aussi le fruit de ces instants.


\newpage
 \pagenumbering{roman}
\thispagestyle{empty} \tableofcontents
\newpage
\thispagestyle{empty} \listoffigures
\newpage
\thispagestyle{empty} \listoftables
\newpage
 \linespread{1.3}
 \normalfont

\pagenumbering{arabic}
 \addcontentsline{toc}{chapter}{Introduction
g\'{e}n\'{e}rale}
\chapter*{Introduction g\'{e}n\'{e}rale}
\markboth{Introduction g\'{e}n\'{e}rale}{Introduction
g\'{e}n\'{e}rale}

Déterminer la façon dont sont organisées les données, les
interpréter et en extraire des connaissances utiles est un
problème ouvert, au regard du nombre croissant des grandes bases
de données. En effet, notre capacité à collecter et à stocker les
données de tout type outrepasse nos possibilités d'analyse, de
synthèse et d'extraction de connaissances à partir de ces données.
Comme exemple, nous pouvons citer le domaine de la biologie
moléculaire, où des millions de séquences génétiques attendent
d'être analysées. Un autre exemple est celui des entreprises de
commerce et de service (banque, assurance, grande distribution,
etc), qui arrivent à stocker, particulièrement, des volumes de
données très larges sur leurs activités, mais qui ne peuvent les
exploiter faute de moyens d'analyse adéquats et performants.

Pour faire face à cette problématique d'analyse de larges volumes
de données, plusieurs travaux de recherche ont été menés pour
mettre au point des systèmes d'extraction de connaissances. Ces
travaux s'intègrent dans un domaine de recherche qui est très
fertile, appelé "Extraction de Connaissances dans les Bases de
Données (ECBD)", ou "Knowledge Discovery in Databases
(KDD)"~\cite{[UFa-GPi-PSm-RUt_96]} en anglais. Ce domaine de
recherche est à la confluence de plusieurs autres disciplines
parmi lesquelles nous retrouvons l'apprentissage automatique,
l'analyse de données, la reconnaissance de formes, les bases de
données, les statistiques, l'intelligence artificielle, les
systèmes experts, et la vision. L'ECBD est un processus,
interactif et itératif, d'analyse d'un {\it grand} ensemble de
données brutes afin d'en extraire des connaissances exploitables
et utiles.

L'ECBD est un processus interactif dans lequel
l'utilisateur-analyste va sélectionner des données, choisir des
outils de fouille de données (ou "data mining" en anglais) pour
construire des modèles expliquant les données. Brachman et Anand
\cite{[RBr-TAn_96]} font remarquer que le processus d'ECBD doit
nous aider à mieux comprendre comment extraire les connaissances,
et surtout comment aider le mieux possible l'utilisateur-analyste.
L'ECBD est aussi un processus itératif comprenant plusieurs étapes
où plusieurs décisions peuvent être prises par
l'utilisateur-analyste, avec retour aux étapes précédentes en cas
de non satisfaction. Ces étapes se résument comme suit
\cite{[UFa-GPi-PSm-RUt_96], [JHa-MKa_01]}:

\begin{enumerate}
\item {\bf Nettoyage et Intégration de bases de données:} \`{A}
partir d'une ou plusieurs bases de données, il s'agit ici de
supprimer les données inconsistantes et/ou de combiner des données
provenant de différentes sources pour générer un entrepôt de
données;

\item {\bf Pré-traitement de données:} Les données d'un entrepôt
sont sélectionnées ou transformées pour être mises sous une forme
qui les rend exploitables avec un outil de fouille de données.

\item {\bf Fouille de données:}  Il s'agit d'appliquer des
méthodes d'extraction de motifs ou de modèles. Le système de
fouille de données doit contenir plusieurs modules fonctionnels
pour assurer différentes tâches telles que la caractérisation, la
recherche de règles associatives, la classification supervisée ou
non supervisée, la prédiction, l'analyse de données évolutives,
etc;

\item {\bf \'{E}valuation et Présentation.} Les motifs extraits
sont évalués pour identifier ceux qui peuvent constituer une
réelle connaissance pour l'analyste-utilisateur. Dans cette étape,
les techniques de représentation de connaissances et d'interface
homme-machine trouvent toute leur importance pour visualiser la
connaissance produite.

\end{enumerate}

Les tâches d'un système de fouille de données utilisent plusieurs
techniques, telles que les règles associatives, les arbres de
décision, les réseaux de neurones, etc. \cite{franco,lefebure}.
L'extraction de règles associatives --qui peut être considérée
comme la technique la plus utilisée actuellement-- vise à
construire un modèle basé sur des règles conditionnelles à partir
d'une base de données. Une règle conditionnelle se définit sous la
forme d'une suite "\texttt{Si Condition(s) alors Résultat}"
\cite{lefebure}. Les parties, \texttt{Condition} et
\texttt{Résultat}, sont formées par un sous-ensemble d'attributs
(ou items) qui décrivent la base de données.

Cependant, les algorithmes de fouille de données fondées sur la
génération des règles associatives sont de nature exponentielle et
ont beaucoup de difficultés à s'adapter à l'exploration  de
données de plus en plus larges. Pour cela, les techniques de
hiérarchisation de données (ou de concepts) fondées sur la notion
de treillis se trouvent être intéressantes pour représenter et
traiter de larges ensembles de données. Le treillis, qui est une
notion très ancienne, trouve son origine vers la moiti\'e du
dix-neuvi\`eme si\`ecle à partir des travaux de Boole
\cite{bool47}. La th\'eorie relative aux treillis a \'et\'e
longtemps n\'eglig\'ee, avant de connaître son essor \`a partir de
l'ann\'ee $1940$ grâce aux travaux de \"{O}re \cite{Ore44},
BirKhoff \cite{birkoff65}, qui l'ont transformée en branche
fertile de l'alg\`ebre.  Par ailleurs, les familles de Moore,
dites aussi syst\`emes de fermeture, se sont av\'er\'ees
fondamentales pour plusieurs domaines de l'informatique (bases de
donn\'ees, analyse formelle de concepts). Les correspondances de
Galois jouent un r\^ole important dans la th\'eorie des ensembles
ordonn\'es depuis leur mise en \'evidence par Birkhoff, en 1940,
sous le nom d'applications polaris\'ees~\cite{birkoff65}.


Cette notion de treillis est pr\'esente dans plusieurs domaines
des math\'ematiques pures et appliqu\'ees et de l'informatique.
Ainsi, la structure compacte des treillis constitue un formalisme
fort intéressant pour d'une part rendre plus compact l'espace
d'exploration en fouille de données et réduire la connaissance
produite sans perte, d'autre part.

C'est dans ce contexte que se situent nos travaux qui ont pour
objectif d'utiliser les fondements mathématiques des treillis pour
l'extraction d'un noyau irréductible et compact de règles
associatives. Ceci donne la possibilité de présenter un ensemble
minimal de règles à l'utilisateur afin de lui permettre de mieux
les visualiser et les exploiter.


\section*{Motivations}
\'{E}tant donné un contexte d'extraction $\mathcal{K}$, le
problème de l'extraction des règles associatives dans
$\mathcal{K}$ consiste à trouver l'ensemble des règles
associatives, dont le support et la confiance sont,
respectivement, au moins égaux à deux mesures statistiques : un
seuil minimal de support (noté \textit{minsup}) et un seuil
minimal de confiance (noté \textit{minconf}). Ce problème peut
alors être décomposé en deux sous-problèmes comme suit
\cite{Agra93}:

\begin{enumerate}
\item Déterminer l'ensemble des itemsets fréquents dans
$\mathcal{K}$, \textit{i.e.}, les itemsets dont le support est
supérieur ou égal à \textit{minsup}.

\item Pour chaque itemset fréquent $I_j$, générer toutes les règles
associatives de la forme $r$: $I_{k} \Rightarrow I_j$ tel que $I_k
\subset I_j$ et dont la confiance est supérieure ou égale à
\textit{minconf}.

\end{enumerate}
\bigskip

Une solution à ce problème a été proposée par Agrawal et
\emph{al.} sous la forme d'un algorithme appelé
\textsc{Apriori}~\cite{Agra94}. Le premier sous-problème a une
complexité exponentielle en nombre d'itemsets. En effet, étant
donné un ensemble d'items de taille $n$, le nombre d'itemsets
fréquents potentiels est égal à $2^{n}$. Le deuxième sous-problème
est exponentiel en la taille des itemsets fréquents. Ainsi, pour
un itemset fréquent $I$, le nombre de règles associatives non
triviales qui peuvent être générées est égal à 2$^{\vert I \vert}$
- 1. Les algorithmes fondés sur l'extraction des itemsets fermés
fréquents~\cite{thesepasq} sont une alternative permettant de
réduire considérablement le coût de l'extraction des itemsets
fréquents, surtout pour des contextes considérés comme denses, et
permettent d'éliminer la redondance au sein de l'ensemble des
règles associatives. Pour extraire des représentations compactes
de règles associatives, connues sous la désignation de
\textit{bases génériques}~\cite{zaki2004}, il faut préalablement
découvrir trois composantes essentielles, à savoir \cite{isi04}:

\begin{enumerate}
    \item Les itemsets fermés fréquents;
    \item La liste des générateurs minimaux associée à chaque
    itemset fermé fréquent;
    \item La relation d'ordre partiel sous-jacente, organisant hiérarchiquement les itemsets fermés
    fréquents.
\end{enumerate}

Les approches existantes se sont focalisées à extraire les
itemsets fermés fréquents -- et au plus leurs générateurs minimaux
associés-- sans se soucier de la relation d'ordre sous-jacente. En
outre, l'apparition de la tendance "\emph{fouille de
connaissances}" a été un tournant dans les intérêts prioritaires
de la communauté de la fouille de données. En effet, les efforts
ne sont plus seulement déployés dans l'amélioration des
performances du processus d'extraction des motifs fréquents mais
de plus en plus de travaux s'intéressent à l'extraction d'une
connaissance de meilleure qualité tout en maximisant sa compacité.
Dans ce registre, nous relevons des travaux visant l'extraction
des représentations concises des itemsets fréquents. Ainsi, parmi
les représentations exactes les plus connues, nous citons celles
fondées sur les itemsets fermés~\cite{pasquier99}, les itemsets
non-dérivables~\cite{calders021} et les itemsets
essentiels~\cite{casaliisi05}. Bien qu'offrant des taux de
compacité assez intéressants, ces représentations concises
exactes, à l'exception de la dernière, présentaient un certain
nombre de limites, dont essentiellement leur incapacité à la
dérivation, avec des coûts acceptables, des supports disjonctifs
et négatifs des itemsets fréquents. Cette limite s'avère très
handicapante dans la génération de règles associatives
généralisées, \textit{i.e.}, incluant les opérateurs de
disjonction et de négation.

Pour remédier à ces insuffisances, nous présentons, dans ce
mémoire, nos principales contributions ayant comme cadre
fédérateur l'extraction des composantes nécessaires à la
génération des bases génériques. Ces contributions peuvent se
résumer dans les trois pistes de recherche complémentaires
suivantes:
\begin{enumerate}
\item Proposition d'une nouvelle base générique informative et
plus compacte que celles définies dans la littérature.

\item Proposition d'approches algorithmiques pour l'extraction de
l'Iceberg du treillis de Galois, dans lequel les noeuds,
représentant les itemsets fermés fréquents, sont étiquetés par
leurs générateurs minimaux. Ceci garantit l'extraction des bases
génériques des règles associatives informatives.


\item Proposition d'une nouvelle représentation concise et exacte
des itemsets fréquents. La définition de cette nouvelle
représentation s'appuie sur la propriété de non-injectivité de
tout opérateur de fermeture, tout en offrant une perspective sur
l'espace disjonctif des itemsets.
\end{enumerate}

Le protocole d'évaluation de ces contributions consiste en des
estimations empiriques basées sur des études expérimentales
effectuées des bases de données benchmark utilisées dans le
domaine de la fouille de données.

%
\section*{Structure du m\'{e}moire}

Ce mémoire est composé de quatre chapitres.

Le \textbf{premier chapitre} décrit brièvement les fondements
mathématiques de l'extraction des règles associatives et leur lien
avec ceux de l'\textit{analyse formelle de concepts}. En outre,
une caractérisation des approches d'extraction des règles
associatives est introduite.


Le \textbf{deuxième chapitre} présente une étude comparative des
approches utilisées pour la formalisation des différentes bases
génériques ainsi que les mécanismes d'inférence associés. Ensuite,
nos différentes contributions sont présentées et discutées par
rapport aux travaux de la littérature.

Le \textbf{troisième chapitre} commence par une étude critique des
différentes approches pour l'extraction des bases génériques de
règles associatives. Ensuite, il présente nos principales
contributions dans le but de collecter toute l'information requise
pour l'extraction efficace des bases génériques des règles
associatives.


Le \textbf{quatrième chapitre} est consacré à la présentation
d'une représentation concise et exacte des itemsets fréquents, qui
est fondée sur l'exploration de l'espace de recherche disjonctif.
La description de cette contribution est précédée par une étude
critique des différents travaux portant sur la formalisation des
représentations concises des itemsets fréquents.

Le mémoire se termine par une conclusion qui résume l'ensemble de
nos travaux et présente quelques perspectives de recherche.

\chapter{Fondements mathématiques}
\section{Introduction}
L'extraction des règles associatives est l'un des principales
techniques de la fouille de données. Ce problème, introduit par
Agrawal \textit{et al.} \cite{Agra93}, a été développé pour
l'analyse de bases de données de transactions de ventes. Chaque
transaction est constituée d'une liste d'articles achetés, afin
d'identifier les groupes d'articles achetés le plus fréquemment
ensemble~\cite{thesepasq}. L'analyse d'associations est alors
appelée, dans ce cas, \textit{analyse du panier de la ménagère}.
L'analyse des associations part des données les plus fines qui
composent une transaction: les ventes des articles élémentaires.
La recherche des associations vise alors à retrouver les
corrélations qui pourraient exister entre $n$ produits (par
exemple, les acheteurs de salade et de tomates achètent de l'huile
d'olive dans 80\% des cas), mais aussi entre les tendances de
vente de produits (quand les ventes de $X$ augmentent alors les
ventes de $Y$ augmentent dans 80\% des cas) \cite{lefebure}.
L'extraction de règles associatives a donc pour objectif
l'identification de relations significatives cachées entre les
données d'une base de données. Les relations obtenues peuvent être
utiles pour les utilisateurs finaux, \emph{e.g.}, experts,
décideurs, qui peuvent les exploiter pour différents objectifs. En
outre, la technique d'extraction des règles associatives a été
appliquée à d'autres secteurs d'activités, dans lesquels il est
intéressant de rechercher des groupements potentiels de produits
ou de services, e.g., services bancaires, services de
télécommunications~\cite{Adrians97}. Elle a également été utilisée
dans le secteur médical pour la recherche de complications dues à
des associations de médicaments ou à la recherche de thérapies
particulières \cite{techniqdataminingag}.

Dans cette quête de connaissances à extraire, la notion du
\textit{treillis de Galois} \cite{barbut70} ou \textit{treillis de
concepts}~\cite{ganter99}, joue un rôle très important. En effet,
elle est à la base d'une famille de méthodes de classification
conceptuelle~\cite{i-dawak06,Wille89}. Cette notion, introduite
par Barbut et Monjardet~\cite {barbut70}, a été utilisée comme
base de l'Analyse Formelle de Concepts (AFC) par Wille
\cite{willee82}. Ce dernier propose de considérer chaque élément
du treillis, apparenté à une hiérarchie, comme un \textit{concept
formel}, et le graphe associé comme une relation de
généralisation/spécialisation. Ainsi, cette hiérarchie de concepts
met en évidence, de façon exhaustive, les regroupements
potentiellement intéressants par rapport aux observations.
L'approche fondée sur le treillis de Galois a initialement trouvé
des applications en intelligence artificielle notamment pour la
représentation et l'acquisition des connaissances~\cite{Wille89},
puis dans la génération des règles associatives à partir du
treillis~\cite{guigues86}. Ainsi, ses fondements ont été d'un
grand apport dans les axes de recherche complémentaires suivants:

\begin{itemize}

\item la définition de nouvelles approches pour la réduction de
l'espace de recherche induit par l'ensemble des itemsets. Ceci a
permis d'améliorer les performances des approches d'extraction des
règles associatives surtout pour des contextes d'extraction
denses;

\item la réduction du nombre de règles redondantes, par le biais
de la formalisation des bases génériques de règles associatives,
permettant une sélection de règles associatives sans perte de
connaissances.
\end{itemize}
\bigskip

\section{Extraction des règles associatives}
\label{secDerRegAssoc}

Dans cette section, nous allons présenter les fondements
mathématiques de l'extraction des règles associatives et leur
liens avec ceux de l'AFC pour la dérivation de bases génériques
des règles associatives.
\subsection{Les règles
associatives} Une base de transactions (ou plus généralement un
contexte d'extraction) peut être formellement représentée sous la
forme d'un triplet
$\mathcal{K}=(\mathcal{O},\mathcal{I},\mathcal{R})$, où
$\mathcal{O}$ et $\mathcal{I}$ sont, respectivement, des ensembles
finis d'objets (ou transactions) et d'attributs (ou items) et
$\mathcal{R}$ $\subseteq$ $\mathcal{O} \times \mathcal{I}$ est une
relation binaire entre les transactions et les items. Un couple
($o$,$i$) appartenant à la relation $\mathcal{R}$ indique que la
transaction $o \in \mathcal{O}$ contient l'item $i \in
\mathcal{I}$.

Une transaction $T$, avec un identificateur appel\'{e}
\textit{TID} (Tuple IDentifier), est un ensemble, non vide,
d'items de $\mathcal{I}$. Un sous-ensemble $X$ de $\mathcal{I}$,
tel que $k$ = $\vert $X$\vert $ est appelé un \textit{$k$-itemset}
ou simplement un \textit{itemset}. Le nombre de transactions de
$\mathcal{K}$ contenant un itemset $X$, $\vert ${\{}$T$ $ \in $
$\mathcal{K}$ $\vert $ $X$ $ \subseteq $ $T$ {\}}$\vert $, est
appel\'{e} \textit{support absolu} de $X$. Le \textit{support
relatif} de $X$ est le rapport entre son support absolu et le
nombre total de transactions de $\mathcal{K}$, \textit{i.e.},
Supp($X$) = $\frac{\vert {\{}T \in \mathcal{K} \vert X \subseteq T
{\}}\vert}{\vert \mathcal{O}\vert}$. Un itemset $X$ est dit
\textit{fr\'{e}quent} si son support relatif est supérieur ou égal
à un seuil minimum \textit{minsup}$^{(}$\footnote{\textit{minsup}
définit le nombre de transactions minimal auxquelles doit
appartenir X pour être qualifié de fréquent.}$^{)}$
sp\'{e}cifi\'{e} par l'utilisateur.

Le problème d'extraction des règles associatives a été formalisé
par Agrawal \textit{et al.} en 1993~\cite{Agra93}. Une règle
associative $R$ est une relation entre itemsets, de la forme $R$:
$X$ $\Rightarrow$ $(Y$-$X)$, dans laquelle $X$ et $Y$ sont des
itemsets fr\'{e}quents, tel que $X \subset Y$. Les itemsets $X$ et
$(Y$-$X)$ sont appel\'{e}s, respectivement, \textit{pr\'{e}misse}
et \textit{conclusion} de la r\`{e}gle $R$. La g\'{e}n\'{e}ration
des r\`{e}gles associatives est r\'{e}alis\'{e}e \`{a} partir d'un
ensemble $\mathcal{F}$ d'itemsets fr\'{e}quents extrait à partir
d'un contexte d'extraction $\mathcal{K}$. Les r\`{e}gles
associatives valides sont celles dont la mesure de confiance,
Conf($R$) = $\frac{Supp(Y)}{Supp(X)}$, est sup\'{e}rieure ou
\'{e}gale \`{a} un seuil minimal de confiance, défini par
l'utilisateur et qui sera noté dans la suite \textit{minconf}. Si
Conf($R$) = 1 alors $R$ est appel\'{e}e \textit{r\`{e}gle
associative exacte}, sinon elle est appel\'{e}e \textit{r\`{e}gle
associative approximative} \cite{pasquier99b}.

Bien qu'elle soit l'une des techniques qui ait le plus suscité
l'intérêt de la communauté de la fouille de données
\cite{techniqdataminingag}, elle souffre de deux problèmes
majeurs, à savoir le coût du processus d'extraction \cite{Agra94},
ainsi que les grandes quantités de règles associatives extraites à
partir de bases de données réelles. Cette quantité énorme de
connaissances-- pouvant atteindre des dizaines de millions de
règles-- rend quasi impossible leur analyse par un expert humain.
Les fondements mathématiques proposés par l'AFC ont permis de
développer de nouvelles approches permettant d'outrepasser ces
limites~\cite{isi04}. Dans ce qui suit, nous allons faire une
brève présentation de ces fondements.

\subsection{L'Analyse formelle de Concepts}

L'\textit{Analyse Formelle de Concepts} (AFC), introduite par
Wille en 1982~\cite{willee82}, traite des concepts formels: un
concept formel est un ensemble d'objets, (ou \textit{extension}),
auquel s'applique un ensemble d'attributs, (ou
\textit{intention}). Ainsi, l'AFC constitue un outil de
classification et d'analyse, dont la notion centrale est le
treillis de Galois (ou de concepts formels). Le treillis de Galois
peut être vu comme un regroupement conceptuel et hiérarchique
d'objets (à travers les extensions du treillis), et interprété
comme une représentation de toutes les implications entre les
items (à travers les intentions) \cite{theserafik}. L'AFC permet
aussi de réduire considérablement le nombre de relations entre
ensembles d'attributs, en ne générant que celles considérées comme
non redondantes. Dans ce qui suit, nous allons brièvement
présenter les fondements mathématiques qui seront utilisés dans la
suite de ce mémoire.

\bigskip

%
%
%
%
%
%
%
%
La définition suivante introduit les différents types de supports
pouvant être associés à un itemset.

\begin{definition} \cite{casaliisi05} \label{supp_all} \textsc{(Supports d'un itemset)}
Soit un contexte d'extraction $\mathcal{K}$ = ($\mathcal{O}$, $\mathcal{I}$, $\mathcal{R}$).
Nous distinguons trois types de supports associés à un itemset $I$ :\\
\textbf{- Support conjonctif :} Supp($I$) =
$\mid\{o\in\mathcal{O}\mid
(\forall\mbox{ } i\in I,(o,i)\in\mathcal{R})\}\mid $\\
\textbf{- Support disjonctif :} Supp($\vee$$I$) =
$\mid\{o\in\mathcal{O}\mid
(\exists\mbox{ } i\in I,(o,i)\in\mathcal{R})\}\mid $\\
\textbf{- Support négatif :} Supp($\neg$$I$) =
$\mid\{o\in\mathcal{O}\mid
(\forall\mbox{ } i\in I,(o,i)\notin\mathcal{R})\}\mid $\\

\end{definition}


\begin{exemple}Considérons le contexte d'extraction de la figure \ref{exemplecontexteformel1}.
Les différents supports que l'on associe à l'itemset
$BC$~\footnote{Les ensembles d'items seront représentés sans
séparateurs, \textit{e.g.}, \textit{BC} représente l'ensemble
\textit{\{B, C\}}.} sont : Supp($BC$) = 2, Supp($\vee$$BC$) = 5,
Supp($\neg$$BC$) = 0.

\begin{figure}[!ht]
\begin{center}
\begin{tabular}{|c|c|c|c|c|c|c|}
  \hline
   & $A$ & $B$ & $C$ & $D$ & $E$ & $F$ \\
  \hline
  $1$ & $\times $ & $\times $ & $\times $ & $\times $ &  &  \\ \hline
  $2$ &  &  & $\times $ & $\times $ & $\times $ &  \\ \hline
  $3$ & $\times $ & $\times $ &  &  & $\times $ & $\times $ \\ \hline
  $4$ & $\times $ & $\times $ & $\times $ & $\times $ & $\times $ & $\times $ \\ \hline
  $5$&  &  & $\times $ & $\times $ &  & $\times $ \\
  \hline
\end{tabular}
\end{center}
\caption{Un exemple de contexte
d'extraction.}\label{exemplecontexteformel1} 
\end{figure}

\end{exemple}

Le lemme suivant établit les relations qui existent entre les
différents supports. Ces relations sont basées sur les identités
d'inclusion-exclusion \cite{galambos2000}.

\begin{lemma}

 \cite{galambos2000} \label{lemmaidentitésinclusionexclusion}\textsc{(Identités d'inclusion-exclusion)} Les identités d'inclusion-exclusion établissent les
liens qui existent entre le support conjonctif, le support
disjonctif et le support négatif.\\
\begin{displaymath}
Supp(I)\mbox{ }=\mbox{ }\sum_{\substack{I_1 \subseteq I \\ I_1\neq
\emptyset}} {(-1)^{\mbox{$\mid I_1\mid$\mbox{ - 1}}} \mbox{
}Supp(\vee I_1)}
\end{displaymath}
\begin{displaymath}
Supp(\vee I)\mbox{ }=\mbox{ }\sum_{\substack{I_1 \subseteq I \\
I_1\neq \emptyset}} {(-1)^{ \mbox{$\mid I_1\mid $\mbox{ - 1}}}
\mbox{ }Supp(I_1)}
\end{displaymath}
\begin{displaymath}
Supp(\neg I)\mbox{ } =\mbox{ } \mid \mathcal{O} \mid \mbox{
}-\mbox{ } Supp(\vee I) \mbox{ }(\mbox{loi de De Morgan})
\end{displaymath}
\end{lemma}

\begin{exemple} Considérons le contexte d'extraction de la figure \ref{exemplecontexteformel1}.
Nous allons montrer comment calculer les différents supports de
l'itemset $BC$ grâce aux identités d'inclusion-exclusion.
\begin{itemize}
    \item Supp($BC$) = $(-1)^{\mid BC \mid\ -\ 1}$
    Supp($\vee$$BC$) + $(-1)^{\mid B \mid\ -\ 1}$
    Supp($\vee$$B$) + $(-1)^{\mid C \mid\ -\ 1}$
    Supp($\vee$$C$) = - Supp($\vee$$BC$) +
    Supp($\vee$$B$) + Supp($\vee$$C$) = - 5 + 3
    + 4 = 2.
    \item Supp($\vee$$BC$) = $(-1)^{\mid BC \mid\ -\ 1}$
    Supp($BC$) + $(-1)^{\mid B \mid\ -\ 1}$
    Supp($B$) + $(-1)^{\mid C \mid\ -\ 1}$
    Supp($C$) = - Supp($BC$) +
    Supp($B$) + Supp($C$) = - 2 + 3 + 4 = 5.
    \item Supp($\neg$$BC$) = $|$$\mathcal{O}$$|$ - Supp($\vee$$BC$) = 5 - Supp($\vee$$BC$)
    = 5 - 5 = 0.
\end{itemize}
\end{exemple}

L'ensemble des itemsets maximaux fréquents constitue la bordure
positive de l'ensemble des itemsets fréquents. Formellement, cet
ensemble est défini comme suit:

\begin{definition} \cite{maxminer} \label{itemsetmaximalfréquent} \textsc{(Itemset maximal fréquent)}
Soit $\mathcal{K}$ = ($\mathcal{O}$, $\mathcal{I}$, $\mathcal{R}$)
un contexte d'extraction. L'ensemble des itemsets maximaux
fréquents extraits à partir de $\mathcal{K}$, noté
$BD^{+}(\mathcal{IF}_\mathcal{K})$, est défini comme suit:
\begin{center}
$BD^{+}(\mathcal{IF}_\mathcal{K})=\{I\subseteq \mathcal{I}\mid
Supp(I)\geq minsup \wedge \forall\mbox{ } I_1 \supset I,
Supp(I_1)<minsup\}$
\end{center}
\end{definition}

La notion d'idéal d'ordre \cite{ganter99} est aussi nécessaire
afin de caractériser les propriétés structurelles des itemsets.

\begin{definition} \cite{ganter99} \label{definitionordre_ideal}\textsc{(Idéal d'ordre)}
 Un ensemble $S$ est un idéal d'ordre s'il vérifie la propriété suivante:

Si $x$ $\in$ $S$, alors $\forall$ $y$ $\subset$ $x$, $y$ $\in$
$S$.

\end{definition}

Notons  que l'ensemble des itemsets fréquents est aussi un idéal
d'ordre~\cite{Agra94}.

\textbf{- Correspondance de Galois :} Soit un contexte
d'extraction $\mathcal{K}$ =
($\mathcal{O}$,$\mathcal{I}$,$\mathcal{R}$). L'application $\phi$,
de l'ensemble des parties de
$\mathcal{O}$$^{(}$\footnote{L'ensemble des parties d'un ensemble
$\mathcal{O}$ est noté 2$^{\mathcal{O}}$.}$^{)}$ dans l'ensemble
des parties de $\mathcal{I}$, associe à un ensemble d'objets $O$
$\subseteq \mathcal{O}$ l'ensemble des items $i$ $\in$
$\mathcal{I}$ communs à tous les objets $o$ $\in$ $O$
\cite{ganter99}:
\begin{center}
$\phi: 2^\mathcal{O}{} \rightarrow 2^\mathcal{I}{}$\\
$\phi(O)=\{i \in \mathcal{I} | \forall o\in O \wedge
o\mathcal{R}i$ $\}$
\end{center}

Dualement, l'application $\psi$, de l'ensemble des parties de
$\mathcal{I}$ dans l'ensemble des parties de $\mathcal{O}$,
associe à tout ensemble d'items $I$ $\subseteq \mathcal{I}$
l'ensemble des objets $o$ $\subseteq \mathcal{O}$ contenant tous
les items $i$ $\in$ $I$ \cite{ganter99}:

\begin{center}
$\psi:2^\mathcal{I}{} \rightarrow 2^\mathcal{O}{}$\\
$\psi(I)=\{o\in \mathcal{O} | \forall i\in I \wedge o\mathcal{R}i$
$\}$
\end{center}

Le couple d'applications ($\phi$,$\psi$) est une
\textit{correspondance de Galois} entre l'ensemble des parties de
$\mathcal{O}$ et l'ensemble des parties de
$\mathcal{I}$~\cite{ganter99}. Étant donné l'opérateur de
fermeture $\omega$ ($\omega=\phi \circ \psi$), un élément $X\in
\mathcal {I}$ est un élément \textit{fermé} si l'image de $X$ par
l'opérateur de fermeture $\omega$ est \'{e}gale \`{a} lui-même,
\textit{i.e.}, $\omega(X)=X$ \cite{davey02}.

\textbf{- Treillis de concepts formels (de Galois):} Étant
donn\'{e} un contexte d'extraction
$\mathcal{K}$=($\mathcal{O}$,$\mathcal{I}$,$\mathcal{R}$),
l'ensemble de concepts formels $\mathcal{C}_{\mathcal{K}}$ est un
treillis complet $\mathcal{L}_{\mathcal{C}_{\mathcal{K}}}$ =
($\mathcal{C}_{\mathcal{K}}$, $ \le $), appel\'{e}
\textit{treillis de Galois}, quand l'ensemble
$\mathcal{C}_{\mathcal{K}}$ est consid\'{e}r\'{e} avec la relation
d'inclusion entre les itemsets \cite{barbut70,ganter99}. La
relation d'ordre partiel entre des concepts formels est
d\'{e}finie comme suit \cite{ganter99}: $\forall $ $c_{_{1}}$=
($O_{_{1}}$,$I_{_{1}})$, $c_{_{2}}$ =($O_{_{2}}$,$I_{_{2}}) \in $
$C_{\mathcal{K}}$, $c_{_{1}}$ $\le$ $c_{_{2}}$
$\Longleftrightarrow$ $I_{_{2}}$ $\subseteq$ $I_{_{1}}$ ($
\Leftrightarrow $ $O_{_{1}}$ $\subseteq $ $O_{_{2}})$ avec
$I_{_{1}}$, $I_{_{2}} \subseteq $ $\mathcal{I}$ et $O_{_{1}}$,
$O_{_{2}} \subseteq $ $\mathcal{O}$.

%
%
%
%
%
%
%
%
La relation d'ordre partiel est utilis\'{e}e pour g\'{e}n\'{e}rer
le graphe du treillis, appel\'{e} \textit{Diagramme de Hasse}.
Dans ce graphe, les sommets correspondent aux \'{e}l\'{e}ments de
l'ensemble $\mathcal{C}_{\mathcal{K}}$ et les arcs aux relations
de couverture entre les sommets$.$ Chaque \'{e}l\'{e}ment c $ \in
$ $\mathcal{C}_{\mathcal{K}}$ est connect\'{e} aussi bien \`{a} un
ensemble de ses successeurs imm\'{e}diats, appel\'{e}
\textit{Couverture sup\'{e}rieure} $(Couv^{s})$, et \`{a} un
ensemble de ses pr\'{e}d\'{e}cesseurs imm\'{e}diats, appel\'{e}
\textit{Couverture inf\'{e}rieure} $(Couv_{i})$.

La construction de diagramme de Hasse se base sur les deux
principes suivants \cite{ganter99}:
\begin{itemize}
\item si $x< y$ alors $c_{x}$ est en dessous de $c_{y}$ ; \item
$c_{x}$  et $c_{y}$  sont li\'es par un segment si et seulement si
$x\prec y$.
\end{itemize}

\textbf{- Itemsets fermés:} Un itemset $l$ $\subseteq \mathcal{I}$
tel que $\omega$($l$) = $l$ est appelé \textit{itemset fermé}.
Ainsi, un itemset fermé est un ensemble maximal d'items communs à
un ensemble d'objets~\cite{thesepasq}.

\textbf{- Iceberg du Treillis de Galois:} Quand nous
consid\`{e}rons seulement l'ensemble des itemsets ferm\'{e}s
\textit{fr\'{e}quents} ordonn\'{e}s par la relation d'inclusion,
la structure obtenue $(\hat{\mathcal{L}},\subseteq)$ est un
sup-demi-treillis sup\'{e}rieur et elle est d\'{e}sign\'{e}e par
\textit{Iceberg du treillis de Galois} \cite{titanic00}.

\textbf{- Générateur minimal:} Un itemset $g$ $\subseteq$
$\mathcal{I}$ est un \textit{g\'{e}n\'{e}rateur minimal}
\footnote{Appelé aussi \textit{itemset libre} dans
\cite{boulicaut_free_sets} et \textit{itemset clé} dans
\cite{titanic02}.} d'un itemset ferm\'{e} $f$ si et seulement si
$\omega(g)$ = $f$ et il n'existe pas $g_{_{1}}$ $\subseteq$
$\mathcal{I}$ tel que $g_{_{1}}$ $\subset$ $g$ et
$\omega(g_{_{1}})$ =$f$~\cite{bastide2000}. L'ensemble $GM_{f}$
des générateurs minimaux d'un itemset fermé $f$ est défini comme
suit:

\begin{center} $GM_{f}$ = {\{} $g$ $\subseteq$ $\mathcal{I}$ | $\omega(g)$=$f$ $\wedge$ $\nexists$ $g_{_{1}}$ $\subset$ $g$ tel que $\omega(g_{_{1}})$
=$f$ {\}}.\end{center}

\begin{definition}\label{définitionreprésentationconcise}\textsc{(Représentation concise exacte/approximative)}

Soit $\mathcal{E}$ un ensemble d'itemsets.

\begin{itemize}
   \item $\mathcal{E}$ représente une \textbf{représentation concise
exacte} de l'ensemble des itemsets fréquents \textit{si}, à partir
de $\mathcal{E}$, nous sommes en mesure de déterminer avec
exactitude le support de chaque itemset fréquent.
    \item $\mathcal{E}$ représente une \textbf{représentation concise
approximative} de l'ensemble des itemsets fréquents \textit{si}, à
partir de $\mathcal{E}$, nous sommes en mesure d'estimer le
support de chaque itemset fréquent.

\end{itemize}
\end{definition}

\bigskip
Dans ce qui suit, nous allons présenter un survol de la
littérature dédiée aux approches d'extraction des fermés
fréquents.
\section{Approches d'extraction des itemsets fermés fréquents}

En général, le critère utilisé pour classer les algorithmes
d'extraction des itemsets (fermés) fréquents est la stratégie
adopt\'{e}e pour l'exploration de l'espace de recherche. Il existe
actuellement trois grandes stratégies
"\textit{Générer-et-tester}", "\textit{Diviser-pour-régner}" et
une stratégie "\textit{hybride}" des deux premières. Néanmoins,
nous allons ajouter une nouvelle catégorie que nous appelons
\textit{hybride sans duplication} pour différentes raisons
expliquées dans la suite.

1. \textbf{La stratégie "\textit{Générer-et-tester}"}: les
algorithmes les plus connus adoptant cette stratégie sont
\textsc{Close} \cite{pasquier99}, \textsc{A-Close}
\cite{pasquier99b} et \textsc{Titanic} \cite{titanic02}. Ces
algorithmes parcourent l'espace de recherche par niveau. Comme
point de départ, ils considèrent l'ensemble des éléments
inf-irréductibles déjà connu, \textit{i.e.}, l'ensemble des items
$\mathcal{I}$. \`{A} chaque itération $k$, un ensemble de
candidats de taille $k$ est g\'{e}n\'{e}r\'{e} en utilisant la
phase combinatoire de l'algorithme \textsc{Apriori-Gen}
\cite{Agra94} appliquée aux $(k$-$1)$-g\'{e}n\'{e}rateurs minimaux
fr\'{e}quents retenus lors de l'itération précédente. Cet ensemble
de candidats est \'{e}lagu\'{e} par la conjonction d'une
m\'{e}trique statistique (\textit{i.e.}, le support) et
d'heuristiques bas\'{e}es essentiellement sur les
propri\'{e}t\'{e}s structurelles des itemsets ferm\'{e}s et/ou des
générateurs minimaux (\textit{e.g.}, l'idéal d'ordre régissant
l'ensemble des générateurs minimaux).

\bigskip

2. \textbf{La stratégie "\textit{Diviser-pour-régner}"}:
l'algorithme représentatif  adoptant cette stratégie est
\textsc{Closet} \cite{closet00}. Ce dernier a introduit
l'utilisation d'une structure de données compacte appelée
\textsc{FP-tree} (Frequent Pattern tree) \cite{han00} dans les
algorithmes permettant l'extraction des itemsets fermés. En
utilisant un parcours en profondeur d'abord de l'espace de
recherche, \textsc{Closet} essaie de diviser le contexte
d'extraction en sous-contextes et d'appliquer le processus de
d\'{e}couverte des itemsets ferm\'{e}s r\'{e}cursivement sur ces
sous-contextes. Ce processus de d\'{e}couverte repose sur un
\'{e}lagage du contexte bas\'{e} essentiellement sur l'utilisation
d'une m\'{e}trique statistique et d'heuristiques \cite{isi04}. Des
améliorations ou des variantes de cet algorithme ont été
proposées, dont principalement \textsc{Closet+} \cite{closet+} et
\textsc{FP-Close} \cite{closet2003}.

\bigskip

3. \textbf{La stratégie "\textit{hybride}"}: les algorithmes
adoptant cette stratégie utilisent les propriétés des deux
stratégies précédentes, dont le plus connu est l'algorithme
\textsc{Charm} \cite{zakii2002}. Contrairement aux autres
algorithmes qui n'exploitent que l'espace de recherche des
itemsets (fermés), \textsc{Charm} explore simultanément l'espace
de recherche des itemsets fermés et celui des transactions en
introduisant une structure de données appelée \textit{IT-tree}
(Itemset-Tidset tree) \cite{zakii2002}. Chaque noeud dans
\textit{IT-tree} contient un candidat itemset fermé fréquent ainsi
que la liste des transactions auxquelles il appartient. Cette
liste est appelée \textit{tidset} \cite{zakii2002}. \textsc{Charm}
explore l'espace de recherche en profondeur d'abord (comme dans le
cas des algorithmes adoptant la stratégie "Diviser-pour-régner")
sans diviser le contexte d'extraction en sous-contextes.
Cependant, il génère un seul $k$-candidat (comme les algorithmes
adoptant la stratégie "Générer-et-tester") et essaie de tester si
ce candidat est un itemset fermé fréquent ou non, en utilisant des
intersections des tidsets ainsi que des tests de couverture. Le
processus d'extraction se base aussi sur des stratégies d'élagage.
Ces dernières consistent à utiliser une m\'{e}trique statistique
en conjonction avec d'autres heuristiques.

\bigskip

4. \textbf{La stratégie "\textit{hybride sans duplication}"}: deux
principaux algorithmes caractérisent cette stratégie, à savoir
\textsc{DCI-Closed} \cite{dciclosedjournal,dciclosedv12} et
\textsc{LCM}~\cite{LCMv22,LCMv21}. Ces algorithmes peuvent
simplement être considérés comme des améliorations de l'algorithme
\textsc{Charm}, étant donné qu'ils héritent de l'utilisation des
tidsets et du parcours hybride  de l'espace de recherche adopté
dans \textsc{Charm}. Cependant, nous avons choisi de placer ces
deux algorithmes dans une nouvelle catégorie étant donné qu'ils
évitent l'inconvénient majeur des algorithmes précédemment
proposés, à savoir le coût élevé des tests de couverture
permettant d'éliminer tout générateur dont la fermeture a été déjà
extraite (cas de la deuxième et de la troisième catégorie). \`{A}
cette fin, \textsc{DCI-Closed} et \textsc{LCM} parcourent l'espace
de recherche en profondeur d'abord. Après la découverte d'un
nouvel itemset fermé fréquent $c$, ils génèrent un nouveau
générateur $g$ en étendant $c$ par un item $i$, $i$ $\notin$ $c$.
Ainsi, ces deux algorithmes extraient
l'ensemble des itemsets fermés fréquents en un temps linéaire de
la taille de cet ensemble \cite{dciclosedjournal,LCMv22}. De plus,
ces algorithmes n'ont nul besoin de stocker, en mémoire centrale,
l'ensemble des itemsets fermés fréquents précédemment extraits,
étant donné qu'ils n'effectuent pas les coûteux tests de
couverture. La différence entre \textsc{DCI-closed} et
\textsc{LCM} réside dans les stratégies adoptées pour calculer les
fermetures, ainsi que les structures de données utilisées pour
stocker les transactions en mémoire centrale.

\bigskip
Outre la stratégie d'exploration adoptée, d'autres éléments de
caractérisation de ces algorithmes peuvent être
retenus~\cite{tareksigkdd06}:

\begin{itemize}

\item  \textit{Choix des g\'{e}n\'{e}rateurs}: certains
algorithmes ont choisi les g\'{e}n\'{e}rateurs minimaux comme
g\'{e}n\'{e}rateurs de chaque classe d'\'{e}quivalence (cas de
l'algorithme \textsc{Close}), tandis que d'autres optent pour une
technique se basant sur le fait qu'\`{a} chaque fois qu'un
g\'{e}n\'{e}rateur est trouv\'{e}, sa fermeture est calcul\'{e}e.
Ceci permettrait de cr\'{e}er de nouveaux g\'{e}n\'{e}rateurs
\`{a} partir des itemsets ferm\'{e}s fr\'{e}quents trouv\'{e}s
(cas de l'algorithme \textsc{Closet}).

\item \textit{Calcul de la fermeture}: le calcul de la fermeture
d'un générateur minimal $X$ varie, d'un algorithme à autre, par le
fait qu'il peut être réalisé soit par des opérations
d'intersection sur les transactions auxquelles appartient $X$
(\textit{i.e.}, son extension), soit d'une manière
incr\'{e}mentale en cherchant les items appartenant \`{a} la
fermeture d'un générateur (minimal) $X$ et qui vérifient la
relation suivante: $\psi(X)$ $ \subseteq $ $\psi(i)$ $ \Rightarrow
$ $i$ $ \in $ $\omega(X)$, avec $i$ un item n'appartenant pas à
$X$ \cite{titanic00}.

\end{itemize}
\bigskip

Notons que le protocole d'évaluation de ces différents algorithmes
est simplement basé sur des comparatifs des performances obtenues
pour des jeux de données \emph{benchmark}, dont les
caractéristiques sont récapitulées dans le tableau
\ref{benchmark}.

\begin{table}
  \centering

\begin{tabular}{|l|l|r|p{3cm}|r|}
\hline
\textbf{Base}&\textbf{Type}&\textbf{ $\#$ Items} &  \textbf{Taille
moyenne des transactions}& \textbf{$\#$ Transactions}\\\hline
\hline \textsc{Connect} &dense& 129 & 43 & 67557\\\hline
\textsc{Pumsb} &dense& 7117 & 74 & 49046\\\hline \textsc{Mushroom}
&dense& 119 & 23 & 8124
\\\hline \textsc{Chess} &dense& 75 & 37 & 3196\\\hline \hline
 \textsc{T10I4D100K} &éparse& 1000 & 10 & 100000\\\hline
 \textsc{T40I4D100K} &éparse& 1000 & 40 & 100000\\\hline
 \textsc{Retail} &éparse& 16470 & 10 & 88162\\\hline
\end{tabular}
 \caption{Caractéristiques des jeux de données \emph{benchmark} utilisées tout le long du mémoire}\label{benchmark}
\end{table}

\section{Conclusion}

Depuis l'apparition de l'approche basée sur l'extraction des
itemsets \textit{fermés} fréquents \cite{lakhal98}, une nouvelle
formulation du probl\`{e}me de l'extraction des r\`{e}gles
associatives, basée sur l'extraction des itemsets ferm\'{e}s
fr\'{e}quents au lieu des itemsets fr\'{e}quents, a été proposée.
Cette nouvelle formulation permet \cite{thesepasq}:
\begin{itemize}
    \item D'am\'{e}liorer les temps de calcul, puisque dans la plupart des
cas, le nombre d'itemsets ferm\'{e}s fr\'{e}quents est largement
inf\'{e}rieur à celui des itemsets fréquents, surtout pour les
contextes d'extraction considérés comme denses.
    \item De ne générer que des règles associatives non redondantes.
\end{itemize}

\bigskip

Cette nouvelle approche -- l'extraction des itemsets fermés
fréquents -- a donné lieu à une s\'{e}lection de sous-ensembles de
r\`{e}gles sans perte d'information. La s\'{e}lection de règles
sans perte d'information repose sur l'extraction d'un
sous-ensemble de toutes les r\`{e}gles associatives, appel\'{e}
\textit{base générique}, \`{a} partir duquel le reste des
r\`{e}gles pourrait \^{e}tre d\'{e}riv\'{e}. C'est sur les
différentes approches de formalisation des bases génériques que se
propose le chapitre 2 de mettre l'accent.

\chapter{Définition de nouvelles bases génériques de règles associatives}
\section{Introduction}

La prolifération des algorithmes fondés sur l'extraction des
itemsets (maximaux/ fermés) fréquents a éludé l'objectif réel de
la technique d'extraction des règles associatives, à savoir
l'extraction d'une connaissance de taille exploitable et
présentant une plus value. Ainsi, cette abondance d'algorithmes
laisse penser que l'objectif de la technique d'extraction des
règles associatives consiste à découvrir des itemsets (fermés)
fréquents. En revanche, peu de travaux se sont concentrés sur la
génération de règles
associatives~\cite{marzena981,thesepasq,zakii2000}. En effet, dans
la pratique, le nombre de règles associatives, pouvant être
extraites à partir de bases de données réelles, est très élevé.
Ceci est dû à la présence de règles redondantes (\textit{i.e.,}
véhiculant la même information). Par conséquent, l'utilisateur ne
peut plus gérer et exploiter efficacement les connaissances utiles
qui lui sont présentées. Ainsi, l'objectif de la technique
d'extraction des règles associatives doit être reformulé en
veillant à :
\begin{enumerate}
\item Déterminer un ensemble minimal et compact de règles
associatives présenté à l'utilisateur tout en maximisant la
quantité d'informations utiles véhiculées;

\item Disposer d'un mécanisme d'inférence qui, suite à une demande
explicite de l'utilisateur, permet de retrouver le reste de règles
associatives, tout en déterminant avec exactitude leurs supports
et leurs confiances sans accéder au contexte d'extraction.
\end{enumerate}

Pour réduire le nombre de règles extraites, une sélection avec
perte d'information peut reposer
sur des patrons définis par l'utilisateur (\textit{user-defined templates%
})~\cite{Keim96,Klemettinen94,liu99}, des opérateurs bool\'{e}ens
ou \textit{SQL-Like} \cite{meo96,ng98,skirant97}. Le nombre de
règles peut \^{e}tre aussi réduit en les \'{e}laguant avec une
information additionnelle, telle qu'une taxonomie
\cite{han95,hipp,inforsid01,skirant95} ou une m\'{e}trique
d'int\'{e}r\^{e}t additionnelle \cite {Brin97a,hullermeir01}
(\emph{e.g.}, corr\'{e}lation de Pearson ou le test-$\chi ^{2}$).

En revanche, dans le cadre d'une sélection sans perte
d'information, certains travaux ont puisé dans les fondements
mathématiques de l'AFC pour proposer des approches de sélection
sans perte d'information des règles associatives. Ces approches
reposent sur l'extraction d'un sous-ensemble de taille compacte,
de toutes les règles associatives, appelé \textit{base générique},
tout en satisfaisant les conditions suivantes~\cite{mariena02}:
\begin{enumerate}
\item \textbf{\textit{Dérivabilité :}} le mécanisme d'inférence
(\textit{e.g.,} un système axiomatique), permettant la dérivation
des règles redondantes doit être \textbf{correct} (\textit{i.e.,}
le système ne permet de dériver que les règles associatives
valides) et \textbf{complet} (\textit{i.e.,} l'ensemble de toutes
les règles valides peut être retrouvé).
    \item \textbf{\textit{Informativité :}} la base générique des règles associatives doit permettre
de retrouver avec exactitude le support et la confiance des règles
dérivées.
\end{enumerate}


\section{Bases génériques de règles associatives: État de l'art}

Dans ce qui suit, nous allons commencer par présenter les
différents mécanismes d'inférence de règles associatives proposés
dans la littérature. Ces mécanismes sont considérés comme étant
des méthodes de dérivation (ou de raisonnement) de règles à partir
d'autres règles~\cite{marzena022}. Un examen de l'état de l'art
montre que ces mécanismes se définissent, généralement, sous la
forme de système axiomatique ou d'opérateurs de dérivation.

\begin{enumerate}
\item \textbf{Les axiomes d'Armstrong ($\mathcal{AA}$):}
le système axiomatique d'Armstrong est utilisé uniquement dans le
cadre des règles exactes~\cite{armstrong74} et il est constitué
des axiomes suivants:
\begin{enumerate}
\item \textit{\textbf{Reflexivité}}: Si $X\subseteq Y $ alors
Conf$(X \Rightarrow Y) =$ 1.
 \item
\textit{\textbf{Augmentation}}: Si Conf$(X \Rightarrow Y) =$ 1
alors, Conf($XZ \Rightarrow Y) =$ 1. \item
\textbf{\textit{Transitivité}}: Si Conf$(X \Rightarrow Y) =$ 1 et
Conf$(Y \Rightarrow Z) = $1 alors Conf$(X \Rightarrow Z) =$ 1.
\end{enumerate}
\item \textbf{La transitivité de la confiance ($\mathcal{TC}$):}
ce mécanisme, introduit par Luxemburger~\cite{luxemburger91},
permet de déterminer la confiance d'une règle à partir de celles
d'autres règles. Contrairement à l'axiome de transitivité
d'Armstrong, $\mathcal{TC}$ est utilisé aussi bien pour les règles
exactes que pour les règles approximatives.

Soient $X$, $Y$ et $Z$ trois itemsets fréquents tels que $X
\subset Y \subset Z$. Le support et la confiance de la règle
$R'':X \Rightarrow (Z-X)$ peuvent être déterminés à partir des
mesures de validité des règles $R:X \Rightarrow (Y-X)$ et $R':Y
\Rightarrow (Z-Y)$ comme suit:
\begin{enumerate}
    \item Supp$(X \Rightarrow Z-X)$ = Supp$(Y \Rightarrow Z-Y)$.
    \item Conf$(X \Rightarrow Z-X)$ = Conf$(X \Rightarrow Y-X) \times$ Conf$(Y \Rightarrow Z-Y)$.
\end{enumerate}
Ainsi, la confiance d'une règle dérivée, \emph{i.e.}, obtenue par
l'application d'un axiome de dérivation, est toujours inférieure
ou égale aux valeurs de confiance respectives des règles à partir
desquelles elle a été inférée.

\item \textbf{L'opérateur de couverture ($\mathcal{C}$):}
dans~\cite{marzena981}, Kryszkiewicz introduit un opérateur de
dérivation de règles associatives défini comme suit:
\begin{definition}

Soient $X$, $Y$, $Z$ et $V$ quatre itemsets fréquents et $R:X
\Rightarrow Y$ une règle valide. L'ensemble des règles dérivées
suite à l'application de l'opérateur de couverture sur une règle
$R$,
noté $\mathcal{C}$($R$), est défini comme suit:\\
$\mathcal{C}$($X \Rightarrow Y$)= $\{X\cup Z \Rightarrow V$ $|$ $Z
\subseteq Y \wedge V \subseteq Y \wedge (Z\cap V) = \emptyset
\wedge V \neq \emptyset\}$.
\end{definition}

\`{A} partir d'une règle $R: X \Rightarrow Y$, nous pouvons
dériver, grâce à l'opérateur de couverture $\mathcal{C}$,
\texttt{3$^{m}$-2$^{m}$} règles, où \texttt{m} désigne la
cardinalité de l'itemset fréquent $Y$. Soit $\mathcal{R}$
l'ensemble de toutes les règles qui couvrent une règle associative
$R'$ (\textit{i.e.,} pour toute règle $R_i \in \mathcal{R}$, nous
avons $R' \in \mathcal{C}(R_i)$). Ainsi, $Conf(R')\geq max
\{Conf(R_i)\}$ et $Supp(R')\geq max \{$Supp$(R_i)\}$.

\item \textbf{L'inférence basée sur l'opérateur de fermeture
($\mathcal{IRF}$)} : ce mécanisme d'inférence est basé sur le fait
que le support d'un itemset fréquent $I$ est égal à celui de sa
fermeture $\omega(I)$~\cite{thesepasq}.

\end{enumerate}
\bigskip
Dans ce qui suit, nous présentons les différentes approches
d'extraction des bases génériques de règles associatives.

\section{Les approches de formalisation de bases génériques de règles associatives}

Les bases d'implication, définies essentiellement dans les
domaines de l'analyse de données et de l'AFC, ont été les
précurseurs des bases génériques de règles associatives. Les
travaux de Guigues-Duquenne (la base
$\mathcal{DG}$~\cite{guigues86}) et Luxemburger (la base
$\mathcal{BP}$~\cite{luxemburger91}) ont largement influencé les
travaux portant sur les bases génériques de règles associatives.
Pour une étude détaillée sur les bases d'implications et les
mécanismes d'inférence associés, prière de se référer
à~\cite{Mastghada}. Un survol de la littérature permet de classer
les travaux dédiés à la formalisation de bases génériques des
règles associatives en deux catégories : bases génériques avec
perte d'informations et bases génériques sans perte
d'informations.

\subsection{Approches de formalisation de bases génériques avec perte d'information}

Les approches de formalisation de bases génériques, qui se font
avec perte d'information~\cite{marzena981,Phan01,zakii2000},
\emph{i.e.}, ne satisfont pas l'une des deux conditions de
dérivabilité ou d'informativité.
\subsubsection{I. La base de règles représentatives ($\mathcal{RR}$)}
En se basant sur l'opérateur de couverture $\mathcal{C}$,
Kryszkiewicz introduit dans~\cite{marzena981} une base générique
de règles associatives, appelée \textit{base de règles
représentatives} ($\mathcal{RR}$), définie comme suit:
\begin{definition}
Soit $\mathcal{AR}$ l'ensemble de toutes les règles valides. La
base $\mathcal{RR}$ des règles représentatives est définie par :
$\mathcal{RR}$=\{R $\in$ $\mathcal{AR}$ $\mid$ $\forall R'\in$
$\mathcal{AR}$ $\wedge~ R\neq R'$, [$ R \in \mathcal{C}(R')
\Rightarrow R'=R$]\}.
\end{definition}

Afin de dériver l'ensemble de règles redondantes, Kryszkiewicz
propose d'utiliser l'opérateur de couverture
$\mathcal{C}$~\cite{marzena981}. Ainsi, dans~\cite{mariena02},
Kryszkiewicz montre que la base générique $\mathcal{RR}$ constitue
l'ensemble le plus réduit de règles génériques couvrant la
totalité des règles associatives valides et véhiculant le maximum
de connaissances utiles. Toutefois, l'extraction de la base se
fait avec une perte d'information~\cite{mariena02}, puisqu'elle
n'est pas informative (\emph{i.e.}, elle ne permet pas de
déterminer avec exactitude le support ainsi que la confiance de
certaines règles associatives dérivées). Bien que l'opérateur de
couverture $\mathcal{C}$ soit correct et complet, l'extraction de
$\mathcal{RR}$ est faite avec perte
d'information~\cite{mariena02}.
\subsubsection{II. La base de règles associatives non redondantes ($\mathcal{RNR}$)}
Dans~\cite{zakii2000}, Zaki redéfinit la notion de redondance
comme suit:
\begin{definition}
\label{redondancezaki} Soit une règle $R_i$: $X_i \Rightarrow Y_i$
et $\mathcal{AR}$ = \{R$_1$, \ldots , R$_n$\}, l'ensemble de
toutes les règles valides pouvant être extraites à partir d'un
contexte d'extraction $\mathcal{K}$. Une règle R$_1$: $X_1
\Rightarrow Y_1$ $\in \mathcal{AR}$ est \emph{plus générale}
qu'une règle R$_2$: X$_2$ $\Rightarrow$ Y$_2$, notée R$_1$ $
\preceq$ R$_2$, si et seulement si les conditions suivantes sont
vérifiées:
\begin{enumerate}
    \item X$_1$ $\subseteq$ X$_2$ $\wedge$ $Y_2$ $\subseteq$ $Y_1$;
    \item Supp(R$_1$) = Supp(R$_2$) et Conf(R$_1$) = Conf(R$_2$).
\end{enumerate}
\end{definition}
Par conséquent, une règle $R_2$ est redondante si et seulement
s'il existe une règle $R_1$ telle que $R_1$ $ \preceq$ $R_2$.
Ainsi, $R_2$ peut être retrouvée en ajoutant des items à la partie
prémisse et/ou à la partie conclusion de la règle $R_1$. En se
basant sur la définition~\ref{redondancezaki}, Zaki introduit une
nouvelle base générique de règles associatives, appelée
$\mathcal{RNR}$, définie comme suit:
\begin{definition}
$\mathcal{RNR}$ = $\{$$R_i$ $\in \mathcal{AR}$ $|$ $\nexists$
R$_j$ $\in \mathcal{AR}$, R$_j$ $ \preceq$ R$_i$ et R$\neq
R'$$\}$.
\end{definition}

Afin de dériver l'ensemble de toutes les règles redondantes, Zaki
propose d'utiliser l'axiome $\mathcal{TC}$ ainsi que l'axiome
d'augmentation. Toutefois, l'augmentation d'une règle de
$\mathcal{RNR}$ ne doit pas se faire avec n'importe quel item,
puisque le risque de générer une règle invalide peut surgir. Pour
se faire, Zaki propose dans~\cite{zakimirage} de présenter à
l'utilisateur les règles génériques sous la forme: $R: X
\Rightarrow Y$ [$I ^\top$, $I ^\bot$], où $I ^\top$ désigne
l'ensemble d'items pouvant être ajoutés à la partie prémisse ou à
la partie conclusion d'une règle $R$. En revanche, $I ^\bot$
désigne l'ensemble d'items pouvant être ajoutés uniquement à la
partie prémisse de $R$ lorsque $Conf(R)$ = 1 ou uniquement à la
partie conclusion lorsque $Conf(R)$$<$ 1. Cette manière de
présenter les règles permet à la base $\mathcal{RNR}$ de devenir
informative~\cite{zakimirage}. Cependant, nous avons noté que
l'approche d'extraction de la base $\mathcal{RNR}$ présente
quelques limites, à savoir:
\begin{itemize}
\item La base $\mathcal{RNR}$ ne couvre pas l'ensemble de toutes
les règles valides, \emph{i.e.}, des règles associatives valides
pourraient ne pas être générées avec le mécanisme d'inférence
proposé. \item En se référant à la
définition~\ref{redondancezaki}, une règle $R_2$ redondante par
rapport à $R_1$, doit avoir les mêmes mesures de support et de
confiance que celles de la règle $R_1$. Néanmoins, la confiance
d'une règle R$_3$, obtenue suite à l'application de l'axiome
$\mathcal{TC}$ sur les deux règles $R_1$ et $R_2$ de
$\mathcal{RNR}$, n'est ni égale à la confiance de $R_1$ ni celle
de $R_2$.

\item L'algorithme qui permet d'extraire la base $\mathcal{RNR}$
n'est pas correct, dans le sens où il ne garantit pas la
génération de toutes les règles non redondantes exactes.

\end{itemize}
Bien que la base $\mathcal{RNR}$ soit informative, l'utilisation
de l'axiome d'augmentation et de $\mathcal{TC}$ ne garantit pas la
génération de toutes les règles valides. Par conséquent, le
mécanisme d'inférence utilisé n'est pas complet et l'extraction de
la base $\mathcal{RNR}$ se fait avec perte d'information.

\subsection{Approches de formalisation de bases génériques sans perte d'information}
Afin d'extraire les bases génériques des règles associatives,
Bastide \emph{et al}. se sont basés sur la définition de
redondance suivante~\cite{bastide2000}:
\begin{definition}
\label{newdefred} Soit $\mathcal{AR}$ l'ensemble des règles
associatives pouvant être extraites à partir d'un contexte
d'extraction $\mathcal{K}$. Une règle R:
X$\overset{\alpha}{\Rightarrow}$Y $\in$ $\mathcal{AR}$ est dite
redondante par rapport à (ou dérivable de) R$_1$:
X$_1$$\overset{\alpha}{\Rightarrow}$Y$_1$ si R satisfait les
conditions suivantes:
\begin{enumerate}
    \item Supp(R) = Supp(R$_1$) $\wedge$ Conf(R) = Conf(R$_1$)= $\alpha$;
    \item X$_1$ $\subseteq$ X $\wedge$ Y $\subseteq$ Y$_1$.
\end{enumerate}
\end{definition}

Bastide \textit{et al.} définissent la base générique des règles
associatives exactes, notée $\mathcal{GBE}$, comme
suit~\cite{bastide2000}:
\begin{definition}
\label{gbe} Soient $\mathcal{IFF}_\mathcal{K}$ l'ensemble
d'itemsets fermés fréquents  et $\mathcal{G}_c$ l'ensemble des
générateurs minimaux d'un itemset fermé fréquent c. La base
$\mathcal{GBE}$ est définie comme suit: $\mathcal{GBE}$ =$\{g
\Rightarrow (c-g) \mid c \in $ $\mathcal{IFF}_\mathcal{K}$
$\wedge$ g $\in \mathcal{G}_c$ $\wedge$ g $\neq$ c$\}$.
\end{definition}

Bastide \textit{et al.} ont également introduit une base
informative des règles approximatives, notée $\mathcal{GBA}$,
définie comme suit~\cite{bastide2000}:
\begin{definition}
\label{gba} Soient $\mathcal{IFF}_\mathcal{K}$ l'ensemble
d'itemsets fermés
fréquents  et $\mathcal{G}$ l'ensemble des générateurs minimaux. La base $\mathcal{GBA}$ est définie comme suit:\\
$\mathcal{GBA}$ = $\{$g $\Rightarrow$ (c-g), c $\in$
$\mathcal{IFF}_\mathcal{K}$ $\wedge$ g $\in \mathcal{G}$ $\wedge$
$\omega$(g) $\subset$ c $\wedge$ Conf(R) $\geq$ minconf\}.
\end{definition}

En se référant aux définitions~\ref{gbe} et~\ref{gba}, les bases
génériques de règles associatives peuvent être obtenues
directement à partir l'iceberg du treillis de
Galois~\cite{dawak04}. En effet, dans une telle structure, chaque
itemset fermé fréquent est étiqueté par la liste de ses
générateurs minimaux. Ainsi, la base $\mathcal{GBA}$ représente
des relations inter-noeuds, alors que la base $\mathcal{GBE}$
représente des relations intra-noeuds~\cite{dawak04}.

Le mécanisme d'inférence $\mathcal{IRF}$ peut servir comme moyen
de dérivation~\cite{thesepasq}, permettant de rendre la base
$\mathcal{GBE}$ informative. Pasquier a montré également qu'en
appliquant le mécanisme d'inférence $\mathcal{IRF}$, il est
possible de déterminer le support et la confiance de toutes les
règles valides approximatives redondantes~\cite{thesepasq}.
Toutefois, Kryszkiewicz a contredit l'affirmation de Pasquier
quant à la validité du mécanisme d'inférence $\mathcal{IRF}$ sur
la base $\mathcal{GBA}$~\cite{mariena02}. L'application du
mécanisme d'inférence $\mathcal{IRF}$  à $\mathcal{GBA}$ ne
garantit pas la dérivation des règles valides. Par conséquent,
l'extraction de la base $\mathcal{GBA}$ est faite avec perte
d'information~\cite{mariena02}. Afin de pallier ce problème,
Kryszkiewicz propose d'appliquer l'opérateur de couverture
$\mathcal{C}$ sur le couple ($\mathcal{GBE}$, $\mathcal{GBA}$).

\section{Discussion}

\begin{table}[!htbp]
 \centering
 \small{
\begin{tabular}{|c|c|c||c|c|c||c|}
\hline
\multirow{2}{*}{\textbf{Base}}                                & \multirow{2}{*}{\textbf{\begin{tabular}[c]{@{}c@{}}Forme des\\ règle\end{tabular}}}        & \multirow{2}{*}{\textbf{Informativité}} & \multicolumn{3}{c||}{\textbf{Dérivabilité}}                                                          & \multirow{2}{*}{\textbf{\begin{tabular}[c]{@{}c@{}}Type de règles\\ dérivées\end{tabular}}} \\ \cline{4-6}
                                                     &                                                                                   &                                & \textbf{\begin{tabular}[c]{@{}c@{}}Mécanisme\\ d'inférence\end{tabular}}        & \textbf{Complet} & \textbf{Correct} &                                                                                    \\ \hline\hline
$\mathcal{DG}^+$                                               & \begin{tabular}[c]{@{}c@{}}pseudo-fermé/\\ fermé\end{tabular}                     & non                            & $\mathcal{AA}$                                                                   & oui     & non     & \begin{tabular}[c]{@{}c@{}}$\mathcal{RA}$\\ exactes\end{tabular} \\ \hline
$\mathcal{BP}$                                                 & fermé/fermé                                                                       & non                            & $\mathcal{IRF}$                                                                  & oui     & non     & \begin{tabular}[c]{@{}c@{}}$\mathcal{RA}$\\ approximatives\end{tabular}                      \\ \hline
($\mathcal{BP}$, $\mathcal{DG}^+$)                                             & \begin{tabular}[c]{@{}c@{}}(pseudo-fermé/\\ fermé); (fermé/\\ fermé)\end{tabular} & oui                            & $\mathcal{IRF}$,$\mathcal{AA}$                                                                 & oui     & oui     & \begin{tabular}[c]{@{}c@{}}$\mathcal{RA}$\\ approximatives\end{tabular}                      \\ \hline
$\mathcal{RR}$                                                 & \begin{tabular}[c]{@{}c@{}}gen.min/\\ fermé\end{tabular}                          & non                            & $\mathcal{C}$                                                                    & oui     & oui     & $\mathcal{RA}$                                           \\ \hline
$\mathcal{RNR}^a$                                              & \begin{tabular}[c]{@{}c@{}}gen.min/\\ gen.min\end{tabular}                        & oui                            & \begin{tabular}[c]{@{}c@{}}Augmentation\\ $+\mathcal{TC}$\end{tabular}           & non     & oui     & $\mathcal{RA}$                                                                               \\ \hline
$\mathcal{BR}^*$                                               & \begin{tabular}[c]{@{}c@{}}gen.min/\\ fermé\end{tabular}                          & non                            & \begin{tabular}[c]{@{}c@{}}Augmentation\\ décomposition$^{**}$\end{tabular} & oui     & oui     & $\mathcal{RA}$                                                                               \\ \hline
$\mathcal{GBE}$                                                & \begin{tabular}[c]{@{}c@{}}gen.min/\\ fermé\end{tabular}                          & oui                            & $\mathcal{IRF}$                                                                  & oui     & oui     & \begin{tabular}[c]{@{}c@{}}$\mathcal{RA}$\\ exactes\end{tabular}                             \\ \hline
$\mathcal{GBA}$                                                & \begin{tabular}[c]{@{}c@{}}gen.min/\\ fermé\end{tabular}                          & non                            & $\mathcal{IRF}$                                                                  & oui     & non     & \begin{tabular}[c]{@{}c@{}}$\mathcal{RA}$\\ approximatives\end{tabular}                      \\ \hline
($\mathcal{GBE}$, $\mathcal{GBA}$) & \begin{tabular}[c]{@{}c@{}}gen.min/\\ fermé\end{tabular}                          & oui                            & $\mathcal{C}$                                                                    & oui     & oui     & $\mathcal{RA}$                                                                               \\ \hline
\end{tabular}

  }
  \caption{Tableau comparatif des bases génériques de règles associatives.}
  \label{comparaisonth}
  \small{($^{*}$): La base $\mathcal{BR}$ est équivalente à la base $\mathcal{RR}$\\
  ($^{**}$): L'application de ces axiomes revient à appliquer l'opérateur de couverture $\mathcal{C}$\\
  ($^{§}$): L'utilisation de l'opérateur de couverture $\mathcal{C}$ ne respecte pas la définition de redondance (\textit{c.f.,} définition~\ref{newdefred}, page~\pageref{newdefred})\\
  ($^{a}$): L'algorithme d'extraction de cette base n'est pas correct.}

\end{table}

Sur la base du tableau~\ref{comparaisonth} -- comparant les
différentes approches de formalisation de bases génériques-- nous
avons pu dégager les constatations suivantes:
\begin{enumerate}
\item Les règles des bases $\mathcal{RR}$, $\mathcal{BR}$ et du
couple ($\mathcal{GBE}$, $\mathcal{GBA}$) véhiculent le maximum
d'informations utiles. En effet, il a été prouvé que les règles
qui traduisent des implications entre les générateurs minimaux et
les itemsets fermés fréquents véhiculent le maximum d'informations
utiles, puisqu'elles possèdent une prémisse minimale et une
conclusion maximale~\cite{marzena022,thesepasq}; \item Les règles
génériques du couple ($\mathcal{DG}^{+}$ \footnote{La base
$\mathcal{DG}^{+}$ désigne la redéfinition de la base de
Guigues-Duquenne adaptée au cadre des règles associatives
exactes~\cite{thesepasq}.}, $\mathcal{BP}$\footnote{La base
$\mathcal{BP}$ désigne la redéfinition de la base de Luxemburger
adaptée au cadre des règles associatives
approximatives~\cite{thesepasq}.}) sont de formes différentes.
Ainsi, cette hétérogénéité peut entraîner une difficulté
supplémentaire pour interpréter les connaissances présentées à
l'utilisateur;

\item Seules les bases génériques $\mathcal{RNR}$,
$\mathcal{GBE}$, ($\mathcal{GBE}$,$\mathcal{GBA}$) et
($\mathcal{DG}^{+}$,$\mathcal{BP}$) sont qualifiées de bases
informatives; \item Seuls les couples
($\mathcal{GBE}$,$\mathcal{GBA}$) et
($\mathcal{DG}^{+}$,$\mathcal{BP}$) sont extraits sans perte
d'information. En effet, la non perte d'information est due, d'une
part au fait que ces deux couples sont informatifs et d'autre
part, au fait que les mécanismes d'inférence utilisés pour la
dérivation des règles associatives sont corrects et complets.

\end{enumerate}

\section {Contributions}

Nous avons pu relever, dans la section précédente, que seule
l'approche ($\mathcal{GBE}$,$\mathcal{GBA}$), tout en étant
homogène, permet de dériver, sans perte d'information, les règles
associatives. Cependant, deux lacunes surgissent:

\begin{itemize}
\item La définition de la redondance donnée par la
définition~\ref{newdefred} n'a pas été respectée lors de
l'utilisation de l'opérateur de couverture $\mathcal{C}$. En
effet, une règle $R': X' \Rightarrow (Y'-X)$ redondante par
rapport à une règle générique $R: X \Rightarrow (Y-X)$ doit avoir
les mêmes mesures de support et de confiance que $R$. Or, en
appliquant l'opérateur de couverture sur une règle générique $R: X
\Rightarrow (Y-X)$, toutes les règles $R': X' \Rightarrow
(Y'-X')$, ayant un support et une confiance supérieurs ou égaux à
ceux de $R$, sont aussi dérivées.

\item L'approche ($\mathcal{GBE}$,$\mathcal{GBA}$) souffre de la
génération d'un nombre important de règles surtout pour des
contextes d'extraction considérées denses. Ce constat est renforcé
par le fait que pour les contextes considérés comme épars,
l'extraction du couple ($\mathcal{GBE}$, $\mathcal{GBA}$)
n'apporte aucun gain en terme de compacité~\cite{Mastghada}.
\end{itemize}

Dans ce qui suit, nous proposons deux systèmes axiomatiques
corrects et complets servant comme mécanisme d'inférence à partir
du couple ($\mathcal{GBE}$,$\mathcal{GBA}$), et qui permettent de
dériver l'ensemble de toutes les règles valides~\cite{dawak04}.

\subsection{Proposition d'un nouveau système axiomatique pour le
couple ($\mathcal{GBE}$,$\mathcal{GBA}$)}

Le couple de bases génériques informatives
($\mathcal{GBE}$,$\mathcal{GBA}$) présente "\emph{les bordures
maximales}" supportant toutes les règles associatives dérivées. En
effet, une règle dérivée ne peut pas présenter une prémisse plus
petite que celle d'une règle générique, \textit{i.e.}, à partir de
laquelle elle peut être dérivée. En outre, une règle dérivée ne
peut pas présenter aussi une conclusion plus large que celle de la
règle générique associée. C'est pour cette raison que nous
remettons en cause l'inclusion large des conclusions donnée dans
la définition ~\ref{newdefred} de la
redondance~\cite{bastide2000}. Pour cela, nous avons redéfini la
notion de redondance comme suit:

\begin{definition}
\label{newdefredsad} Soit $\mathcal{AR}$ l'ensemble de règles
associatives pouvant être extrait du contexte d'extraction
$\mathcal{K}$. Une règle $R: X\overset{\alpha}{\Rightarrow}Y$
\footnote{Si Conf(R: X$\overset{\alpha}{\Rightarrow}$Y) =1, alors
$\alpha$ est omis, \textit{i.e.}, R: X$\Rightarrow$Y.} $\in$
$\mathcal{AR}$ est dite redondante par rapport à (dérivée à partir
de) la règle non redondante R$_1$:
X$_1$$\overset{\alpha}{\Rightarrow}$Y$_1$ si elle remplit les
conditions suivantes:
\begin{enumerate}
    \item Supp(R)=Supp(R$_1$) et Conf(R)=Conf(R$_1$)
    \item (X$_1$ $\subset$ X $\wedge$ Y $\subset$ Y$_1$) ou (X$_1$ = X $\wedge$ Y $\subset$ Y$_1$)
\end{enumerate}
\end{definition}

Dans ce qui suit, nous allons montrer comment dériver les règles
associatives à partir du couple de bases génériques
($\mathcal{GBE}$,$\mathcal{GBA}$), grâce à l'utilisation de deux
mécanismes d'inférence. Puisque, la prémisse d'une règle générique
est minimale, alors suite à l'augmentation de cette prémisse, il
est possible de dériver une autre règle associative. De plus,
comme la conclusion d'une règle générique est maximale, alors il
est possible de dériver une autre règle associative suite à une
décomposition de cette conclusion.

\subsubsection{Axiomatisation de la dérivation à partir de la base $\mathcal{GBE}$}
La proposition \ref{propgbe} ci-dessous introduit un ensemble
d'axiomes permettant la dérivation de toutes les règles
associatives exactes à partir de la base générique de règles
associatives exactes $\mathcal{GBE}$.
\begin{proposition} \label{propgbe}
Soient $\mathcal{ERB}$ et $\mathcal{EAR}$ respectivement
l'ensemble des règles génériques exactes et l'ensemble de toutes
les règles associatives exactes pouvant être extraits d'un
contexte d'extraction $\mathcal{K}$. Alors, les axiomes suivants
sont valides~\cite{dawak04}:
\begin{description}
    \item[A1.~Décomposition:]Si X$\Rightarrow$Y $\in$
    $\mathcal{ERB}$ alors X$\Rightarrow$Z $\in$
    $\mathcal{EAR}$, $\forall$ Z $\subset$ Y
    \item[A2.~Augmentation:]Si X$\Rightarrow$Y $\in$
    $\mathcal{ERB}$ alors X $\cup$ Z$\Rightarrow$Y-$\{$Z$\}$ $\in$
    $\mathcal{EAR}$, $\forall$ Z $\subset$ Y
\end{description}
\end{proposition}
%
%
%
%
%
%
%
%
%
%
%
%
%
%

\subsubsection{Axiomatisation de la dérivation de la base $\mathcal{GBA}$}
La proposition \ref{propgba} suivante caractérise un ensemble
d'axiomes permettant la dérivation de toutes les règles
approximatives à partir de la base générique $\mathcal{GBA}$.
\begin{proposition}
\label{propgba} Soient $\mathcal{ARB}$ et $\mathcal{AAR}$
respectivement l'ensemble des règles génériques approximatives et
l'ensemble de toutes les règles approximatives pouvant êtres
extraits à partir du contexte d'extraction
 $\mathcal{K}$.
Les axiomes d'inférence suivants sont valides~\cite{dawak04}:
\begin{description}

    \item[A1.~Décomposition:]Si X$\overset{\alpha}{\Rightarrow}$Y $\in$
    $\mathcal{ARB}$ alors X$\overset{\alpha}{\Rightarrow}$Z $\in$
    $\mathcal{AAR}$ si $|$XZ$|$=$|$XY$|$ avec Z $\subset$
    Y.
    \item[A2.~Augmentation]Si X$\overset{\alpha}{\Rightarrow}$Y $\in$
    $\mathcal{ARB}$ alors X $\cup$ Z$\overset{\alpha}{\Rightarrow}$Y-$\{$Z$\}$ $\in$
    $\mathcal{AAR}$ si $|$XZ$|$=$|$X$|$ avec Z $\subset$
    Y.
\end{description}
\end{proposition}

En outre, et au delà de la notion d'informativité et de compacité,
l'approche d'extraction du couple
($\mathcal{GBE}$,$\mathcal{GBA}$) proposée par Bastide \textit{et
al.} \cite{bastide2000} soulève un certain nombre de questions
d'ordre sémantique:
 \begin{enumerate}
\item  A-t-on atteint l'objectif de l'extraction des bases génériques,
à savoir la présentation d'un sous-ensemble réduit de règles
associatives exploitables par l'utilisateur?
\item Le couple ($\mathcal{GBE}$, $\mathcal{GBA}$) fournit-il vraiment à l'utilisateur la "bonne" information dont il a besoin?
 \end{enumerate}

Dans ce quit suit, nous allons essayer de porter quelques éléments
de réponse à ces interrogations, en proposant une nouvelle base
générique informative, et en revisitant la sémantique associée à
la notion de règle générique.
\subsection{Proposition d'une nouvelle base générique $\mathcal{IGB}$} Les
limites du couple ($\mathcal{GBE}$, $\mathcal{GBA}$) nous ont
motivés pour introduire une nouvelle base générique, appelée
$\mathcal{IGB}$, que nous définissons comme
suit~\cite{ghadacap05,pakdd05b}:
\begin{definition}
\label{def_1} Soient $\mathcal{IFF}_\mathcal{K}$ l'ensemble des
itemsets fermés fréquents et $\mathcal{G}_{c}$ l'ensemble des
générateurs minimaux de tous les itemsets fermés fréquents inclus
ou égaux dans un itemset fermé fréquent $I$.
La base $\mathcal{IGB}$ est définie comme suit:\\
 $\mathcal{IGB}$ = \{R : g$_s$ $\Rightarrow$ (I-g$_s$) $\mid$ I
$\in$ $\mathcal{IFF}_\mathcal{K}$ $\wedge$ I-g$_s$ $\neq$
$\emptyset$ $\wedge$ g$_{s}$ $\in$ $\mathcal{G}_{c}$, c $\in$
$\mathcal{IFF}_\mathcal{K}$ $\wedge$ c $\subseteq$ I $\wedge$
Conf(R) $\geq$ minconf $\wedge$ $\nexists$ $g'$ $\subset$ g$_s$
$\wedge$ Conf(g$'$ $\Rightarrow$ I-g$'$)$\geq$ minconf \}.
\end{definition}

Ainsi, les règles génériques de $\mathcal{IGB}$ représentent des
implications entre des prémisses minimales et des conclusions
maximales (selon l'ordre d'inclusion). En effet, il a été prouvé
dans la littérature que ce type de règles est le plus général
(\textit{i.e.,} convoyant le maximum
d'information)~\cite{marzena022,thesepasq}.
\begin{proposition}
\label{prop_1} Soit I un itemset fermé fréquent non vide. Si
Supp(I)$\geq$ minconf, alors nous pouvons extraire la règle
générique suivante: $\emptyset \Rightarrow$I~\cite{pakdd05b}.
\end{proposition}

\`{A} partir de la proposition~\ref{prop_1}, nous constatons que
les prémisses de certaines règles génériques de la base
$\mathcal{IGB}$ peuvent être vides. Pour cela, nous distinguons
deux types de règles génériques~\cite{pakdd05b}:
\begin{enumerate}
\item \textit{\textbf{Règles factuelles}}, qui ont une prémisse
vide. Bien que telles règles aient été considérées dans les
travaux de Kryszkiewicz~\cite{marzena981} et de
Luong~\cite{Phan01}, aucune interprétation sémantique n'a été
fournie pour ce type de connaissance. \item \textit{\textbf{Règles
implicatives}}, qui ont une prémisse non vide.
\end{enumerate}

Afin de montrer l'intérêt pratique de la base générique
introduite, supposons qu'un gestionnaire d'une surface commerciale
cherche à déterminer le lot d'items minimal pour arriver à pousser
les ventes des lots d'items qui se vendent actuellement avec une
première valeur, pour atteindre une autre valeur plus élevée. Sur
un exemple pratique et en utilisant les règles génériques du
couple ($\mathcal{GBE}$, $\mathcal{GBA}$), il a été montré que ces
règles n'apportent aucune plus value et ne sont donc d'aucune
utilité pour le gestionnaire~\cite{Mastghada}. Cependant, les
règles génériques de la base $\mathcal{IGB}$ rejoignent l'objectif
du gestionnaire, et ceci grâce aux raisons
suivantes~\cite{Mastghada}:
\begin{itemize}
\item d'une part, par la catégorisation des règles de
$\mathcal{IGB}$. En effet, \textit{les règles factuelles}
traduisent des faits (\textit{i.e.,} l'objectif matérialisé par le
seuil $minconf$ est déjà atteint). En d'autres termes, ces règles
contiennent les items sur lesquels aucune activité promotionnelle
n'est requise. En revanche, les règles implicatives traduisent des
conditions qu'il faut remplir afin d'atteindre l'objectif fixé par
l'utilisateur. \item d'autre part, l'exploitation de règles
comportant les prémisses minimales seulement.
\end{itemize}

Afin de dériver l'ensemble de toutes les règles associatives
valides, nous proposons, dans ce qui suit, un système axiomatique
et nous prouvons qu'il est \textbf{correct} et \textbf{complet}.

La correction du système axiomatique est donnée par la proposition
suivante :
\begin{proposition}
\label{valide} Soient $\mathcal{IGB}$ la base générique et
$\mathcal{AR}$ l'ensemble de toutes les règles valides pouvant
être extraites à partir d'un contexte d'extraction $\mathcal{K}$.
Le système axiomatique suivant est correct~\cite{pakdd05b}:
\begin{description}
\item[A0.~Réfléxivité conditionnelle:]Si $X
\overset{\alpha}{\Rightarrow} Y$ $\in$
    $\mathcal{IGB}$ et $X \neq \emptyset$ alors $X \overset{\alpha}{\Rightarrow} Y$ est valide.
\item[A1.~Augmentation:]Si $R:X \overset{\alpha}{\Rightarrow} Y$
$\in$
    $\mathcal{IGB}$ et $Z \subset Y$, alors $R':XZ \overset{\alpha'}{\Rightarrow} Y-Z$ est valide avec une confiance $\alpha'\geq \alpha$.
\item[A2.~Décomposition:]Soit $R:X \overset{\alpha}{\Rightarrow}
Y$ une règle valide obtenue suite à l'application de l'un des
axiomes de réflexivité conditionnelle ou d'augmentation sur l'une
des règles génériques de $\mathcal{IGB}$, $Z \subset Y$ et
$\omega(XZ)$ = XY. La règle $R':X \overset{\alpha}{\Rightarrow} Z$
est valide.

\end{description}
\end{proposition}

La complétude du système axiomatique, qui revient à montrer qu'il
permet de dériver \textbf{toutes} les règles valides pouvant être
extraites d'un contexte d'extraction, est donnée par la
proposition suivante~\cite{pakdd05b}.
\begin{proposition}
\label{complet} Toutes les règles pouvant être extraites du
contexte d'extraction $\mathcal{K}$ sont dérivables de
$\mathcal{IGB}$ moyennant le système axiomatique donné par la
proposition \ref{valide}.
\end{proposition}
Au-delà de la validité des règles dérivées, la proposition
suivante montre l'informativité de la base $\mathcal{IGB}$.
\begin{proposition}
La base $\mathcal{IGB}$ est informative, \textit{i.e.,} les
mesures de support et de confiance des règles redondantes peuvent
être déterminées avec exactitude \cite{ghadacap05}.
\end{proposition}

Comparons maintenant le nombre de règles de la base
$\mathcal{IGB}$ par rapport à celui des règles des différentes
bases génériques (\textit{i.e.}, $\mathcal{RR}$, ($\mathcal{GBE}$,
$\mathcal{GBA}$) et $\mathcal{RNR}$), puis les taux de compacité
des différentes bases génériques par rapport au nombre de toutes
les règles valides $\mathcal{AR}$ pouvant être extraites en
utilisant l'algorithme \textsc{Apriori}~\cite{Agra94} par exemple.
Le taux de compacité est mesuré comme suit: \texttt{comp$_{base} =
1-(\frac{taille~Base}{taille~\mathcal{AR}})$}.

\begin{itemize}
\item \textbf{Contextes épars}: Le tableau~\ref{nbreeparse} montre
l'évolution du nombre de règles des différentes bases génériques
lorsque nous considérons des contextes de test épars. Le
tableau~\ref{compacteparse} illustre le taux de compacité pour des
contextes considérés épars.

\begin{table}[!ht]
    \centering

        \begin{tabular}{|l|r|r|r|r|r||r|r|}
        \hline
        Base&$minconf$ ($\%$)&$\mathcal{RR}$&$\mathcal{IGB}$&$\mathcal{RNR}$&($\mathcal{GBE}$, $\mathcal{GBA}$)&$\mathcal{AR}$\\\hline

            \textsc{T10I4D100K}&0,50& 585& 1074&1231& 2216&2216\\
            &1,00&890&1595&1231& 2216&2216\\
            &10,00&497&1188&1105& 2172&2172\\
            &50,00&215&606&608& 1145&1145\\
            &100,00&0&0&0&0&0\\\hline
            \textsc{Retail}&00,50& 284& 580&861& 1382&1382\\
            &1,00&459&757&838&1334&1334\\
            &10,00&214&427&553&770&770\\
            &50,00&305&353&402&438&438\\
            &100,00&0&0&0&0&0\\\hline
            \textsc{Accidents}&70,00& 23&529&1911& 8226&8226\\
            &80,00&83&864&1822&6445&6445\\
            &90,00&50&588&1318&3407&3407\\
            &100,00&0&0&0&0&0\\\hline

        \end{tabular}
    \caption{Variation du nombre de règles associatives extraites de contextes considérés épars \textit{vs.} la variation de $minconf$}
    \label{nbreeparse}
\end{table}
\`{A} la lecture des tableaux~\ref{nbreeparse}
et~\ref{compacteparse}, nous constatons que:

\begin{itemize}
\item L'extraction du couple ($\mathcal{GBE}$, $\mathcal{GBA}$)
n'apporte aucun gain en terme de compacité. Ceci est dû au fait
que l'ensemble des itemsets (fermés) fréquents est confondu avec
celui des générateurs minimaux. Ainsi, construire des implications
entre les générateurs minimaux et les itemsets fermés fréquents
revient à construire des implications entre les itemsets
fréquents.

\item Pour $minconf = 100\%$, la cardinalité de toutes les bases
génériques est égale à 0. En effet, pour une telle valeur de
$minconf$, les bases $\mathcal{RR}$, $\mathcal{IGB}$ et
($\mathcal{GBE}$, $\mathcal{GBA}$) sont constituées de règles
représentant des implications entre les générateurs minimaux d'une
même classe d'équivalence et l'itemset fermé fréquent associé.
\'{E}tant donné que l'ensemble des itemsets fermés fréquents est
confondu avec celui des générateurs minimaux et que la prémisse et
la conclusion d'une règle doivent être disjointes, le nombre de
règles générées est alors égal à 0. Quant à la base
$\mathcal{RNR}$, les règles représentent des implications entre
les générateurs minimaux d'un itemset fermé fréquent $I$ et ceux
d'un itemset fermé fréquent $J$ prédécesseur immédiat de $I$.
Puisque, pour les contextes considérés épars, un générateur
minimal de $J$ est inclus dans celui de $I$, alors le nombre de
règles est lui aussi égal à 0. \item Les taux de compacité des
différentes bases génériques peuvent être ordonnés comme suit:
comp$_{\mathcal{RR}}$ $\leq$ comp$_{\mathcal{IGB}}$ $\leq$
comp$_{\mathcal{RNR}}$ $\leq$ comp$_{(\mathcal{GBE},
\mathcal{GBA})}$.

\end{itemize}

\begin{table}[!ht]
    \centering
        \begin{tabular}{|l|r|r|r|r|r|}
        \hline
        Base&$minconf$ ($\%$)&comp$_{\mathcal{RR}}$&comp$_{\mathcal{IGB}}$&comp$_{\mathcal{RNR}}$&comp$_{(\mathcal{GBE}, \mathcal{GBA})}$\\\hline
        \textsc{T10I4D100K}&0,50&73,61$\%$&51,54$\%$&44,45$\%$&00,00$\%$\\
            &1,00&59,84$\%$&28,03$\%$&44,45$\%$&00,00$\%$\\
            &10,00&77,12$\%$&45,41$\%$&49,13$\%$&00,00$\%$\\
            &50,00&81,23$\%$&47,08$\%$&46,90$\%$&00,00$\%$\\
            &100,00&00,00$\%$&00,00$\%$&00,00$\%$&00,00$\%$\\\hline
            \textsc{Retail}&0,50&79,46$\%$&58,04$\%$&37,7$\%$&00,00$\%$\\
            &1,00&65,60$\%$&43,26$\%$&37,18$\%$&00,00$\%$\\
            &10,00&72,21$\%$&44,55$\%$&28,19$\%$&00,00$\%$\\
            &50,00&30,37$\%$&19,41$\%$&8,22$\%$&00,00$\%$\\
            &100,00&00,00$\%$&00,00$\%$&00,00$\%$&00,00$\%$\\\hline
            \textsc{Accidents}&70,00&99,73$\%$&93,57$\%$&76,77$\%$&00,00$\%$\\
            &80,00&98,72$\%$&86,60$\%$&71,40$\%$&00,00$\%$\\
            &90,00&98,54$\%$&82,75$\%$&61,32$\%$&00,00$\%$\\
            &100,00&00,00$\%$&00,00$\%$&00,00$\%$&00,00$\%$\\\hline

        \end{tabular}
    \caption{Comparatif des taux de compacité des bases génériques extraites à partir de contextes considérés épars.}
    \label{compacteparse}
\end{table}

\item \textbf{Contextes denses}:  Le tableau~\ref{nbredense} met
en évidence l'évolution du nombre de règles des différentes bases
génériques lorsque nous considérons des contextes considérés
denses. Le tableau~\ref{compactdense} illustre le taux de
compacité des différentes bases génériques par rapport au nombre
de toutes les règles valides $\mathcal{AR}$ extraites à partir des
contextes considérés denses.

\begin{table}[!ht]
    \centering

        \begin{tabular}{|l|r|r|r|r|r||r|r|}
        \hline
        Base&$minconf ($\%$)$&$\mathcal{RR}$&$\mathcal{IGB}$&$\mathcal{RNR}$&($\mathcal{GBE}$, $\mathcal{GBA}$)&$\mathcal{AR}$\\\hline

            \textsc{Mushroom}&30&63& 427&1829&7623&94894\\
            &50&318&967&1732&5761&79437\\
            &70&429&966&1501&2159&58010\\
            &90&520&799&933&2159&24408\\
            &100&558&558&32&557&8450\\\hline
            \textsc{Chess}&87& 71& 1194&4873& 31538&42740\\
            &89&430&2423&4873& 29704&40451\\
            &91&486&2655&4799&26147&36098\\
            &93&616&2766&4734&21350&29866\\
            &95&768&2754&4260&14373&20312\\
            &97&467&1286&2857& 7695&10830\\
            &100&342&342&3&342&418\\\hline
            \textsc{Connect}&95& 97&809&3054&25336&77816\\
            &96&1003&2140&3054&18470&73869\\
            &97&1178&2438&3051&18470&60101\\
            &98&1404&2598&2804&11717&41138\\
            &99&1667&2284&2473&5250&19967\\
            &100&682&682&10&682&2260\\\hline
            \textsc{Pumsb}&90&260& 1466&8385& 36640&71488\\
            &92&1567&3696&8375&35142&68465\\
            &94&2572&5448&8304&29028&56349\\
            &96&258&5208&7466&18350&35013\\
            &98&1753&3340&5201&8014&14692\\
            &100&1118&1118&74&1118&1190\\\hline
        \end{tabular}
    \caption{Variation du nombre de règles associatives extraites à partir de contextes considérés denses \textit{vs.} la variation de $minconf$}
    \label{nbredense}
\end{table}

%
%

\begin{table}[!ht]
    \centering
        \begin{tabular}{|l|r|r|r|r|r|}
        \hline
        Base&$minconf ($\%$)$&comp$_{\mathcal{RR}}$&comp$_{\mathcal{IGB}}$&comp$_{\mathcal{RNR}}$&comp$_{(\mathcal{GBE}, \mathcal{GBA})}$\\\hline
            \textsc{Mushroom}&30&99,94$\%$&99,56$\%$&98,08$\%$&91,97$\%$\\
            &50&99,96$\%$&98,79$\%$&97,82$\%$&92,75$\%$\\
            &70&99,27$\%$&98,34$\%$&97,42$\%$&92,21$\%$\\
            &90&97,87$\%$&96,73$\%$&96,18$\%$&91,16$\%$\\
            &100&93,40$\%$&93,40$\%$&99,63$\%$&93,40$\%$\\\hline
            \textsc{Chess}&87&99,84$\%$&97,21$\%$&88,6$\%$&26,20$\%$\\
            &89&98,94$\%$&94,02$\%$&87,96$\%$&27,57$\%$\\
            &91&98,66$\%$&92,71$\%$&86,71$\%$&27,57$\%$\\
            &93&97,6$\%$&90,74$\%$&84,15$\%$&28,52$\%$\\
            &95&96,12$\%$&86,45$\%$&79,03$\%$&29,24$\%$\\
            &97&95,69$\%$&88,13$\%$&73,62$\%$&28,94$\%$\\
            &100&18,19$\%$&18,19$\%$&99,29$\%$&18,19$\%$\\\hline
            \textsc{Connect}&95&99,88$\%$&98,97$\%$&96,08$\%$&67,45$\%$\\
            &96&98,65$\%$&97,11$\%$&95,87$\%$&67,81$\%$\\
            &97&98,04$\%$&95,95$\%$&94,93$\%$&69,27$\%$\\
            &98&96,59$\%$&93,69$\%$&93,19$\%$&71,52$\%$\\
            &99&91,66$\%$&88,57$\%$&87,62$\%$&26,28$\%$\\
            &100&69,83$\%$&69,83$\%$&99,56$\%$&69,83$\%$\\\hline
            \textsc{Pumsb}&90&99,64$\%$&97,95$\%$&88,28$\%$&48,74$\%$\\
            &92&97,89$\%$&94,61$\%$&87,77$\%$&48,68$\%$\\
            &94&95,44$\%$&90,34$\%$&85,27$\%$&48,49$\%$\\
            &96&99,27$\%$&85,13$\%$&78,68$\%$&47,60$\%$\\
            &98&88,07$\%$&77,34$\%$&64,60$\%$&45,46$\%$\\
            &100&6,06$\%$&6,06$\%$&93,79$\%$&6,06$\%$\\\hline

        \end{tabular}
    \caption{Comparatif des taux de compacité des bases génériques pour des contextes considérés denses}
    \label{compactdense}
\end{table}

L'analyse des tableaux~\ref{nbredense} et~\ref{compactdense} nous
permet de relever ce qui suit:
\begin{itemize}

\item Pour $minconf = 100\%$, le nombre de règles des bases
$\mathcal{RR}$, $\mathcal{IGB}$ et ($\mathcal{GBE}$,
$\mathcal{GBA}$) est identique avec une singularité pour
($\mathcal{GBE}$, $\mathcal{GBA}$) que nous expliciterons par la
suite. Ceci est justifié par le fait que, pour cette valeur de
$minconf$, les bases sont constituées de règles représentant des
implications entre les générateurs minimaux et l'itemset fermé
fréquent d'une même classe d'équivalence. \item Pour le contexte
\textsc{Mushroom}, le nombre de règles du couple ($\mathcal{GBE}$,
$\mathcal{GBA}$) est différent de celui des bases $\mathcal{RR}$
et $\mathcal{IGB}$ lorsque $minconf = 100\%$. En effet, l'item
$'85'$ apparaît dans toutes les transactions; par conséquent, les
bases $\mathcal{RR}$ et $\mathcal{IGB}$ vont contenir la règle
$"\emptyset \Rightarrow 85"$. Cependant, cette règle ne peut pas
apparaître dans le couple ($\mathcal{GBE}$, $\mathcal{GBA}$)
puisque les règles ayant une prémisse vide ne sont pas prises en
considération.

\item Pour $minconf = 100\%$, le taux de  compacité de la base
$\mathcal{RNR}$ est très important. Ceci peut s'expliquer par le
fait que les règles génériques représentent des implications entre
les générateurs minimaux d'un itemset fermé fréquent $I$ et ceux
d'un itemset fermé fréquent $J$ prédécesseur immédiat de $I$.
Ainsi, le nombre de règles exactes est déterminé par l'ensemble
des générateurs minimaux de $J$ non inclus dans les générateurs
minimaux de $I$.

\item Pour $minconf < 100\%$, les taux de compacité des
différentes bases génériques sont ordonnés comme suit:
comp$_{\mathcal{RR}}$ $\leq$ comp$_{\mathcal{IGB}}$ $\leq$
comp$_{\mathcal{RNR}}$ $\leq$ comp$_{(\mathcal{GBE},
\mathcal{GBA})}$.

En revanche, pour $minconf = 100\%$, les taux sont classés comme
suit: compacité$_{\mathcal{RNR}}$ $\leq$
compacité$_{\mathcal{RR}}$ $\leq$ compacité$_{\mathcal{IGB}}$
$\leq$ compacité$_{(\mathcal{GBE}, \mathcal{GBA})}$.
\end{itemize}

\end{itemize}

Il est à noter que l'évolution du nombre de règles des différentes
bases génériques n'est pas déterminée par la nature du contexte.
En effet, le nombre de règles génériques évolue de la même manière
pour les contextes considérés denses ou épars.

\section{Conclusion}

Le développement d'algorithmes d'extraction d'itemsets fermés
fréquents de plus en plus efficaces a permis d'améliorer les
performances du processus d'exploration de bases de données très
volumineuses. Cependant, la taille de la connaissance présentée à
l'utilisateur, en terme du nombre de règles associatives, reste
prohibitive. Ceci a motivé des travaux pour la proposition
d'approches d'extraction de sous-ensembles réduits de règles,
appelés \textit{bases génériques}, contenant les règles qui
véhiculent le maximum d'informations utiles. Une étude critique
des différentes approches a permis de dégager le fait que seul le
couple de bases génériques ($\mathcal{GBE}$, $\mathcal{GBA}$),
proposé par Bastide \textit{et al.}, peut être qualifié d'approche
sans perte d'information. Toutefois, nous avons montré que ce
couple de bases ne peut pas constituer un outil d'aide à la
décision présentant une plus-value cognitive, \textit{i.e.} qu'il
contient des règles associatives qui n'ajoutent aucune
connaissance additionnelle pour l'utilisateur. Ainsi, comme
première contribution, nous avons proposé deux systèmes
axiomatiques corrects et complets qui peuvent servir de mécanismes
d'inférence à partir de ce couple de bases. Ensuite, nous avons
proposé une nouvelle base générique informative $\mathcal{IGB}$
permettant de déterminer exactement le support et la confiance des
règles dérivées via un système axiomatique correct et complet.
Contrairement à la catégorisation classique des règles
associatives (\textit{i.e.,} exactes et approximatives), nous
avons proposé deux nouveaux types de règles, factuelles et
implicatives, et nous avons montré leur intérêt dans le processus
de prise de décision. L'étude empirique menée sur des bases
benchmarks a fait ressortir que la base $\mathcal{IGB}$ apporte
d'importants gains en terme de compacité par rapport aux bases
génériques extraites sans perte d'information. Nous avons
également montré que l'exploitation des règles associatives
peuvent être améliorées, d'une part grâce à la compacité de la
base $\mathcal{IGB}$ et d'autre part grâce à la catégorisation des
règles génériques en \textit{règles factuelles} et \textit{règles
implicatives}.

\chapter{Extraction de bases génériques de règles associatives}
\section{Introduction}

Les approches classiques basées sur l'extraction des itemsets
fréquents font ressortir que le problème de la pertinence et de
l'utilité des règles extraites, est d'une importance capitale dans
un processus d'extraction des règles associatives. En effet, pour
un itemset fréquent $I$, le nombre de règles associatives non
triviales qui peuvent être générées est égal à (2$^{\vert I
\vert}$ - 1). Ainsi, dans la plupart des contextes d'extraction
réels, des milliers et même des millions de règles associatives
peuvent être générées \cite{stumme01,zaki2004}, dont une large
partie est redondante \cite{dawak04}. Pour remédier à ce problème
de redondance, les approches basées sur l'extraction des itemsets
fermés ont proposé une autre alternative en fournissant -- du
moins théoriquement -- les éléments nécessaires pour extraire des
bases génériques de règles associatives informatives et compactes.
Ces éléments sont \cite{isi04}: (\emph{i}) les itemsets fermés
fréquents; (\emph{ii}) la liste des générateurs minimaux associés
à chaque
    itemset fermé fréquent; et (\emph{iii}) la relation d'ordre entre les itemsets fermés
    fréquents.

Un survol critique des algorithmes fondés sur l'extraction des
itemsets fermés fréquents montre que la quasi-totalité de ces
algorithmes s'est focalisée sur l'extraction de la première
composante, \textit{i.e.}, les itemsets fermés fréquents, et
omettent les deux dernières \cite{isi04}. Seuls les algorithmes
adoptant la stratégie "\textit{Générer-et-tester}" font mieux en
permettant aussi l'extraction des générateurs minimaux, sans se
soucier de la construction de la relation d'ordre partiel. Par
conséquent, les algorithmes existants ne sont pas en mesure de
générer les bases génériques de règles associatives.

\section{Discussion}
\label{secDis-chap3}

Le problème de découverte des règles associatives, considéré sous
l'optique de la découverte des itemsets fermés, pourrait être
reformulé comme suit~\cite{isi04}:
\begin{enumerate}
    \item Découvrir deux "\textit{systèmes de fermeture}" distincts, \textit{i.e.}, des ensembles fermés sous l'opérateur d'intersection,
    à savoir l'ensemble des itemsets fermés et l'ensemble des
    générateurs minimaux. En outre, la relation d'ordre sous-jacente devrait être
    déterminée;
   \item Dériver des bases génériques de règles
   associatives à partir de ces deux "\textit{systèmes de fermeture}" et de la relation d'ordre
   sous-jacente.

   \end{enumerate}

\bigskip

\`{A} la lumière de cette reformulation et en tenant compte de
l'objectif de réduire le nombre de règles associatives, nous
considérons les caractéristiques (ou dimensions) suivantes
permettant de d\'{e}terminer les buts r\'{e}alis\'{e}s
(construction de la relation d'ordre) et de montrer les
diff\'{e}rences majeures qui pourraient exister entre les
algorithmes d'extraction des itemsets fermés (stratégie adoptée,
éléments générés, etc)~\cite{tareksigkdd06} :

1. \textbf{Stratégie d'exploration}: trois stratégies principales
sont utilisées afin d'explorer l'espace de recherche:
"\textit{Générer-et-tester}", "\textit{Diviser-pour-générer}" et
une stratégie \textit{hybride} qui combine les deux précédentes;

2. \textbf{Espace de recherche}: la valeur "exhaustive" de cette
caractéristique indique que toutes les transactions seront prises
en compte lors du premier balayage. En revanche, la valeur
"échantillon" signifie que seul un échantillon du contexte
d'extraction sera considéré du moins pour un premier balayage du
contexte d'extraction~\cite{Toi96};

3. \textbf{Type du format des donn\'{e}es}: cette caractéristique
indique le type des données prises en entrée de l'algorithme. Elle
peut prendre comme valeurs: format horizontal (étendu), format de
vertical, relationnel, texte brut, données multimédia;

4. \textbf{Stockage des informations}: différentes structures de
données peuvent être utilisées pour stocker les informations
nécessaires à l'exécution d'un algorithme (pour stocker les
candidats, le contexte d'extraction, etc). La structure la plus
privilégiée semble être la structure \textit{trie}, qui est  un
arbre de recherche, dont les éléments sont stockés d'une manière
condensée;

5. \textbf{Choix des générateurs} : cette caractéristique indique
si les générateurs sont choisis parmi ceux considérés comme
minimaux;

6. \textbf{Calcul de la fermeture}: nous distinguons
principalement deux manières pour calculer la fermeture d'un
générateur (minimal): soit \textit{via} des intersections des
transactions auxquelles il appartient (\textit{i.e.}, son
extension), soit d'une manière \textit{incrémentale},
\textit{i.e.}, \textit{via} la recherche des items appartenant à
la fermeture en question. Nous distinguons aussi le fait que le
calcul des fermetures est réalisé sous-ensemble par sous-ensemble
(calcul partiel) ou non (calcul global);

7. \textbf{Calcul redondant des fermetures} : étant donné la
propriété anti-matroide des classes d'équivalence, \textit{i.e.},
un itemset fermé fréquent peut avoir plus qu'un seul générateur
minimal, il peut être calculé plus qu'une fois sauf si des tests
sont effectués. Aussi, cette caractéristique précise s'il y a une
possibilité de calcul redondant des fermetures. Dans le cas
contraire, elle montre comment ce calcul redondant peut être
évité;

8. \textbf{Type des règles associatives générées en sortie} :
cette caractéristique fixe le type des règles associatives
dérivées : \textit{e.g.}, booléen, flou, spatial, temporel,
généralisé, etc;

9. \textbf{Ordre partiel}:  cette caractéristique indique si la
relation d'ordre sous-jacente des itemsets fermés fréquents a
\'{e}t\'{e} ou non d\'{e}termin\'{e}e;

10. \textbf{El\'{e}ments g\'{e}n\'{e}r\'{e}s en sortie}: grâce à
cette caract\'{e}ristique, nous pouvons préciser les connaissances
extraites par un algorithme donné. La valeur "RAs génériques"
signifie que seul un sous-ensemble, non-redondant et sans perte
d'information, de toutes les règles a été dérivé. Par contre, la
valeur "RAs redondantes" indique que seules toutes les règles
associatives redondantes peuvent être dérivées des él\'{e}ments
g\'{e}n\'{e}r\'{e}s en sortie. \medskip

Le tableau~\ref{tab1} résume les principales caract\'{e}ristiques
des algorithmes de g\'{e}n\'{e}ration d'itemsets ferm\'{e}s
fr\'{e}quents par rapport aux caractéristiques pr\'{e}sent\'{e}es
ci-dessus. Une première constatation qui se dégage de ce tableau
est qu'aucun algorithme n'est arrivé à tenir les promesses
avancées par cette nouvelle alternative d'algorithmes en matière
de réduction de la redondance des règles associatives. En effet,
la construction de la relation d'ordre est une condition
\textit{sine qua non} pour l'obtention de la base générique des
règles associatives approximatives \cite{isi04}.

Une deuxième constatation concerne la notion de générateur
minimal. En effet, malgré l'importance de cette dernière dans les
bases génériques, étant donné qu'elle permet d'avoir des prémisses
minimales, seuls les algorithmes adoptant la stratégie
"\textit{Générer-et-tester}" se sont intéressés à  l'extraction
des générateurs minimaux (\textit{cf.} dernière ligne du tableau
\ref{tab1}). Pour ces algorithmes, l'obtention des règles
génériques exactes peut se faire d'une manière directe, étant
donné qu'ils extraient les générateurs minimaux fréquents. Pour
pouvoir extraire les règles génériques approximatives, ils
nécessitent l'exécution en aval d'un algorithme permettant de
construire la relation d'ordre partiel, tel que celui proposé par
Valtchev \textit{et al.} \cite{petko00}. Les algorithmes,
appartenant aux trois autres catégories, se sont focalisés sur
l'énumération des itemsets fermés fréquents. Pour obtenir les
bases génériques de règles associatives, ces algorithmes
nécessitent d'être couplés avec deux autres algorithmes. Le
premier permettra de construire l'ordre partiel et le second aura
pour objectif de retrouver les générateurs minimaux fréquents tel
que l'algorithme \textsc{Jen} \cite{jen03}. Ce dernier permet de
retrouver les générateurs minimaux à condition que l'ordre partiel
soit déjà construit.

\begin{table}[htbp]
\begin{center}

 \small{
    \begin{tabular}
{|p{95pt}|p{90pt}|p{90pt}|p{90pt}|p{90pt}|} \hline &
\textbf{\textsc{1$^{\grave{e}re}$ catégorie}}&
\textbf{\textsc{2$^{\grave{e}me}$ catégorie}}&
\textbf{\textsc{3$^{\grave{e}me}$ catégorie}}& \textbf{\textsc{4$^{\grave{e}me}$ catégorie}}\\
\hline \textbf{Stratégie
d'exploration}&Générer-et-tester&Diviser-pour-régner&Hybride&Hybride\\\hline
\textbf{Espace de recherche}
&exhaustive&exhaustive&exhaustive&exhaustive\\\hline \textbf{Type
du format des donn\'{e}es} &horizontal&horizontal&horizontal,
vertical&horizontal, vertical\\\hline \textbf{Stockage des
informations} &\textit{trie}&\textit{trie}&\textit{trie}&matrice
de bits pour \textsc{DCI-Closed} et de simples tableaux pour
\textsc{LCM}\\\hline \textbf{Choix des générateurs} &générateurs
minimaux &générateurs&générateurs&générateurs\\\hline
\textbf{Calcul de la fermeture} &intersections des transactions,
global pour \textsc{Close} et \textsc{Titanic} et  partiel pour
\textsc{A-Close} & calcul incrémental et partiel &calcul
incrémental et partiel&calcul incrémental et partiel\\\hline
\textbf{Calcul redondant des fermetures} &oui &non, avec des tests
de couverture&non, avec des tests de couverture&non, avec des
tests préservant l'ordre total\\\hline \textbf{Type des règles
associatives} &booléen &booléen&booléen&booléen\\\hline
\textbf{Ordre partiel} &non &non&non&non\\\hline
\textbf{El\'{e}ments g\'{e}n\'{e}r\'{e}s en sortie} &itemsets
fermés fréquents, leurs générateurs minimaux associés et les RAs
redondantes &itemsets fermés fréquents et les RAs redondantes
&itemsets fermés fréquents et les RAs redondantes&itemsets fermés
fréquents et les RAs redondantes\\\hline
\end{tabular}
}

 \caption{Tableau comparatif des caractéristiques des quatre catégories d'approches
d'extraction des itemsets fermés fréquents.} \label{tab1}
\end{center}
\end{table}

Une troisième constatation concerne le coût induit par le calcul
de la fermeture. En effet, bien que ces algorithmes constituaient
une alternative intéressante par rapport aux algorithmes
d'énumération des itemsets fréquents dans le cas de contextes
considérés denses, leurs performances sont faibles pour les
contextes considérés épars. Ceci se justifie par le fait que, dans
ce type de contexte, l'espace de recherche des itemsets fermés
fréquents tend à se confondre avec celui des itemsets fréquents.

Une étude approfondie de ces algorithmes nous a permis de dégager
les remarques suivantes :
\begin{itemize}
    \item Seuls les algorithmes \textsc{Titanic}, \textsc{DCI-Closed} et \textsc{LCM} considèrent la classe d'équivalence dont l'ensemble vide est le générateur
    minimal. Ainsi, ce sont les seuls algorithmes dont les éléments générés
    permettent de construire un \textit{Iceberg du treillis de Galois} en utilisant par
    exemple l'algorithme proposé par Valtchev \textit{et
    al.}~\cite{petko00};

\item Les stratégies d'élagage adoptées par \textsc{Titanic} sont
une amélioration de celles utilisées dans \textsc{A-Close}, grâce
à l'utilisation du support estimé qui permet d'éviter des parcours
coûteux de l'ensemble des générateurs minimaux déjà extraits.
Celles de \textsc{Closet} et de \textsc{Charm} (resp. de
\textsc{DCI-Closed} et de \textsc{LCM}) peuvent être considérées
comme étant les mêmes;

\item \textsc{Close}, \textsc{A-Close} et \textsc{Titanic} ont
pour inconvénient de calculer la même fermeture plusieurs fois,
dans le cas où la classe d'équivalence associée admet plusieurs
générateurs minimaux. Cette faiblesse est évitée par
\textsc{Closet} et \textsc{Charm} grâce aux tests de couverture.
Cependant, afin d'accélérer ces tests, \textsc{Closet} et
\textsc{Charm} sont contraints de maintenir, en mémoire centrale,
l'ensemble des itemsets fermés fréquents trouvés.
\textsc{DCI-Closed} et \textsc{LCM} sont alors principalement
dédiés à résoudre ce problème. En effet, ils essaient d'extraire
les itemsets fermés fréquents en un temps linéaire et sans les
maintenir en mémoire centrale en utilisant un test préservant
l'ordre total adopté.
\end{itemize}


\section {Contributions}

\`{A} la lumière des insuffisances que nous avons pu déceler dans
la section \ref{secDis-chap3}, nous allons présenter deux
nouvelles approches visant à extraire toute l'information
nécessaire pour la génération des bases génériques de règles
associatives.

\subsection{Une nouvelle approche pour la construction de structures partiellement ordonnées}

Dans ce qui suit, nous allons introduire une nouvelle approche
pour l'extraction des bases g\'{e}n\'{e}riques de r\`{e}gles
associatives, dont les propri\'{e}t\'{e}s principales
sont~\cite{egc2004}:
\begin{itemize}
 \item \'{E}viter les op\'{e}rations d'entrée/sortie intensives: pour cela, nous introduisons une nouvelle structure
   de donn\'{e}es compacte, appel\'{e}e \emph{\textsc{Itemset-Trie}}, pour stocker un contexte d'extraction.
\item L'usage de la stratégie "Diviser-pour-g\'{e}n\'{e}rer" pour
r\'{e}duire le co\^{u}t de la construction de l'iceberg du
treillis de Galois. En utilisant cette stratégie, nous sommes en
mesure de construire, en parall\`{e}le, des sous-structures
partiellement ordonn\'{e}es. Ensuite, ces sous-structures sont
parcourues pour d\'{e}river, d'une mani\`{e}re directe, des bases
g\'{e}n\'{e}riques locales de r\`{e}gles associatives. Finalement,
ces bases g\'{e}n\'{e}riques locales sont fusionnées pour
constituer une base g\'{e}n\'{e}rique globale.
\end{itemize}

L'arbre \textsc{Itemset-Trie}~\cite{egc2004}, que nous avons
proposé, est une extension de l'id\'{e}e propos\'{e}e par les
auteurs de l'arbre \textsc{FP-Tree}~\cite{han00} et
\textsc{Cats}~\cite{cats03}, qui ont cherch\'{e} \`{a} réduire le
co\^{u}t de stockage en proposant une structure de donn\'{e}es
compacte. Ainsi, cette structure pourrait \^{e}tre charg\'{e}e
totalement en m\'{e}moire centrale et permettre ainsi la recherche
d'itemsets (fermés) fr\'{e}quents en minimisant la
g\'{e}n\'{e}ration d'itemsets candidats.

L'algorithme de construction de l'\textsc{Itemset-Trie}, que nous
avons développé, parcourt un contexte d'extraction transaction par
transaction. Chaque transaction du contexte est ins\'{e}r\'{e}e
dans l'\textsc{Itemset-Trie}, en respectant la r\`{e}gle li\'{e}e
au partage d'un item en pr\'{e}fixe. Pour cela, nous comparons
deux listes \`{a} chaque fois: la liste de la transaction courante
et la liste du noeud courant dans l'\textsc{Itemset-Trie}. S'il y
a un item commun entre ces deux listes, nous le laissons dans le
noeud courant et nous ins\'{e}rons le reste de chaque liste en
tant que fils de ce noeud. Dans le cas contraire, nous
ins\'{e}rons le reste dans les fr\`{e}res de ce noeud. Le
traitement se r\'{e}p\`{e}te jusqu'\`{a} la fin des transactions
composant le contexte.

\begin{figure}[htbp]
    \centering

\includegraphics[scale=0.5]{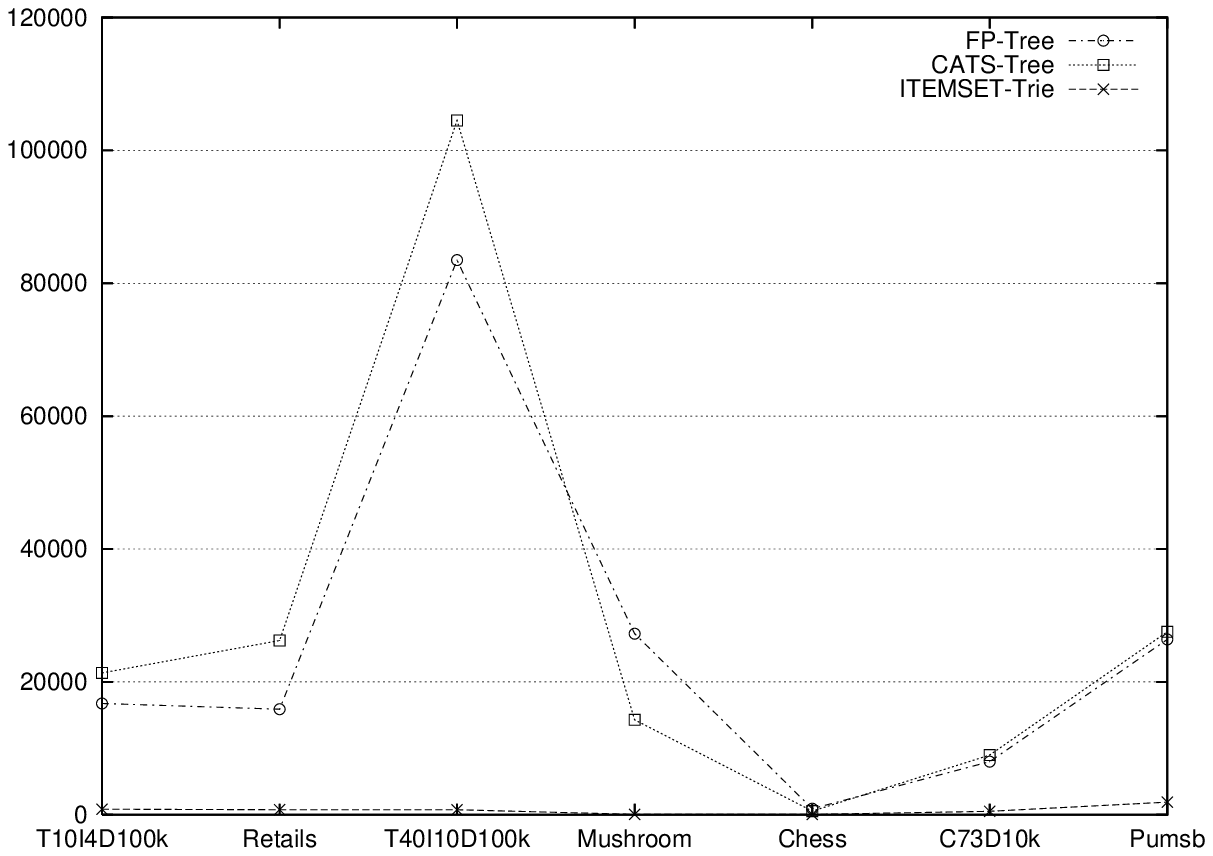}
\includegraphics[scale=0.5]{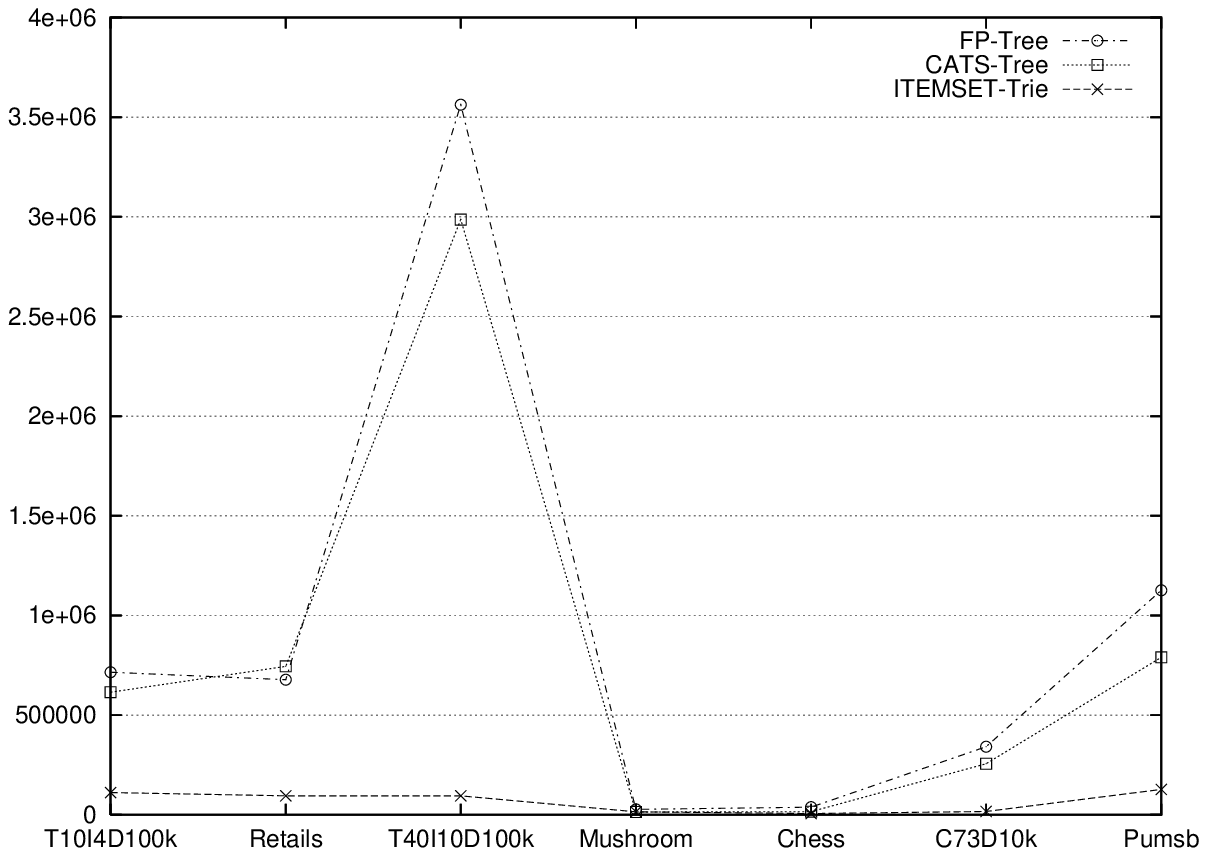}
\caption{\small{(\textbf{Gauche}) Nombre de noeuds
l'\textsc{Itemset-Trie} \emph{vs.} ceux respectivement de
\textsc{FP-tree} et \textsc{Cats}. (\textbf{Droite}) Taille de
l'\textsc{Itemset-Trie} \emph{vs.} celles, respectivement, de
\textsc{FP-tree} et \textsc{Cats}.}}

    \label{fig:size}
\end{figure}

La complexité de la construction de l'\textsc{Itemset-Trie} est
lin\'{e}aire en fonction du nombre de transactions
\cite{pakdd05b}. 
L'étude expérimentale de la structure proposée, en matière de
réduction de l'espace de stockage en mémoire centrale, a donné des
résultats encourageants~\cite{jihen041}. En effet, à la lecture de
la figure \ref{fig:size} (Droite), nous remarquons que la
structure \textsc{itemset-trie} est plus compacte que les
structures de données \textsc{FP-tree} et \textsc{Cats}. En effet,
cette compacité s'illustre sur le double aspect du nombre de
noeuds et la quantité de mémoire centrale nécessaire pour stocker
un contexte d'extraction\footnote{Toutes ces structures ont été
construites pour \textit{minsup}=1.}. Ainsi, les contextes
considérés épars sont au plus représentés par 24,4\% de leur
taille réelle, tandis que les contextes considérés denses sont au
plus représentés par 18,8\% de leur taille réelle. La figure
\ref{fig:size} met en exergue que ces performances sont fortement
corrélées au nombre de niveaux de la structure, qui lui même est
intimement lié à la densité du contexte. Ainsi, plus le contexte
est dense, plus le nombre de niveaux a tendance à croître (par
exemple, le nombre de niveaux de la base \textsc{Chess} atteint
25)~\cite{jihen041}.

Une fois que la structure \textsc{Itemset-Trie} est construite,
l'algorithme que nous avons développé, appel\'{e}
\textsc{Sub-Trie}~\cite{egc2004}, permet de construire les
sous-structures ordonn\'{e}es locales. Ensuite, il fusionne les
bases génériques locales -- extraites à partir des sous-structures
locales -- pour d\'{e}river une base g\'{e}n\'{e}rique globale de
r\`{e}gles associatives. L'algorithme, opérant selon la stratégie
"Diviser-pour-régner", permet la d\'{e}couverte des itemsets
ferm\'{e}s fr\'{e}quents avec leurs g\'{e}n\'{e}rateurs minimaux
associ\'{e}s, ainsi que la relation d'ordre sous-jacente.
L'id\'{e}e cl\'{e} est de construire des sous-structures
partiellement ordonn\'{e}es pour chaque 1-itemset fr\'{e}quent.
Ensuite, ces sous-structures, partiellement ordonn\'{e}es, sont
explor\'{e}es pour d\'{e}river, d'une mani\`{e}re directe, des
bases g\'{e}n\'{e}riques locales de r\`{e}gles associatives.
Finalement, les bases g\'{e}n\'{e}riques locales sont
regroup\'{e}es pour g\'{e}n\'{e}rer une base g\'{e}n\'{e}rique
globale~\cite{pakdd05a}.


Partant de l'objectif fixé, les itemsets ferm\'{e}s fr\'{e}quents
sont à extraire ainsi que leurs g\'{e}n\'{e}rateurs minimaux
associ\'{e}s pour étiqueter les noeuds de l'Iceberg du treillis de
Galois. Ainsi, contrairement aux algorithmes
\textsc{FP-Growth}~\cite{han00} et
\textsc{Closet}~\cite{closet00}, nous supposons que les items sont
uniquement ordonnés selon l'ordre lexicographique. En outre, un
item d\'{e}j\`{a} trait\'{e} ne va pas \^{e}tre exclu du
traitement, sous peine de ne pas être en mesure de déterminer
l'ensemble de tous les g\'{e}n\'{e}rateurs minimaux associ\'{e}s
\`{a} un itemset ferm\'{e} donn\'{e}.


Pour mesurer le passage à l'échelle de l'algorithme
\textsc{Sub-Trie}, nous avons mesuré les performances obtenues en
faisant varier le nombre de transactions par des incréments de
taille fixe (\textit{e.g.}, 1000 transactions pour la base
\textsc{T10I4D100K}). Les performances obtenues sont linéaires en
fonction du nombre de transactions et cette augmentation dans les
temps d'exécution a tendance à stagner au delà d'un certain seuil.
Pour les contextes denses, et à cause de contraintes matérielles,
nous avons fait varier le nombre d'items. Dans ce cas, les
performances dépendent linéairement de ce paramètre. La densité du
contexte est aussi un facteur non négligeable dans les
performances. Par exemple, les temps d'exécution obtenus pour la
base \textsc{C73D10K} sont moins importants que ceux de la base
\textsc{Chess}. Le facteur densité, matérialisé par le nombre de
niveaux de la structure \textsc{Itemset-Trie}, explique à lui seul
cette différence, \textit{e.g.}, 15 niveaux pour la base
\textsc{C73D10K} \textit{vs.} 25 niveaux pour la base
\textsc{Chess}~\cite{jihen041}.

\subsection {L'algorithme \textsc{Prince}}
L'objectif principal visé en proposant un nouvel algorithme pour
l'extraction de bases génériques de règles associatives, appelé
\textsc{Prince}, est de pallier la principale lacune des
algorithmes dédiés à l'extraction des itemsets fermés fréquents,
\textit{i.e.}, ne pas construire la relation d'ordre partiel.
L'originalité de l'algorithme \textsc{Prince} est qu'il construit
le \textit{treillis des générateurs minimaux} \cite{egc2003}, et
met en place un mécanisme de gestion des classes d'équivalence
permettant de générer la liste intégrale des itemsets fermés
fréquents sans duplication et sans recours aux tests de
couverture. Il permet aussi de réduire, d'une manière notable, le
coût du calcul des fermetures en utilisant les notions de
\textit{bloqueur minimal} et de \textit{face} introduites par
Pfaltz and Taylor \cite{pfatz02}.

\bigskip

L'algorithme \textsc{Prince} prend en entrée un contexte
d'extraction $\mathcal{K}$, le seuil minimal de support
\textit{minsup} et le seuil minimal de confiance \textit{minconf}.
Il donne en sortie la liste des itemsets fermés fréquents, leurs
générateurs minimaux associés ainsi que les bases génériques de
règles. \`{A} cette fin, il opère en trois étapes
successives~\cite{Masttarek}:
\begin{enumerate}
    \item \textbf{Détermination des générateurs minimaux}: en adoptant la stratégie "\textit{Générer-et-tester}",
l'algorithme \textsc{Prince} parcourt l'espace de recherche par
niveau pour déterminer l'ensemble des générateurs minimaux
fréquents, noté $\mathcal{GMF}_{\mathcal{K}}$, ainsi que la
bordure négative des générateurs minimaux, notée
$\mathcal{GB}$d$^{-}$. Il utilise, dans cette étape, les mêmes
stratégies d'élagage que \textsc{Titanic}, \textit{i.e.},
\textit{minsup}, l'idéal d'ordre  régissant des générateurs
minimaux fréquents et le support estimé.

    \item \textbf{Construction du treillis des générateurs
    minimaux}: l'objectif de cette étape est d'organiser les générateurs minimaux
fréquents sous forme d'un \textit{treillis} \textit{sans effectuer
un accès supplémentaire au contexte d'extraction}. Pour atteindre
cet objectif, les listes des successeurs immédiats seront
modifiées d'une manière itérative. Ainsi, nous parcourons
l'ensemble trié $\mathcal{GMF}_{\mathcal{K}}$ en introduisant un
par un les générateurs minimaux fréquents dans le \textit{treillis
des générateurs minimaux}. Chaque générateur minimal fréquent $g$
de taille $k$ $(k\geq1)$ est introduit dans le \textit{treillis
des générateurs minimaux} en le comparant avec les successeurs
immédiats de ses sous-ensembles de taille $(k$-$1)$. La
détermination de la relation d'ordre entre deux générateurs est
effectuée en utilisant la proposition suivante :
\begin{proposition}\label{proprelation}\cite{tarekdawak05}
Soient deux générateurs minimaux $X$, $Y$ $\in$
$\mathcal{GMF_{\mathcal{K}}}$ et $\mathcal{C}_{X}$ et
$\mathcal{C}_{Y}$ leurs classes d'équivalence
respectives:\\
(a). Si Supp($X$) = Supp($Y$) = Supp($X$ $ \cup $ $Y$) alors $X$
et $Y$ appartiennent à la même classe d'équivalence.\\(b). Si
Supp($X$) $<$ Supp($Y$) et Supp($X$) = Supp($X$ $ \cup $ $Y$)
alors $\mathcal{C}_{X}$ (resp. $\mathcal{C}_{Y}$) est un
successeur (resp. pr\'{e}d\'{e}cesseur) de $\mathcal{C}_{Y}$
(resp. $\mathcal{C}_{X}$).
\end{proposition}

\item \textbf{Extraction des bases génériques informatives de
règles}: Dans cette étape, l'algorithme \textsc{Prince} extrait
les bases génériques de règles associatives. Dans son implantation
actuelle, il est en mesure d'extraire la base $\mathcal{IGB}$, ou
conjointement la base $\mathcal{GBE}$ et la réduction transitive
de la base $\mathcal{GBA}$. Pour cela, il retrouve les itemsets
fermés fréquents. Cet objectif est facilement atteint grâce à la
proposition \ref{propiff}, qui indique comment retrouver, pour
chaque classe d'équivalence, l'itemset fermé fréquent associé.
\begin{proposition} \label{propiff}\cite{tarekdawak05} Soient
deux itemsets fermés fréquents $f$ et $f_{_{1}}$ $\in$
$\mathcal{IFF}_{\mathcal{K}}$ tels que $f \prec$ $f_{_{1}}$ dans
l'Iceberg du treillis de Galois. Soit $GM_{f}$ l'ensemble des
générateurs minimaux de $f$. Alors, l'itemset fermé fréquent $f$
est égal à: $f$ = $\cup \{g | g \in GM_{f}\} \cup f_{_{1}}$.
\end{proposition}

\end{enumerate}

\begin{figure}
\begin{center}
\parbox{3.cm}{
\small{
  \begin{tabular}{|c|c|c|c|c|c|}
  \hline
   & A & B & C & D & E \\\hline
  1 & $\times$ &  & $\times$ & $\times$ &  \\\hline
  2 &  & $\times$ & $\times$ & & $\times$ \\\hline
  3 & $\times$ & $\times$ & $\times$ &  & $\times$ \\\hline
  4 &  & $\times$ &  &  & $\times$ \\\hline
  5 & $\times$ & $\times$ & $\times$ &  & $\times$ \\\hline
\end{tabular}
  }}
\caption{Contexte d'extraction $\mathcal{K}$} \label{runfig}

\bigskip

\parbox{5cm}{
\small{
\includegraphics[scale=.6]{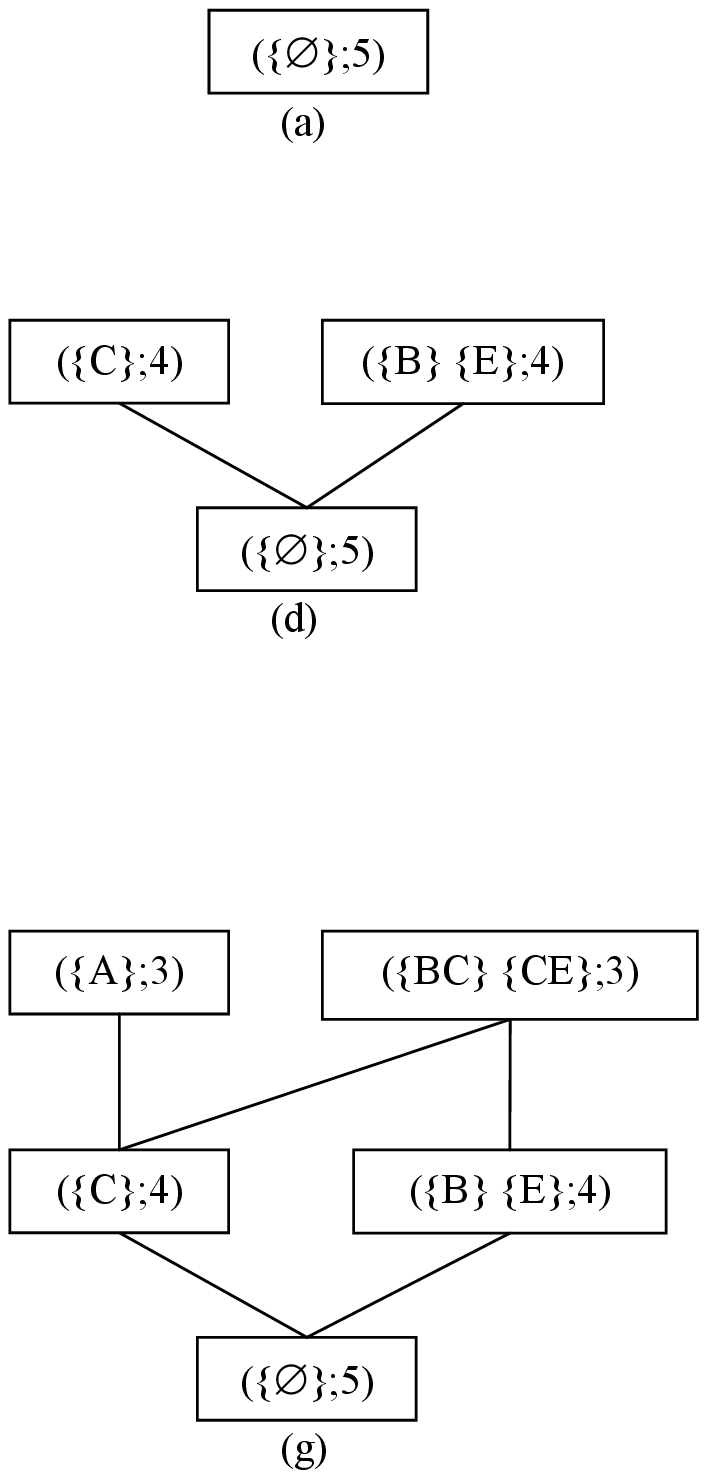}}}
\hspace{0cm}
\parbox{5cm}{
\small{
\includegraphics[scale=.6]{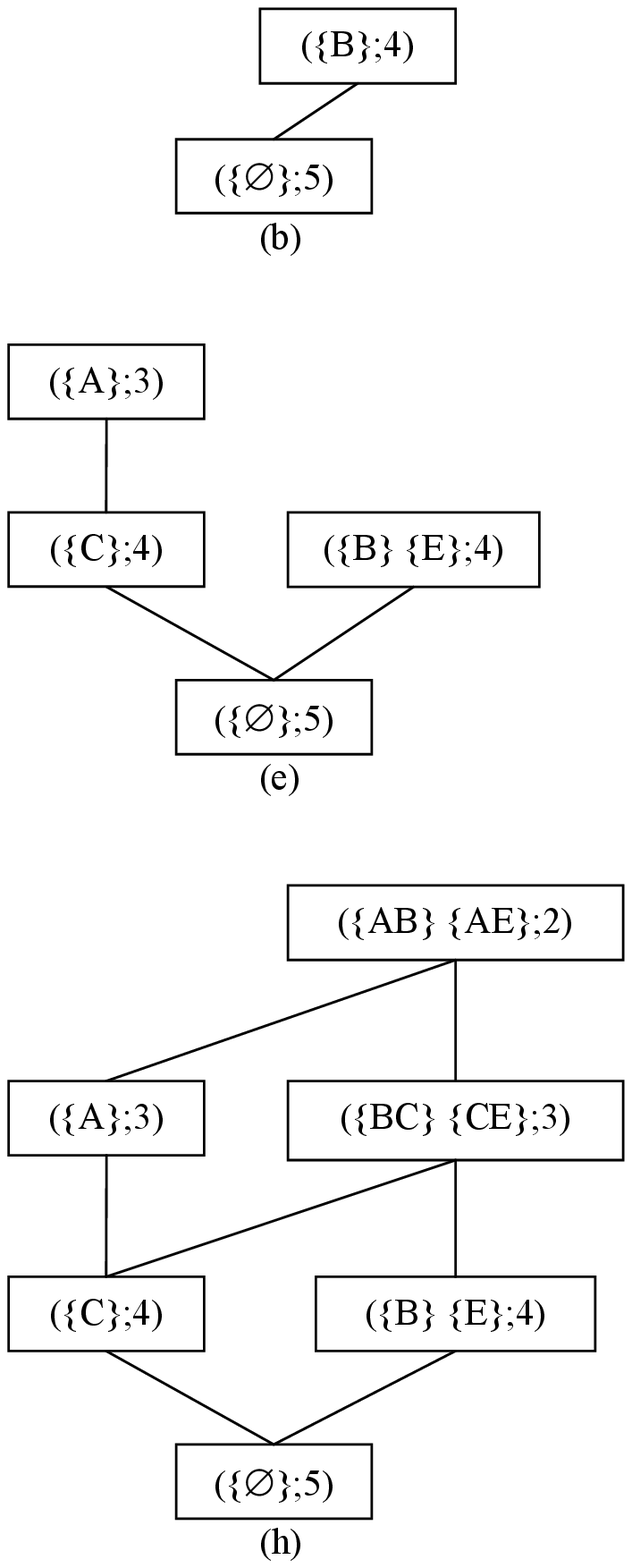}}}
\parbox{5cm}{
\small{
\includegraphics[scale=.6]{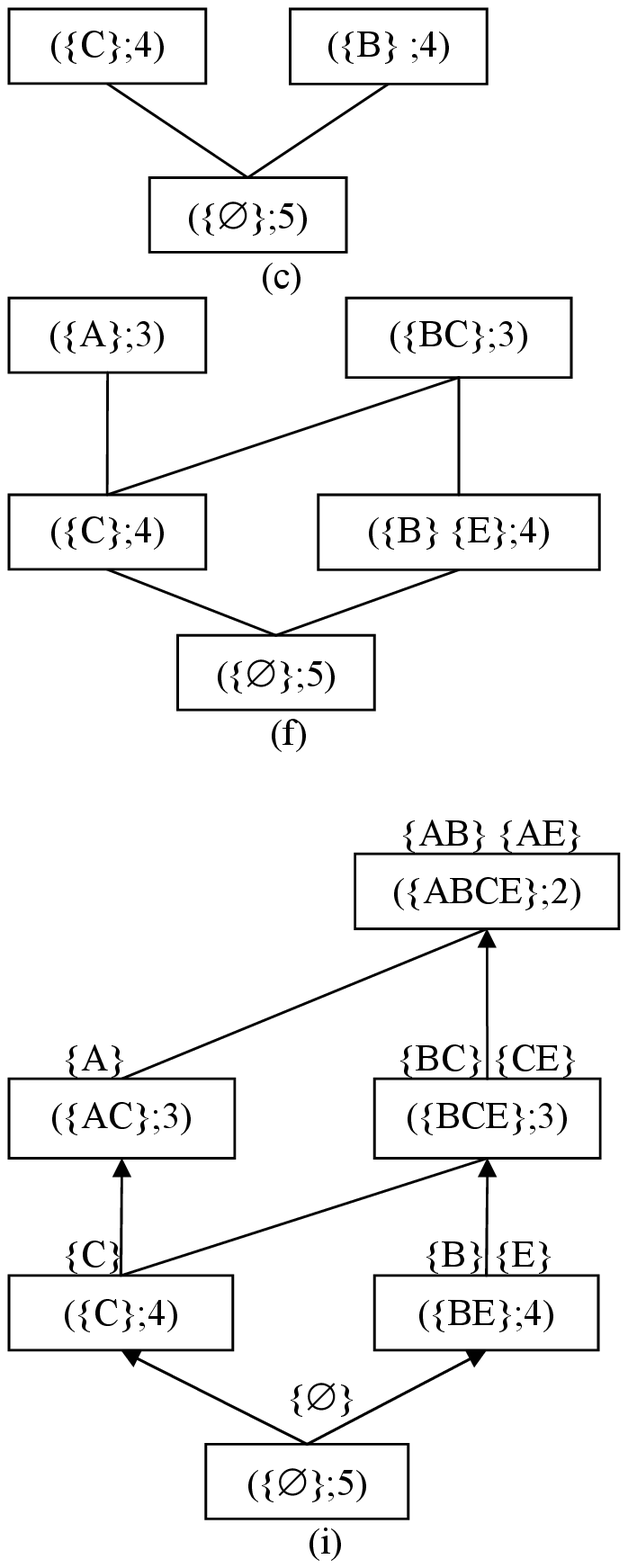}}}
\caption{Construction du \textit{treillis des générateurs
minimaux} et de l'Iceberg du treillis de  Galois associés au
contexte d'extraction donné par la Figure \ref{runfig} pour
\textit{minsup}=2}\label{figexemple}

\end{center}
\end{figure}

\begin{exemple} Afin d'illustrer le déroulement de l'algorithme \textsc{Prince},
considérons le contexte d'extraction $\mathcal{K}$ donné par la
Figure \ref{runfig} pour \textit{minsup} = 2 et \textit{minconf} =
0,5. La première étape permet de déterminer l'ensemble des
générateurs minimaux $\mathcal{GMF}\mathcal{_{K}}$ trié, ainsi que
la bordure négative $\mathcal{GB}$d$^{-}$.
$\mathcal{GMF}_{\mathcal{K}}$ = \{($\emptyset$,5),(B,4),(C,4),
(E,4), (A,3), (BC,3),(CE,3), (AB,2), (AE,2)\} et
$\mathcal{GB}$d$^{-}$=\{(D,1)\}. Dans la deuxième étape,
l'ensemble $\mathcal{GMF}_{\mathcal{K}}$ est parcouru en comparant
chaque générateur minimal fréquent $g$ de taille k $(k\geq1)$ aux
listes des successeurs immédiats de ses sous-ensembles de taille
$(k$-$1)$. L'ensemble vide, n'ayant aucun sous-ensemble, est
introduit directement dans le \textit{treillis des générateurs
minimaux} (\textit{cf.} Figure \ref{figexemple}.a). Ensuite, B est
ajouté à $\emptyset$.\texttt{\textit{succs-immédiats}}
(\textit{cf.} Figure \ref{figexemple}.b), la liste des successeurs
immédiats de l'ensemble $\emptyset$, étant initialement vide.
Ensuite, C sera comparé à B. BC étant un générateur minimal,
$\mathcal{C}_{B}$ et $\mathcal{C}_{C}$ sont alors incomparables et
C est ajouté à $\emptyset$.\texttt{\textit{succs-immédiats}}
(\textit{cf.} Figure \ref{figexemple}.c). E est alors comparé à
cette liste. En comparant E à B, E.\texttt{\textit{support}} =
B.\texttt{\textit{support}} = Supp(BE) et donc E $\in$
$\mathcal{C}_{B}$ dont B est le représentant (\textit{cf.} Figure
\ref{figexemple}.d). Ainsi, le traitement continue en remplaçant
les occurrences de E par B dans des listes des successeurs
immédiats (dans ce cas, il n'y a aucune occurrence) et en
poursuivant les comparaisons avec B au lieu de E (dans ce cas, il
n'y a plus de comparaisons à faire \textit{via} E). Les
traitements s'arrêtent alors pour E. \`{A} cette étape,
$\emptyset$.\texttt{\textit{succs-immédiats}} = \{B,C\}. A est
alors comparé à B. Comme AB $\in$ $\mathcal{GMF}\mathcal{_{K}}$,
$\mathcal{C}_{B}$ et $\mathcal{C}_{A}$ sont incomparables. Par
contre, en comparant A et C, A.\texttt{\textit{support}} <
C.\texttt{\textit{support}} et A.\texttt{\textit{support}} =
Supp(AC) et donc $\mathcal{C}_{A}$ est un successeur de
$\mathcal{C}_{C}$. A est tout simplement ajouté à
C.\texttt{\textit{succs-immédiats}} étant donné qu'elle est encore
vide (\textit{cf.} Figure \ref{figexemple}.e).
Le même traitement est appliqué pour les reste des générateurs
minimaux. Le \textit{treillis des générateurs minimaux} obtenu est
donné par la Figure \ref{figexemple}.h. Pour la dérivation des
règles génériques, le \textit{treillis des générateurs minimaux}
est parcouru d'une manière ascendante à partir de
$\mathcal{C}_{\emptyset}$. Comme $\omega(\emptyset)=\emptyset$, il
n'y a donc pas de règle exacte informative relative à
$\mathcal{C}_{\emptyset}$.
$\emptyset$.\texttt{\textit{succs-immédiats}}=\{B,C\}. L'itemset
fermé fréquent correspondant à $\mathcal{C}_{B}$ est alors
retrouvé et est égal à BE (\textit{cf.} Figure
\ref{figexemple}.i)\footnote{Dans la Figure \ref{figexemple}.i,
les flèches indiquent les relations de précédence utilisées pour
retrouver les itemsets fermés fréquents.}. La règle approximative
informative $\emptyset\Rightarrow BE$ de support 4 et de confiance
0,8 sera alors extraite. De même avec $\mathcal{C}_{C}$. De la
même manière et à partir de $\mathcal{C}_{B}$ et
$\mathcal{C}_{C}$, le parcours du treillis se fait d'une façon
ascendante jusqu'à extraire toutes les règles associatives
informatives valides. \`{A} la fin de l'exécution de l'algorithme,
nous obtenons l'Iceberg du treillis de Galois associé au contexte
d'extraction $\mathcal{K}$ (\textit{cf.} Figure
\ref{figexemple}.i).

\end{exemple}

\begin{figure}[!ht]
\begin{center}
\parbox{7.cm}{\includegraphics[scale=1]{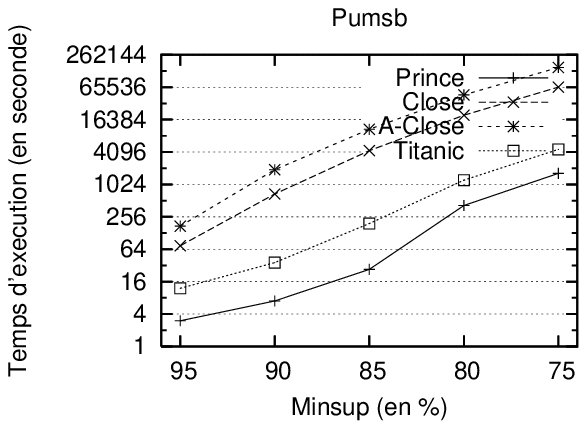}}
\parbox{7.cm}{\includegraphics[scale=1]{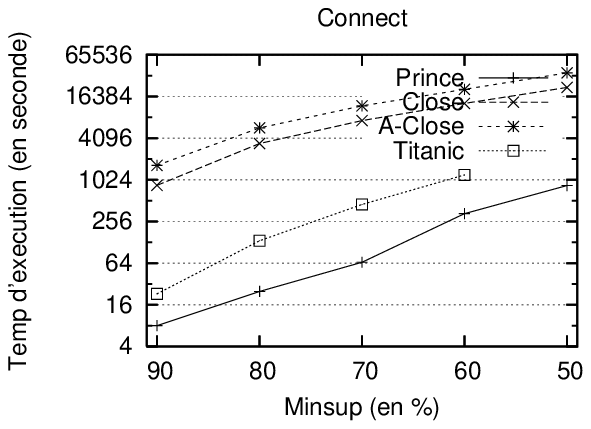}}
\parbox{7.cm}{\includegraphics[scale=1]{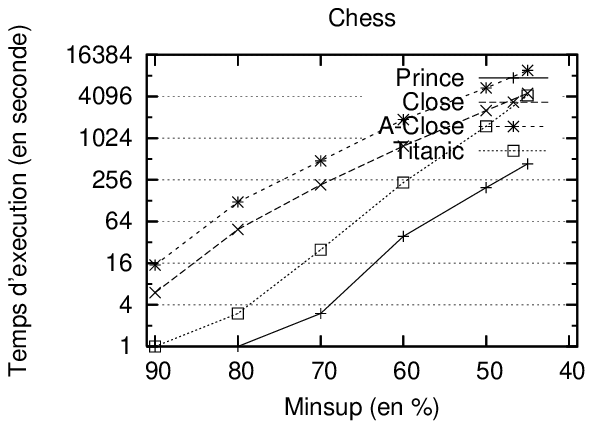}}
\parbox{7.cm}{\includegraphics[scale=1]{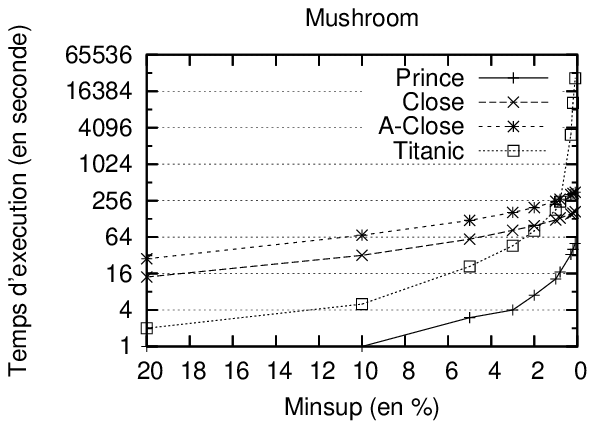}}
\end{center}
\caption{Performances de \textsc{Prince} vs \textsc{Close},
\textsc{A-Close} et \textsc{Titanic} pour les contextes considérés
comme denses} \label{fig_tps_exec_denses}
\end{figure}

L'étude des performances de l'algorithme \textsc{Prince}
\textit{vs.} respectivement celles des algorithmes les plus connus
adoptant la stratégie "Générer-et-tester", \emph{i.e.},
\textsc{Close}, \textsc{A-Close} et \textsc{Titanic} a donné des
résultats encourageants~\cite{Masttarek}. Les contextes utilisés
lors de cette étude sont les contextes benchmark que l'on retrouve
dans la littérature associée au domaine de fouille de donnée. Les
temps d'exécution de l'algorithme \textsc{Prince} comparés,
respectivement, aux algorithmes \textsc{Close}, \textsc{A-Close}
et \textsc{Titanic} sur des contextes considérés comme denses sont
présentés dans la Figure \ref{fig_tps_exec_denses}. De plus, le
tableau \ref{princeratios} présente des ratios comparatifs des
performances de l'algorithme \textsc{Prince} \emph{vs.} celles des
autres algorithmes considérés.

Ainsi, l'algorithme \textsc{Prince} s'avère performant sur les
contextes considérés comme denses et pour toutes les valeurs de
\textit{minsup}. La différence entre les performances de
\textsc{Prince} et celles de \textsc{Close} et \textsc{A-Close}
atteint son maximum pour le contexte \textsc{Pumsb}. En effet,
l'algorithme \textsc{Prince} est environ 160 (resp. 403) fois plus
rapide que \textsc{Close} (resp. \textsc{A-Close}) pour un support
de 85\%. De même, l'algorithme \textsc{Prince} est environ 538
fois plus rapide que \textsc{Titanic} pour le contexte
\textsc{Mushroom} et pour un support égal à 0,01\%. Pour ces
contextes denses, \textsc{Prince} est en moyenne 42 (resp. 89 et
41) fois plus rapide que \textsc{Close} (resp. \textsc{A-Close} et
\textsc{Titanic})~\cite{Masttarek}.

\begin{table}
\begin{center}
\small{
   \begin{tabular}{|l||c|c|c|c|c|c|}
      \hline  & \multicolumn{2}{|c|}{ $\frac{\textsc{Close}}{\textsc{Prince}}$}  &\multicolumn{2}{|c|}{ $\frac{\textsc{A-Close}}{\textsc{Prince}}$}&\multicolumn{2}{|c|}{ $\frac{\textsc{Titanic}}{\textsc{Prince}}$}      \\\hline
Contexte& min &max  &min &max &min &max     \\ \hline \hline

\textsc{Pumsb}&24,67&160,44&57,33&402,78&2,78&7,11    \\\hline
\hline \textsc{Connect}&26,17&137,12&42,39&230,84&2,87&6,85
\\\hline \hline \textsc{Chess}&6,00&72,33&15,00&160,33&1,00&9,74
\\\hline \hline \textsc{Mushroom}&3,36&32,00&7,04&69,00&2,00&537,54
\\\hline \hline
\textsc{T10I4D100K}&0,88&6,50&1,03&10,00&2,33&17,40    \\\hline
\hline \textsc{T40I10D100K}&9,31&47,60&5,68&19,87&3,47&202,12
\\\hline \hline \textsc{Retail} &2,05&114,86&1,26&29,18&8,78&261,30
\\\hline

\end{tabular}
} \caption{Récapitulatif des performances de l'algorithme
\textsc{Prince} \emph{vs.} celles de \textsc{Close},
\textsc{A-Close} et \textsc{Titanic}.} \label{princeratios}
\end{center}
\end{table}

L'algorithme \textsc{Prince} s'avère aussi performant sur des
contextes considérés épars et pour toutes les valeurs de
\textit{minsup}. La différence entre les performances de
\textsc{Prince} et celles de \textsc{Close} et \textsc{Titanic}
atteint sa valeur maximale pour le contexte \textsc{Retail}. En
effet, l'algorithme \textsc{Prince} est environ 115 (resp. 261)
fois plus rapide que \textsc{Close} (resp. \textsc{Titanic}) pour
un support de 0,004\%. Pour ces contextes épars, \textsc{Prince}
est en moyenne 31 (resp. 13 et 32) fois plus rapide que
\textsc{Close} (resp. \textsc{A-Close} et \textsc{Titanic}).

\subsection{L'algorithme \textsc{IMG-Extractor}}
Pour permettre une plus grande réduction informative du nombre de
règles associatives extraites, la notion de générateur minimal
semble jouer un rôle central. Le peu d'attention accordé aux
générateurs minimaux nous a encouragé à une exploration
approfondie de leur structure intrinsèque. Ainsi, cette piste,
concernant la détection des éléments redondants par la
substitution de sous-ensembles par d'autres, a été initialement
amorcée par Dong et \emph{al.}~\cite{dong05}. Ces derniers ont
constaté que quelques générateurs minimaux peuvent être déduits à
partir d'autres par substitution de leurs sous-ensembles
respectifs. Ils ont alors proposé le Système Succinct des
Générateurs Minimaux (SSGM) comme une représentation concise des
générateurs minimaux. Ainsi, seuls les générateurs minimaux
utiles, ou \emph{succincts}, seront maintenus, et ceux dits
"\emph{redondants}" seront alors éliminés. L'algorithme
\textsc{IMG-Extractor}, que avons proposé dans~\cite{tarekcap06},
nous permettait une extraction efficace de ce SSMG tel que défini
dans~\cite{dong05}\footnote{Le prototype \textsc{IMG-Extractor}
est disponible à l'adresse suivante
www.cck.rnu.tn/sbenyahia/software\_release.htm.}. Cependant, suite
à une étude approfondie de cette définition, nous avons relevé que
ce système souffre d'une perte d'information, \emph{i.e.}, que
quelques générateurs minimaux redondants peuvent ne pas être
dérivés à partir du système. Ainsi, nous avons proposé une
nouvelle définition~\cite{IJFCS08}, permettant de remédier à
l'inconvénient de la définition originelle. Néanmoins, cette
nouvelle définition proposée a fait perdre au SSGM sa structure
intéressante à savoir un idéal d'ordre~\cite{ganter99}, dont
bénéficiait la définition originelle. Cette dernière facilite
énormément la localisation des éléments à retenir et réduit ainsi
le coût des traitements à effectuer. Récemment, le travail que
nous avons proposé dans~\cite{icfca07} avait pour but de concilier
les avantages des deux définitions initiales à savoir la non perte
d'information et le maintien de la structure d'idéal d'ordre. Une
nouvelle famille d'éléments a été alors introduite en se basant
sur l'axiome de pseudo-transitivité
d'Armstrong~\cite{armstrong74}. Les expérimentations réalisées ont
montré que cette famille permet d'éliminer un nombre important de
générateurs minimaux redondants, et par conséquent de règles
associatives génériques redondantes. Une piste primordiale qui
reste à étudier concerne l'étude du coût de la dérivation des
générateurs minimaux redondants à partir de cette famille, en
s'inspirant des travaux réalisés dans des domaines liés à la
fouille de données, \emph{e.g.}, la théorie des graphes. En effet,
la séquence de substitutions à suivre pour atteindre un générateur
minimal redondant donné mériterait d'être étudiée afin d'éviter la
redondance dans la dérivation, puisque différentes séquences
peuvent mener à un même générateur minimal redondant.

\section{Conclusion}
Dans ce chapitre, nous avons présenté une revue de l'état de l'art
sur les algorithmes d'extraction des itemsets fermés fréquents. Le
constat majeur est que ces algorithmes se sont focalisés sur
l'extraction des itemsets fermés fréquents et ils ont négligé la
composante relation d'ordre partiel sous-jacente. Or celle-ci est
primordiale pour l'extraction des règles génériques
approximatives. Pour cela, nous avons proposé deux approches pour
le calcul de la relation d'ordre. Le cadre fédérateur de nos
propositions a consisté à définir de nouvelles approches visant à
extraire toute l'information nécessaire pour trouver des bases
génériques de règles associatives.

En outre, il serait d'une part intéressant d'étudier la
possibilité d'intégrer dans \textsc{Prince} l'approche proposée
dans~\cite{cal07}. Cette approche, applicable à tout ensemble
vérifiant la propriété d'idéal d'ordre, peut être utilisée pour
l'ensemble des générateurs minimaux utilisé dans notre cas. Il
permet de réduire le nombre de candidats auxquels un accès au
contexte d'extraction est nécessaire pour calculer le support.
Étant donné que cette étape est la plus coûteuse de la première
étape, une telle intégration permettra de réduire un tel coût.
D'autre part, la mise en place d'un choix adaptatif de la bordure
à ajouter à l'ensemble des générateurs minimaux fréquents est
envisagée. En effet, pour certains contextes, il est plus
avantageux du point de vue cardinalité de maintenir une autre
bordure que celle actuellement maintenue, à savoir
$\mathcal{GB}d^-$~\cite{liu06}. Ceci dépend étroitement des
caractéristiques des contextes et qui seront donc à étudier pour
bien sélectionner la bordure à maintenir.



\chapter{Les itemsets fermés disjonctifs : une nouvelle représentation concise}

\section{Introduction}
L'apparition de la "fouille de connaissances" a été un tournant
dans les intérêts prioritaires de la communauté de la fouille de
données. En effet, les efforts ne sont plus seulement déployés
dans la réduction des temps d'extraction des motifs fréquents mais
de plus en plus de travaux s'intéressent à l'extraction d'une
connaissance de meilleure qualité tout en maximisant sa compacité.
Dans ce registre, nous relevons des travaux visant l'extraction
des représentations concises des itemsets fréquents. Ainsi, parmi
les représentations exactes les plus connues, nous citons celles
fondées sur les itemsets fermés~\cite{pasquier99}, les itemsets
non-dérivables~\cite{calders021} et les itemsets
essentiels~\cite{casaliisi05}. Un excellent état de l'art des
principales représentations concises des itemsets fréquents qui
ont été proposées dans la littérature se trouve dans
\cite{calderssurvey}.

Les itemsets de la représentation basée sur les essentiels
fréquents sont caractérisés à travers des relations qui manipulent
deux métriques, à savoir le support disjonctif et le support
conjonctif. Le support disjonctif d'un itemset $I$, noté
$Supp(\vee I)$, désigne le nombre de transactions contenant au
moins un item de $I$. Une telle caractérisation des itemsets de la
représentation fondée sur les essentiels fréquents lui confère le
pouvoir de dériver directement les supports disjonctifs des
itemsets fréquents et d'offrir un mécanisme correct de dérivation
de leurs supports conjonctifs. Cette caractéristique est d'une
importance capitale dans l'évaluation efficace des supports et des
confiances de certaines formes de règles d'association
généralisées. Par contre, les éléments des représentations par les
itemsets fermés fréquents et les itemsets non dérivables fréquents
sont caractérisés à travers des relations qui manipulent
uniquement le support conjonctif. De telles caractérisations
rendent impossible la dérivation directe du support disjonctif des
itemsets fréquents à partir de ces représentations concises
exactes. Bien qu'offrant un taux de compacité intéressant, la
représentation fondée sur les itemsets essentiels souffrait de son
association avec la bordure positive afin de la rendre exacte.

Dans ce chapitre, nous présentons une contribution visant la
proposition d'une nouvelle représentation concise des itemsets
fréquents, qui exploite la propriété de non-injectivité de tout
opérateur de fermeture. Deux particularités sont à mettre au
crédit de cette nouvelle représentation :
\begin{itemize}
    \item La dérivation directe des supports disjonctifs et négatifs des
itemsets ;
    \item L'élimination de la bordure positive
puisque l'ensemble des itemsets fermés disjonctifs associés aux
itemsets essentiels fréquents, augmenté d'un autre ensemble
d'itemsets que nous avons caractérisé, constitue à lui seul une
représentation concise exacte des itemsets fréquents.
\end{itemize}
 Les expérimentations que nous avons menées sur des contextes benchmark ont
montré que la nouvelle représentation concise présente un taux de
compacité largement supérieur à ceux des autres représentations
concises. En particulier, nous arrivons même à réduire les
représentations des contextes considérés comme épars.

\section{Les représentations concises : revue des travaux antérieurs}

Dans ce qui suit, nous allons présenter une étude succincte des
principales représentations concises exactes des itemsets
fréquents~\cite{calderssurvey}. Nous allons commencer par nous
focaliser sur les propriétés structurelles de la représentation
concise fondée sur les itemsets essentiels, ensuite celle basée
sur les itemsets non-dérivables//.

\subsection{Représentation par itemsets essentiels fréquents}

La représentation concise exacte fondée sur les itemsets
essentiels possède deux avantages majeurs : d'une part, elle offre
à l'utilisateur la possibilité de calculer les différents types de
supports et d'autre part, elle présente un taux de compacité
intéressant~\cite{casaliisi05}.

\begin{definition}~\cite{casaliisi05} \label{définitionmotifessentiel}\textsc{(Itemset essentiel fréquent)}
Soit $\mathcal{K}$ = ($\mathcal{O}$, $\mathcal{I}$, $\mathcal{R}$)
un contexte d'extraction et $I$ $\subseteq$ $\mathcal{I}$. $I$ est
appelé un itemset \emph{essentiel} \textit{si} :
Supp($\vee$$I$)$\mbox{ }$ $>$
$\max\{$Supp($\vee$$I$$\backslash$$i$) $|$ $i$$\in$ $I\}$.
\end{definition}

\begin{exemple}Considérons le contexte d'extraction de la Figure \ref{exemplecontexteformel1} (page \pageref{exemplecontexteformel1})
pour \textit{minsup} = 1. L'itemset $AB$ n'est pas un itemset
essentiel puisque Supp($\vee$$AB$) = $\max\{$Supp($\vee$$A$),
Supp($\vee$$B$)$\}$= Supp($\vee$$A$) = 3 alors que $AC$ est un
itemset essentiel puisque Supp($\vee$$AC$) $>$
$\max\{$Supp($\vee$$A$), Supp($\vee$$C$)$\}$ (car
$\max\{$Supp($\vee$$A$), Supp($\vee$$C$)$\}$ = Supp($\vee$$C$) = 4
et Supp($\vee$$AC$) = 5). De plus, $AC$ est fréquent puisque
Supp($AC$) = 2 $\geq$ \textit{minsup}.
\end{exemple}

La proposition suivante montre que l'ensemble des itemsets
essentiels fréquents vérifie la propriété intéressante d'idéal
d'ordre, \emph{i.e.}, que si $I$ est un itemset essentiel
fréquent, alors tous ses sous-ensembles sont des itemsets
essentiels fréquents.

\begin{proposition}~\cite{casaliisi05} \label{propositionordreidéal}
L'ensemble des itemsets essentiels fréquents est un idéal d'ordre.
\end{proposition}
Cette propriété donne la possibilité aux algorithmes d'extraction,
par niveau, d'extraire efficacement cet ensemble. Ainsi,
l'algorithme \textsc{MEP}~\cite{casalidawak105}, permettant
d'extraire les itemsets essentiels fréquents, est une adaptation
de l'algorithme \textsc{Apriori}~\cite{Agra94}.

Dans ce qui suit, nous désignerons par $\mathcal{IEF}_\mathcal{K}$
(resp. $\mathcal{IF}_\mathcal{K}$) l'ensemble des itemsets
essentiels fréquents (resp. itemsets fréquents) qui peuvent être
extraits à partir d'un contexte d'extraction $\mathcal{K}$. Le
lemme suivant montre comment nous pouvons obtenir le support
disjonctif d'un itemset fréquent à partir de l'ensemble
$\mathcal{IEF}_\mathcal{K}$.

\begin{lemma} \cite{casaliisi05}\label{lemmasupportdisjonctifitemsetmaximalessentiel}
$\forall$ $I$ $\in$ $\mathcal{IF}_\mathcal{K}$, Supp($\vee$$I$) =
max$\{$Supp($\vee$$I_1$) $|$ $I_1$ $\subseteq$ $I$ $\wedge$ $I_1$
$\in$ $\mathcal{IEF}_\mathcal{K}\}$.
\end{lemma}
Le théorème qui suit caractérise la représentation concise fondée
sur les itemsets essentiels fréquents.

\begin{theorem}~\cite{casaliisi05} \label{concise-repre-closed-itemsets} L'ensemble
$\mathcal{IEF}_\mathcal{K}$ des itemsets essentiels fréquents
augmenté par l'ensemble $BD^{+}(\mathcal{IF}_\mathcal{K})$ des
itemsets maximaux fréquents est une représentation concise exacte
de l'ensemble des itemsets fréquents.
\end{theorem}
Le théorème qui suit indique comment dériver le
support conjonctif d'un itemset fréquent une fois que nous avons
extrait l'ensemble des itemsets essentiels fréquents.

\vspace{-5pt}
\begin{theorem} (\cite{casaliisi05}) \label{théorèmesupportconjonctifdesitemsetsnonessentielsfréquents}
$\forall$ $X$ $\in$ $\mathcal{IF}_\mathcal{K}$. Soit $Y$ $\in$
Argmax($\{$Supp($\vee$$X_1$) $|$ $X_1$ $\subseteq$ $X$ et $X_1$
$\in$ $\mathcal{IEF}_\mathcal{K}$$\}$). Alors nous avons :
\begin{displaymath}
Supp(X)=\sum_{\substack{X_1\subseteq X\\X_1\neq\emptyset\\X_1
\not\subseteq Y}}{(-1)^{\mid X_1\mid -1}Supp(\vee X_1)}
\end{displaymath}
\end{theorem}

\medskip

\subsection{Repr{\'e}sentation par les itemsets non d{\'e}rivables
fr{\'e}quents}\label{ndipresentation}

La notion d'itemset non d{\'e}rivable a {\'e}t{\'e} introduite par
Calders et Goethals \cite{CALDERSChapter04,calders021}. Il s'est
av{\'e}r{\'e} que dans des applications r{\'e}elles, le nombre
d'itemsets non d{\'e}rivables fr{\'e}quents est largement
inf{\'e}rieur à celui des itemsets fr{\'e}quents ainsi qu'à celui
des itemsets ferm{\'e}s fr{\'e}quents \cite{CALDERSChapter04}.
Dans ce qui suit, nous désignons l'ensemble des itemsets non
dérivables fréquents par $\mathcal{INDF}$.
La notion d'itemset non d{\'e}rivable repose principalement sur
les \emph{règles de déduction} que nous pr{\'e}sentons dans la
sous-section ci-dessous.

\bigskip
\textbf{\ref{ndipresentation}.1 Règles de déduction}

\bigskip
Les règles de déduction représentent un ensemble de r{\`e}gles
permettant de d{\'e}duire les bornes du support conjonctif d'un
itemset $I$. Pour ce faire, les itemsets qui ne sont pas inclus
dans $I$ seront ignor{\'e}s et ils seront même exclus du contexte
d'extraction. L'op{\'e}ration de suppression de ces itemsets se
fait grâce à une op{\'e}ration appelée \emph{$I$-projection}
pr{\'e}sent{\'e}e dans la d{\'e}finition suivante.
\begin{definition}
\textbf{$I$-Projection}\mbox{ }\\
Soit $I$ $\subseteq$ $\mathcal{I}$ un itemset.

\smallskip
$\bullet$ L'ensemble r{\'e}sultant de la $I$-projection d'une
transaction $T$, noté {\small$\Pi_I$}$T$, est d{\'e}fini par
{\small$\Pi_I$}$T$ := $\{ i \mid i \in T\cap I\}$.

\smallskip
$\bullet$ L'ensemble résultant de la $I$-projection d'une base de
transactions $\mathcal{K}$, noté {\small$\Pi_I$}, est l'ensemble
des $I$-projections des transactions de $\mathcal{K}$.
\end{definition}
Le lemme qui suit définit la relation qui existe entre le support
conjonctif d'un itemset dans une base de transactions
$\mathcal{K}$ et le support conjonctif dans {\small$\Pi_I$}.
\begin{lemma}
Soient $I$ et $J$ deux itemsets tels que $I$ $\subseteq J$
$\subseteq$ $\mathcal{I}$. Pour chaque transaction de
$\mathcal{K}$, nous avons l'égalité suivante:
\begin{center}
$Supp(I)$ $=$ $Supp(I$,{\small$\Pi_J$})
\end{center}
\end{lemma}

Avant d'entamer les r{\`e}gles de d{\'e}duction, les notions de
fraction et de couverture doivent être introduites.
\begin{definition}
(\textbf{$I$-Fraction})\mbox{ }\\
Soient $I$ et $J$ deux itemsets tels que $I$ $\subseteq$ $J$
$\subseteq \mathcal{I}$. Alors, la $I$-fraction de
{\small$\Pi_J$}, notée $f_{I}^{J}$, est {\'e}gale au nombre de
transactions de {\small$\Pi_J$} qui sont {\'e}gales à $I$. Ainsi,
le support conjonctif d'un itemset est {\'e}gal à $\sum_{I
\subseteq I' \subseteq \mathcal{I}}f_{I'}$.\\

\end{definition}
%

\begin{exemple}
Considérons le contexte d'extraction d{\'e}fini par la figure
 \ref{exemplecontexteformel1} (page \pageref{exemplecontexteformel1}). La projection
du contexte $\mathcal{K}$ sur l'itemset $\textsc{AB}$, not{\'e}e
${\small\Pi_{\textsc{AB}}}$, est {\'e}gale à $\{
\textsc{AB},\emptyset,\textsc{AB},\textsc{AB},\emptyset\}$. Ainsi,
nous obtenons $f_{\textsc{A}}^{\textsc{AB}} = 0$.
\end{exemple}
\begin{definition}
(\textbf{Couverture})\mbox{ }\\
Soit $I$ $\subseteq$ $\mathcal{I}$ un itemset. La couverture de
$I$ dans la base de transactions $\mathcal{K}$, not{\'e}e
$Couv(I,\mathcal{K})$, est l'ensemble des transactions de
$\mathcal{K}$ qui contiennent $I$.
\end{definition}

Soient $I$ et $J$ deux itemsets tels que $I\subseteq J $
$\subseteq$ $\mathcal{I}$ avec $J$ $=$ $I$ $\cup$
\{$A_1,\dots,A_n$\}. Remarquons que $Couv(J)$ $=$ $\cap_{i=1}^n
Couv(I\cup \{ A_i\})$ et que $\mid \cup_{i=1}^n Couv(I\cup\{A_i
\})\mid$ $=$ $\mid Couv(I)\mid$ - $f_I^J$. En utilisant le
principe d'inclusion-exclusion \cite{galambos2000},
l'{\'e}galit{\'e} pr{\'e}c{\'e}dente se r{\'e}{\'e}crit comme suit
: {\setlength\arraycolsep{5pt}
\begin{eqnarray*}
\mid Couv(I)\mid - f_I^J & = & \sum_{\substack{1\leq i \leq n}}
\mid Couv(I\cup \{A_i\})\mid - \sum_{\substack{1\leq i \leq j \leq
n}} \mid Couv(I\cup \{A_i,A_j\})\mid \\ & & + \dots + (-1)^n \mid
Couv(J)\mid
\end{eqnarray*}}
Puisque $Supp(I\cup \{A_{i_1},\ldots,A_{i_l}\})$ $=$ $\mid
Couv(I\cup \{A_{i_1},\ldots,A_{i_l}\}) \mid$, nous obtenons :
{\setlength\arraycolsep{5pt}
\begin{eqnarray*}
(-1)^{\mid J\setminus I\mid} Supp(J) - f_I^J & = & Supp(I) -
\sum_{\substack{1\leq i \leq n}} Supp(I\cup \{A_i\}) \\ & & +
\sum_{\substack{1\leq i \leq j \leq n}} Supp(I\cup \{A_i,A_j\}) +
\ldots \\  & & + (-1)^{\mid J\setminus I \mid -
1}\sum_{\substack{1\leq i \leq n}} Supp(J - \{A_i\})
\end{eqnarray*}}
Dans ce qui suit, nous noterons le terme à droite de
l'{\'e}galit{\'e} ci-dessus par $\sigma(I,J)$. Puisque $f_I^J$ est
toujours positif, nous obtenons le th{\'e}or{\`e}me suivant.
\begin{theorem}\label{theosigmamajorantminorant}
Pour chaque couple d'itemsets $I \subseteq J$ $\subseteq$
$\mathcal{I}$, $\sigma(I,J)$ est un minorant $($\textit{resp.}
majorant$)$ de $Supp(J)$ si $\mid J\setminus I\mid$ est pair
$($resp. impair$)$. De plus, $\mid Supp(J) - \sigma(I,J)\mid$ =
$f_I^J$.
\end{theorem}
Si pour chaque $I$ $\subseteq$ $J$, $Supp(I)$ est connu, les
r{\`e}gles $\{\mathcal{R}_J(I)\mid I \subseteq J\}$ permettent de
calculer le majorant et le minorant de $Supp(J)$.\\

Les r{\`e}gles de d{\'e}duction partitionnent l'ensemble des
itemsets fr{\'e}quents en deux sous-ensembles disjoints: (i)
\emph{les itemsets dérivables}, dont le support conjonctif associé
peut être calcul{\'e} de façon exacte à partir des r{\`e}gles de
d{\'e}duction; (ii) \emph{les itemsets non dérivables}, dont le
support conjonctif ne peut être calcul{\'e} de façon exacte à
partir des r{\`e}gles de d{\'e}duction qui lui sont associées.


Ainsi, si un itemset $I$ est non d{\'e}rivable, alors tous ses
sous-ensembles sont non d{\'e}rivables. Pour chaque itemset $I$,
nous d{\'e}signons par $l_I$ (\textit{resp.} $u_I$) le plus grand
minorant (\textit{resp.} le plus petit majorant) pouvant être
d{\'e}duit à partir des r{\`e}gles de d{\'e}duction.
\begin{lemma}\label{lemmendimonotonicit{\'e}}

\textbf{Monotonicit{\'e}} \cite{calders021}

\medskip
Soit $I$ $\subseteq$ $\mathcal{I}$ un itemset, et $i$ $\in$
$\mathcal{I}\setminus I$ un item. Alors nous avons:
\begin{displaymath}
2\mid u_{I\cup \{i\}} - l_{I\cup \{i\}}\mid \leq 2 \min\{\mid
Supp(I)-l_{I}\mid , \mid Supp(I) - u_{I}\mid \} \leq \mid u_{I} -
l_{I}\mid
\end{displaymath}
En particulier, si $I$ est un itemset d{\'e}rivable, alors ses
sur-ensembles sont des itemsets d{\'e}rivables.
\end{lemma}

Grâce au th{\'e}or{\`e}me suivant, qui est une cons{\'e}quence du
Lemme \ref{lemmendimonotonicit{\'e}},
l'ensemble des itemsets non d{\'e}rivables est montr{\'e} qu'il
constitue une repr{\'e}sentation concise exacte.
\begin{theorem} \cite{calders021}
\textbf{Repr{\'e}sentation par les itemsets non d{\'e}rivables fr{\'e}quents }\\
Soient $\mathcal{K}$ un contexte d'extraction, \textit{minsup} un
seuil minimal du support conjonctif d{\'e}fini par l'utilisateur
et
\textsc{NDI}$($$\mathcal{K}$,\textit{minsup}$)$ l'ensemble
d{\'e}fini comme suit:
\begin{displaymath}
\textsc{NDI}(\mathcal{K},\textit{minsup}) = \{(I,Supp(I))\mid l_I
\neq u_I \wedge Supp(I) \geq  minsup\}
\end{displaymath}
\textsc{NDI}$($$\mathcal{K}$,\textit{minsup}$)$ est une
repr{\'e}sentation concise exacte de l'ensemble des itemsets
fr{\'e}quents.
%
\end{theorem}

\subsection{Discussion}
\`{A} la lumi{\`e}re de ce qui a {\'e}t{\'e} pr{\'e}c{\'e}demment
pr{\'e}sent{\'e} dans ce chapitre, nous remarquons que les
représentations concises exactes pr{\'e}sentent quelques éléments
de comparaison, qui peuvent être organis{\'e}es selon les
dimensions suivantes:
\begin{enumerate}
    \item \textbf{Itemsets constituant la repr{\'e}sentation:} cette dimension décrit la nature
    des itemsets contenus dans la représentation concise exacte.

    \item \textbf{Outils de caractérisation des itemsets constituant la
    repr{\'e}sentation:} cette dimension d{\'e}termine les moyens de caractérisation des itemsets de la représentation concise, tels que les m{\'e}triques (\textit{i.e.} les supports), les r{\`e}gles de d{\'e}duction et la taille des itemsets.
    \item \textbf{Algorithmes d'extraction:} afin d'extraire chaque
    repr{\'e}sentation concise exacte, un ou plusieurs algorithmes ont
    {\'e}t{\'e} d{\'e}finis. \`{A} travers cette dimension, nous voulons mettre en évidence les algorithmes les plus connus permettant l'extraction de chaque
    repr{\'e}sentation.
    \item \textbf{M{\'e}thodes de r{\'e}g{\'e}n{\'e}ration:} cette
    dimension précise, pour chaque représentation, la méthode sp{\'e}cifique de r{\'e}g{\'e}n{\'e}ration des itemsets fr{\'e}quents.
    \item \textbf{Avantages:} cette dimension pr{\'e}sente les principaux avantages des
    repr{\'e}sentations concises propos{\'e}es dans la litt{\'e}rature, qui peuvent se résumer comme suit:
    \begin{enumerate}
        \item \emph{Dérivation des supports disjonctifs et n{\'e}gatifs}: cette
        caract{\'e}ristique est tr{\`e}s importante vu qu'elle permet
        d'{\'e}viter l'{\'e}valuation des identit{\'e}s d'inclusion-exclusion
        exprimant le support disjonctif et n{\'e}gatif en fonction du support
        conjonctif. Ces identit{\'e}s peuvent être tr{\`e}s lourdes à
        {\'e}valuer lorsque le nombre d'itemsets fr{\'e}quents est tr{\`e}s
        grand (pour des contextes denses).
        \item \emph{Homog{\'e}n{\'e}it{\'e}}: à travers cette caractéristique, il est possible de connaître si les itemsets
        utilisés dans la représentation concise sont de même nature.
        \item \emph{Dérivation des règles associatives génériques}: les
        r{\`e}gles associatives g{\'e}n{\'e}riques permettent de g{\'e}n{\'e}rer la totalit{\'e} des r{\`e}gles associatives valides \cite{thesepasq}.
        Elles permettent aussi de r{\'e}duire le nombre de r{\`e}gles manipul{\'e}es et de supprimer la redondance existante au sein de ces
        r{\`e}gles.

        \item \emph{D{\'e}rivation directe des règles associatives g{\'e}n{\'e}ralis{\'e}es}: contrairement aux r{\`e}gles associatives classiques, les
        r{\`e}gles associatives g{\'e}n{\'e}ralis{\'e}es constituent une mine d'informations établissant les liens entre les conjonctions,
        les disjonctions et les n{\'e}gations d'items. Elles r{\'e}v{\`e}lent à l'utilisateur des corr{\'e}lations plus fines existantes entre les items du contexte d'extraction.
        Ainsi, elles aident les utilisateurs à d{\'e}gager les constatations les plus raffin{\'e}es sur le contexte d'extraction.
        \end{enumerate}
\end{enumerate}
Le tableau \ref{tabcomparatifCR} r{\'e}sume l'{\'e}tude
comparative relative aux principales repr{\'e}sentations concises
propos{\'e}es dans la litt{\'e}rature. Les constatations majeures
relatives à cette {\'e}tude sont les suivantes:
\begin{enumerate}
    \item La d{\'e}rivation du support conjonctif, à partir de la
    repr{\'e}sentation par les ferm{\'e}s fr{\'e}quents, est quasi-imm{\'e}diate.
    N{\'e}anmoins, la d{\'e}rivation directe des supports disjonctifs et n{\'e}gatifs
    des itemsets fr{\'e}quents est impossible.
    \item La repr{\'e}sentation par les itemsets ferm{\'e}s
    fr{\'e}quents est la seule à offrir la possibilit{\'e} de g{\'e}n{\'e}rer les
    r{\`e}gles associatives g{\'e}n{\'e}riques.
    \item La r{\'e}g{\'e}n{\'e}ration des itemsets fr{\'e}quents à partir des
    itemsets non d{\'e}rivables est tr{\`e}s coûteuse en temps d'ex{\'e}cution.
    En effet, le calcul du support conjonctif d'un itemset d{\'e}rivable
    n{\'e}cessite l'{\'e}valuation de $2^{n}$ r{\`e}gles de d{\'e}duction. En contre
    partie, la repr{\'e}sentation bas{\'e}e sur les
    essentiels fr{\'e}quents ne n{\'e}cessite que l'{\'e}valuation d'une
    identit{\'e} d'inclusion-exclusion afin de d{\'e}terminer le support
    conjonctif des itemsets fr{\'e}quents.
    \item La repr{\'e}sentation bas{\'e}e sur les essentiels
    fr{\'e}quents est la seule à offrir la possibilit{\'e} de d{\'e}river directement diverses formes de
    r{\`e}gles associatives, vu qu'elle permet la d{\'e}rivation imm{\'e}diate
    des supports disjonctifs et n{\'e}gatifs des itemsets fr{\'e}quents.
\end{enumerate}
\begin{table}[!ht]
\begin{center}
{\footnotesize
\begin{tabular}{|l|c|c|c|}
    \hline
    \textbf{Repr{\'e}sentation} & \texttt{Repr{\'e}sentation}  & \texttt{Repr{\'e}sentation}   & \texttt{Repr{\'e}sentation bas{\'e}e } \\
    \textbf{concise exacte} & \texttt{par les ferm{\'e}s} & \texttt{par les non d{\'e}rivables} & \texttt{sur les
    essentiels} \\
     & \texttt{fr{\'e}quents} & \texttt{fr{\'e}quents} & \texttt{fr{\'e}quents}\\
    \hline

    \textbf{Itemsets constituant} & ferm{\'e}s fr{\'e}quents & non d{\'e}rivables fr{\'e}quents & essentiels fr{\'e}quents + \\
    \textbf{la repr{\'e}sentation}& & & maximaux fr{\'e}quents\\
    \hline
    \textbf{Outils de caract{\'e}ri-} & op{\'e}rateur de  & r{\`e}gles de d{\'e}duction + & support disjonctif + \\
    \textbf{sation des itemsets}  & fermeture de  & support conjonctif & support conjonctif + \\
    \textbf{constituant la} & Galois + & & taille des itemsets\\
    \textbf{repr{\'e}sentation} & support conjonctif & & \\
    \hline
    \textbf{Algorithmes d'extraction} & \textsc{CLOSE}, & \textsc{NDI}, df\textsc{NDI} & \textsc{MEP} \\
     & \textsc{CHARM}, etc & & \\
    \hline
    \textbf{M{\'e}thodes de r{\'e}g{\'e}n{\'e}ration} & le support d'un itemset   & {\'e}valuation de $2^{\mid X\mid}$ r{\`e}gles & {\'e}valuation d'une identit{\'e} \\
    \textbf{des itemsets fr{\'e}quents} & fr{\'e}quent est {\'e}gal à celui&  de d{\'e}duction pour chaque & d'inclusion-exclusion par\\
     &  du plus petit ferm{\'e} & itemset non dérivable $X$ &  itemset\\
     &  fr{\'e}quent le contenant& & \\
     \hline
     \textbf{D{\'e}rivation directe des}&  non & non &  oui\\
     \textbf{supports disjonctifs et}&   &  &  \\
     \textbf{n{\'e}gatifs}&   &  &  \\
     \hline
     \textbf{Homog{\'e}n{\'e}it{\'e}} & oui & oui & non\\
     \hline
     \textbf{D{\'e}rivation des r{\`e}gles}  & oui & non & non\\
     \textbf{associatives g{\'e}n{\'e}riques}&   &  &  \\
     \hline
     \textbf{D{\'e}rivation directe des }  & non & non & oui\\
     \textbf{r{\`e}gles associatives}&   &  &  \\
     \textbf{g{\'e}n{\'e}ralis{\'e}es}& & & \\
     \hline
\end{tabular}
}\caption{Tableau comparatif des principales représentations
concises exactes propos{\'e}es dans la
littérature}\label{tabcomparatifCR} \end{center}
\end{table}

\section{Contribution : Représentation concise exacte fondée sur la fermeture
disjonctive} Dans cette section, nous présentons une nouvelle
représentation concise exacte des itemsets fréquents. L'idée de
base de cette nouvelle représentation concise est d'appliquer un
opérateur de fermeture sur les itemsets essentiels fréquents afin
d'obtenir une représentation concise plus compacte. Cependant, cet
opérateur est différent de celui appliqué pour le représentation
concise fondée sur les itemsets fermés
fréquents~\cite{pasquier99}. En effet, les itemsets essentiels
sont plutôt caractérisés par leurs supports \emph{disjonctifs} que
par leurs supports \emph{conjonctifs}. Pour cela, nous avons
besoin d'un nouvel opérateur de fermeture, que nous avons dénommé
\emph{opérateur de fermeture disjonctive}. L'intérêt de cette
nouvelle représentation concise fondée sur la fermeture
disjonctive est double :

\begin{enumerate}
\item Avoir une représentation plus compacte que celle fondée sur
les essentiels fréquents. En effet, la fermeture disjonctive,
comme tout opérateur de fermeture, est une fonction non injective.
Le nombre des itemsets fermés disjonctifs sera, dans tous les cas,
inférieur ou égal au nombre des itemsets essentiels fréquents. De
plus, cette représentation concise va préserver l'avantage d'avoir
l'information relative aux différents supports et non seulement le
support conjonctif.

\item Éviter le besoin d'ajouter une information supplémentaire
afin de vérifier si un itemset est fréquent ou non, telle que
$BD^{+}(\mathcal{IF}_\mathcal{K})$ dans le cas de la
représentation concise fondée sur les itemsets essentiels
fréquents.
\end{enumerate}

\subsection {Les opérateurs disjonctifs et leurs propriétés}

Afin de présenter la fermeture disjonctive, nous avons besoin de
définir les applications correspondantes qui assurent le lien
entre ${\cal P}$($\mathcal{I}$) et ${\cal P}$($\mathcal{O}$) et
vice-versa.

\begin{definition}\label{définitionfg} Soit $\mathcal{K}$ = ($\mathcal{O}$, $\mathcal{I}$,
$\mathcal{R}$) un contexte d'extraction. Les opérateurs assurant
la connexion entre ${\cal P}$($\mathcal{I}$) et ${\cal
P}$($\mathcal{O}$) sont les suivants :\\
$\large f_d:
\mathcal{P}\textsc{(}\mathcal{O}\textsc{)}\rightarrow\mathcal{P}\textsc{(}\mathcal{I}\textsc{)}$

$\mbox{ }O\mbox{ }\mapsto f_d\textsc{(}O\textsc{)}\mbox{ }=\mbox{
}\{i\in\mathcal{I}\mid \textsc{(}\exists\mbox{ } o \in
O\textsc{)}\mbox{ }\textsc{(}\textsc{(}\textsc{(}o,i\textsc{)}\in
\mathcal{R}\textsc{)}\mbox{ }\wedge
\textsc{(}\textsc{(}\forall\mbox{ } o_{1} \in \mathcal{O}
\backslash O\textsc{)}\textsc{(}\textsc{(}o_{1},i\textsc{)}
\not\in \mathcal{R}\textsc{)}\textsc{)}\textsc{)}\}$\\\\$\large
g_{d}:
\mathcal{P}\textsc{(}\mathcal{I}\textsc{)}\rightarrow\mathcal{P}\textsc{(}\mathcal{O}\textsc{)}$

\ \ $I\mbox{ }\mapsto g_d\textsc{(}I\textsc{)}\mbox{ }=\mbox{
}\{o\in\mathcal{O}\mid \textsc{(}\exists\mbox{ } i \in
I\textsc{)}\textsc{(}\textsc{(}o,i\textsc{)}\in\mathcal{R}\textsc{)}\}$
\end{definition}
La sémantique des opérateurs de correspondance peut se définir
comme suit. L'opérateur $f_{d}(O)$ détermine l'ensemble des items
qui apparaissent au moins dans une transaction de $O$ et qui
n'apparaissent que dans les transactions de $O$. Dualement,
l'opérateur $g_{d}(I)$ est l'ensemble des transactions qui
contiennent au moins un item $i$ de $I$.

Après avoir énoncé les opérateurs de la correspondance, nous
pouvons introduire leurs composées.
\begin{definition}~\cite{tarekjordan06} \label{definitionfoggofdenden}Soient $\mathcal{K}$ = ($\mathcal{O}$, $\mathcal{I}$,
$\mathcal{R}$) un contexte d'extraction, et $f_d$ et $g_d$ les
opérateurs de la définition \ref{définitionfg}. Les opérateurs
composés relatifs à ces deux opérateurs se définissent comme
suit :\\

$h_d$ = $\large f_d\circ g_d:
\mathcal{P}\textsc{(}\mathcal{I}\textsc{)}\rightarrow\mathcal{P}\textsc{(}\mathcal{I}\textsc{)}$\\$\mbox{
}\hspace{1.5cm}I\hspace{0.42cm}\mapsto
h_d\textsc{(}I\textsc{)}\mbox{ }=\mbox{ }\{i\in\mathcal{I}\mid
\textsc{(}\forall\mbox{ } o \in\mathcal{O}\textsc{)}\mbox{
}\textsc{(}\textsc{(}\textsc{(}o,i\textsc{)}\in \mbox{ }
\mathcal{R}\textsc{)}\Rightarrow \textsc{(}\exists\mbox{ }
i_{1}\in
I\textsc{)}\textsc{(}\textsc{(}o,i_{1}\textsc{)}\in\mathcal{R}\textsc{)}\textsc{)}\}$

$h'_d$ = $\large g_d\circ f_d:
\mathcal{P}\textsc{(}\mathcal{O}\textsc{)}\rightarrow\mathcal{P}\textsc{(}\mathcal{O}\textsc{)}$\\$\mbox{
}\hspace{1.5cm}O\hspace{0.42cm}\mapsto
h'_d\textsc{(}O\textsc{)}\mbox{ }=\mbox{ }\{o\in\mathcal{O}\mid
\textsc{(}\exists\mbox{ } i\in\mathcal{I}\textsc{)}\mbox{
}\textsc{(}\textsc{(}\textsc{(}o,i\textsc{)}\in \mbox{
}\mathcal{R}\textsc{)} \wedge \textsc{(}\textsc{(}\forall\mbox{ }
o_{1} \in \mathcal{O}\backslash O\textsc{)}$
$\textsc{(}\textsc{(}o_{1},i\textsc{)}\not\in\mathcal{R}\textsc{)}\textsc{)}\textsc{)}\}$

\end{definition}

Maintenant, nous présentons les principales propriétés des
opérateurs (composés) introduits ci-dessus.
\begin{proposition}\label{properties-f-g-h}~\cite{tarekjordan06}
Soit $f_d$ (resp. $g_d$) l'opérateur assurant le lien entre
$\mathcal{P}(\mathcal{I})$ (resp. $\mathcal{P}(\mathcal{O})$) et
$\mathcal{P}(\mathcal{O})$ (resp.
$\mathcal{P}(\mathcal{I})$). 
Soient $I$, $I_{1}$, $I_{2}$ $\in$ $\mathcal{P}(\mathcal{I})$ et
$O$, $O_{1}$, $O_{2}$ $\in$ $\mathcal{P}(\mathcal{O})$. Nous avons
les propriétés suivantes données par la table \ref{tableprop}.
\begin{table}[htbp]
\begin{center}
\begin{tabular}{ll}
  (1) $O_{1}$ $\subseteq$ $O_{2}$ $\Rightarrow$ $f_{d}$($O_{1}$) $\subseteq$ $f_{d}$($O_{2}$) & (1') $I_{1}$ $\subseteq$ $I_{2}$ $\Rightarrow$ $g_{d}$($I_{1}$) $\subseteq$ $g_{d}$($I_{2}$)\\
   (2) $I$ $\subseteq$ $h_{d}$($I$) & (2') $h'_{d}$($O$) $\subseteq$ $O$ \\
   (3) $I_{1}$ $\subseteq$ $I_{2}$ $\Rightarrow$ $h_{d}$($I_{1}$) $\subseteq$ $h_{d}$($I_{2}$) & (3')
   $O_{1}$ $\subseteq$ $O_{2}$ $\Rightarrow$ $h'_{d}$($O_{1}$) $\subseteq$ $h'_{d}$($O_{2}$) \\
   (4) $f_{d}$($O$) = $h'_{d}$($f_{d}$($O$)) & (4') $g_{d}$($I$) = $h_{d}$($g_{d}$($I$))\\
   (5) $h_d$($I$) = $h_d$($h_d$($I$))& (5') $h'_{d}$($O$) = $h'_d$($h'_d$($O$))\\
\multicolumn{2}{c}{(6) $g_{d}$($I$) $\subseteq$ $O$
$\Leftrightarrow$ $I$ $\subseteq$ $f_{d}$($O$)}
\end{tabular}
 \caption{Les propriétés des opérateurs
disjonctifs} \label{tableprop}
\end{center}

\end{table}
\end{proposition}
$h_d$ est un opérateur de fermeture puisqu'il vérifie les
conditions requises, \textit{i.e.}, isotonie, idempotence et
extensivité. En effet, il est extensif (\textit{cf.} Propriété
(2)), isotone (\textit{cf.} Propriété (3)) et idempotent
(\textit{cf.} Propriété (5)) \footnote{$h'_d$ n'est pas un
opérateur de fermeture puisqu'il n'est pas extensif (\textit{cf.}
Propriété (2')). $h'_d$ est dit opérateur d'ouverture
(\cite{ganter99}).}. La définition de la fermeture disjonctive va
nous permettre de proposer une nouvelle représentation concise de
l'ensemble des itemsets fréquents.
\begin{definition}~\cite{tarekjordan06}\label{def-fdci} \textsc{(Itemset fermé disjonctif)} Un itemset $I$ $\subseteq$
$\mathcal{I}$ est dit \textit{fermé disjonctif} \textit{si et
seulement si} $h_{d}$($I$) = $I$. Un itemset fermé disjonctif est
l'ensemble maximum des items contenus uniquement dans l'ensemble
des transactions dans lesquelles apparaît au moins un item de $I$
et qui n'apparaissent nul part ailleurs.
\end{definition}


L'ensemble des itemsets fermés disjonctifs est défini comme suit :
\begin{definition}~\cite{tarekjordan06}\label{set-IFFs} \textsc{(Ensemble des itemsets fermés disjonctifs essentiels)}
Soit $\mathcal{K}$ un contexte d'extraction et $h_{d}$ l'opérateur
de fermeture disjonctive. L'ensemble $\mathcal{IFDE}_\mathcal{K}$
($\mathcal{I}$temsets $\mathcal{F}$ermés $\mathcal{D}$isjonctifs
$\mathcal{E}$ssentiels), extrait à partir de $\mathcal{K}$, est
défini comme suit : $\mathcal{IFDE}_\mathcal{K}$ =
$\{$$h_{d}$($I$) $|$ $I$ $\in$ $\mathcal{IEF}_\mathcal{K}$$\}$.

\end{definition}

\subsection{Repr{\'e}sentation concise exacte fondée sur les
ferm{\'e}s disjonctifs}

Dans cette section, nous introduisons, à l'aide d'une suite de
propositions, la nouvelle repr{\'e}sentation concise exacte
bas{\'e}e sur les ferm{\'e}s disjonctifs.
\begin{proposition}\label{propositionitemsetfr{\'e}quentcouvertparunferm{\'e}}\cite{tarekcla07}
Soit \textit{I} un itemset, alors nous avons:
\begin{center}
$(I\in \mathcal{IF}_\mathcal{K})\Rightarrow(\exists\mbox{ }F
\mbox{ un ferm{\'e} disjonctif relatif à un essentiel
fr{\'e}quent}\mid I\subseteq F)$
\end{center}
\end{proposition}

\begin{proposition}\label{propositionfermetureitemsetsfr{\'e}quent}\cite{tarekcla07}
Soit $I\in \mathcal{IF}_\mathcal{K}$, alors $h_d(I)$ est le plus
petit ferm{\'e} disjonctif relatif à un essentiel fr{\'e}quent
contenant I. En outre, le support disjonctif d'un itemset $I$ est
égal à celui du plus petit itemset fermé disjonctif le contenant.
\end{proposition}


L'ensemble $\mathcal{IFDE}_\mathcal{K}$ à lui seul ne peut pas
constituer une repr{\'e}sentation concise exacte de l'ensemble des
itemsets fr{\'e}quents. Pour montrer cela, supposons que avons
disposons d'une m{\'e}thode de r{\'e}g{\'e}n{\'e}ration, qui
consiste à affecter au support disjonctif de chaque itemset
candidat $X_i$ le support disjonctif du plus petit ferm{\'e}
disjonctif contenant $X_i$ et à calculer le support conjonctif de
$X_i$ en utilisant les identit{\'e}s d'inclusion-exclusion. Ainsi,
il s'est av{\'e}r{\'e} que, dans certains cas, cette m{\'e}thode
g{\'e}n{\`e}re un ensemble diff{\'e}rent de l'ensemble des
itemsets fr{\'e}quents. En effet, consid{\'e}rons le contexte
illustr{\'e} par la table \ref{exemplecontexteformel10}.
\begin{table}[!ht]
\begin{center}
\begin{tabular}{|c|c|c|c|c|c|c|}
  \hline
   & $\textsc{A}$ & $\textsc{B}$ & $\textsc{C}$ & $\textsc{D}$ \\
  \hline
  $1$ & $\times $ &  &  &   \\ \hline
  $2$ & $\times $ & $\times $ &  &   \\ \hline
  $3$ & $\times $ &  & $\times $&  \\ \hline
  $4$ & $\times $ &  & & $\times $ \\ \hline
  $5$ & $\times $ &$\times $  & $\times $ &  \\ \hline
  $6$ & $\times $ &$\times $  &  &  $\times $\\ \hline
  $7$ & $\times $ &  & $\times $ &  $\times $\\
  \hline
\end{tabular}
\end{center}
\caption{Contexte d'extraction.}\label{exemplecontexteformel10}
\end{table}
Pour $minsup = 1$, nous avons $\mathcal{IFDE}_\mathcal{K}$ $=$
$\{$$(\textsc{B},3)$, $(\textsc{C},3)$, $(\textsc{D},3)$,
$(\textsc{BC},5)$, $(\textsc{BD},5)$, $(\textsc{CD},5)$,
$(\textsc{ABCD},7)$$\}$. Dans cet exemple, le support disjonctif
de n'importe quel itemset de taille inf{\'e}rieure à 2 peut être
correctement calcul{\'e} à partir de $\mathcal{IFDE}_\mathcal{K}$.
Par contre, si nous voulons calculer le support de $\textsc{BCD}$
en utilisant la m{\'e}thode pr{\'e}c{\'e}demment d{\'e}crite, nous
affectons au support disjonctif de $\textsc{BCD}$ la valeur $7$
alors que son support disjonctif r{\'e}el est {\'e}gal à $6$. En
calculant le support conjonctif de $\textsc{BCD}$ par le biais des
identit{\'e}s d'inclusion-exclusion, nous obtenons
$Supp(\textsc{BCD}) = 7 - 5 - 5 - 5+ 3 + 3 + 3 = 1$ alors que le
support conjonctif exact de $\textsc{BCD}$ est {\'e}gal à $0$.
Ainsi, $\textsc{BCD}$ sera ins{\'e}r{\'e} dans l'ensemble des
itemsets fr{\'e}quents alors qu'en r{\'e}alit{\'e}, il est
infr{\'e}quent.

Afin de pallier cette lacune, une m{\'e}thode directe consiste à
enrichir l'ensemble $\mathcal{IFDE}_\mathcal{K}$ avec un autre
ensemble d'itemsets. La caract{\'e}risation de ces itemsets
d{\'e}pend {\'e}troitement de la m{\'e}thode selon laquelle le
support conjonctif sera calcul{\'e} ainsi que de la
caract{\'e}risation des itemsets dont le support conjonctif est
calcul{\'e} de façon erron{\'e}e. Dans ce qui suit, nous supposons
que notre m{\'e}thode de r{\'e}g{\'e}n{\'e}ration des itemsets
fr{\'e}quents parcourt l'espace de recherche des itemsets
fr{\'e}quents en largeur. Cette contrainte est impos{\'e}e par
l'application des égalités des identit{\'e}s
d'inclusion-exclusion, qui n{\'e}cessite la connaissance des
supports disjonctifs des sous-ensembles d'un itemset afin de
pouvoir calculer son support conjonctif.

\medskip

Ainsi, nous nous proposons de substituer l'ensemble des essentiels
fr{\'e}quents par un autre ensemble. Cet ensemble, not{\'e} par la
suite $\mathcal{IFDA}_\mathcal{K}$ ($\mathcal{I}$temsets
$\mathcal{F}$erm{\'e}s $\mathcal{D}$isjonctifs
$\mathcal{A}$jout{\'e}s), est introduit dans ce qui suit.

%
%
%
\begin{definition}~\cite{tarekcla07}\label{definition_fda}
Soit $BD^-(\mathcal{IEF}_\mathcal{K})=\{\min_{\subseteq}\{Y\in
\cal{P}(\mathcal{I})\setminus
\mathcal{\mathcal{IEF}_\mathcal{K}}\}\}$ l'ensemble des itemsets
non essentiels ou infr{\'e}quents, tels
que tous leurs sous-ensembles stricts sont des itemsets essentiels fr{\'e}quents.\\
Nous d{\'e}finissons l'ensemble $\mathcal{IFDA}_\mathcal{K}$ comme
suit: {\setlength\arraycolsep{5pt}
\begin{center}
$\mathcal{IFDA}_\mathcal{K} = \{h_d(X) \mid (X\in
BD^-(\mathcal{IEF}_\mathcal{K})\cap \mathcal{IE}) \ \wedge \
((-1)^{\mid X\mid} = -1) \ \wedge \ $\\$ (\forall X'\subseteq
\mathcal{I}, \mbox{ }h_d(X') = h_d(X) \Rightarrow Supp(X') <
minsup) \ \wedge \ (\exists f \in \mathcal{IFDE}_\mathcal{K} \
\textit{s.t.} \ h_d(X) \subset f)\}$
\end{center}}
\end{definition}

Autrement dit, $\mathcal{IFDA}_\mathcal{K}$ est l'ensemble des
ferm{\'e}s disjonctifs ayant tous leurs essentiels infréquents et
dont au moins un appartient à $BD^-(\mathcal{IEF}_\mathcal{K})\cap
\mathcal{IE}$ et est de taille impaire. Chaque fermeture de cet
ensemble doit être couverte par au moins un itemset fermé
disjonctif faisant partie de $\mathcal{IFDE}_\mathcal{K}$ pour
être maintenue.

En ajoutant $\mathcal{IFDA}_\mathcal{K}$ à l'ensemble
$\mathcal{IFDE}_\mathcal{K}$, nous pouvons garantir un processus
correct de r{\'e}g{\'e}n{\'e}ration des itemsets fr{\'e}quents à
partir de ce nouvel ensemble. Cette affirmation est formellement
décrite par le th{\'e}or{\`e}me suivant.
\begin{theorem}\label{th{\'e}or{\`e}me_repr{\'e}sentation_bas{\'e}e sur fd}
L'ensemble des itemsets de $\mathcal{IFDE}_\mathcal{K}\cup
\mathcal{IFDA}_\mathcal{K}$ munis de leurs supports disjonctifs
est une représentation concise exacte~\cite{tarekcla07}.
\end{theorem}

Le lemme suivant permet la dérivation du support disjonctif des
itemsets fréquents à partir de ceux des éléments de l'ensemble
$\mathcal{IFDE}$.
\vspace{-0.2cm}
\begin{lemma}\label{derivatin_support_disjonctif_ifde}\textsc{\textsc{(}}\cite{tarekcla07}\textsc{\textsc{)}} - Soient $\mathcal{IF}$ l'ensemble des itemsets fréquents, $I$
$\subseteq$ $\mathcal{I}$ un itemset et $I_{min}$ $=$
$\min{}_\subseteq\{X\in \mathcal{IFDE}\mid I\subseteq X\}$ s'il
existe. Alors, nous avons: \vspace{-0.4cm}
$$I\in \mathcal{IF}~ \Rightarrow ~\textsc{\textsc{(}}\exists X\in \mathcal{IFDE}\mid I\subseteq X\textsc{\textsc{)}} \wedge \textsc{\textsc{(}}Supp\textsc{\textsc{\textsc{(}}}\vee I\textsc{\textsc{\textsc{)}}} = Supp\textsc{\textsc{\textsc{(}}}\vee I_{min}\textsc{\textsc{\textsc{)}}}\textsc{\textsc{)}}.$$
\end{lemma}

Le statut de fréquence d'un itemset quelconque $I$ ainsi est
déterminé son support conjonctif avec exactitude sont déterminés
comme suit. Nous débutons par la dérivation des supports
disjonctifs et le calcul des supports conjonctifs des
sous-ensembles stricts avec la même méthode utilisée pour traiter
$I$. Dans le cas où $I$ contient un sous-ensemble strict qui est
infréquent, nous déduisons son infréquence \textsc{\textsc{(}}car
les itemsets fréquents constituent un idéal
d'ordre\textsc{\textsc{)}}. Dans le cas échéant
\textsc{\textsc{(}}\textit{i.e.}, tous les sous-ensembles stricts
de $I$ sont fréquents\textsc{\textsc{)}}, deux possibilités se
présentent:
\begin{enumerate}
\item \textbf{$I$ est fréquent}: Dans ce cas, nous pouvons
déterminer correctement $Supp\textsc{\textsc{\textsc{(}}}\vee
I\textsc{\textsc{\textsc{)}}}$ à partir de ceux des éléments de
l'ensemble $\mathcal{IFDE}$ grâce au lemme
\ref{derivatin_support_disjonctif_ifde}.
    Ainsi, en évaluant la somme de l'égalité 1 du lemme
    \ref{lemmaidentitésinclusionexclusion} par rapport à $I$, la valeur obtenue correspond au support conjonctif exact de
    $I$.
    \item \textbf{$I$ est infréquent}: Soit $I_{min}$ $=$ $\min{}_\subseteq\{X\in\mathcal{IFDE}\mid I\subseteq X\}$. Alors, deux sous-cas se
    présentent:
    \begin{enumerate}
    \item \textbf{$I$ est de taille paire}: Dans ce cas, nous attribuons à
    $Supp\textsc{\textsc{\textsc{(}}}\vee I\textsc{\textsc{\textsc{)}}}$ le support disjonctif de $I_{min}$.
    Étant donné que $I$ est infréquent \textsc{\textsc{(}}\textit{i.e.}, $Supp\textsc{\textsc{\textsc{(}}}I\textsc{\textsc{\textsc{)}}}$ $<$ \textit{minsup}\textsc{\textsc{)}}, alors $Supp\textsc{\textsc{\textsc{(}}}\vee I\textsc{\textsc{\textsc{)}}}$ peut être strictement inférieur à
    celui de $I_{min}$ \textsc{\textsc{\textsc{(}}}car $I\subseteq I_{min}$\textsc{\textsc{\textsc{)}}}. En dépit de cette estimation erronée de $Supp\textsc{\textsc{\textsc{(}}}\vee
    I\textsc{\textsc{\textsc{)}}}$, nous
    pouvons détecter l'infréquence de $I$. En effet, lorsque la taille de $I$ est paire, nous avons $Supp\textsc{\textsc{\textsc{(}}}I\textsc{\textsc{\textsc{)}}}$
    $=$ $\sum_{I'\subseteq I}\textsc{\textsc{(}}-1\textsc{\textsc{)}}^{\mid I' - 1\mid}Supp\textsc{\textsc{\textsc{(}}}\vee
    I'\textsc{\textsc{\textsc{)}}}$ $=$ $-Supp\textsc{\textsc{\textsc{(}}}\vee I\textsc{\textsc{\textsc{)}}}$ $+$ $\sum_{I'\subset I}\textsc{\textsc{(}}-1\textsc{\textsc{)}}^{\mid I'\mid - 1}Supp\textsc{\textsc{\textsc{(}}}\vee
    I'\textsc{\textsc{\textsc{)}}}$. Étant donné que $Supp\textsc{\textsc{\textsc{(}}}\vee I\textsc{\textsc{\textsc{)}}}$ $\leq$ $Supp\textsc{\textsc{\textsc{(}}}\vee I_{min}\textsc{\textsc{\textsc{)}}}$, nous obtenons $-Supp\textsc{\textsc{\textsc{(}}}\vee I_{min}\textsc{\textsc{\textsc{)}}}$ $+$
$\sum_{I'\subset I}\textsc{\textsc{(}}-1\textsc{\textsc{)}}^{\mid
I'\mid - 1}Supp\textsc{\textsc{\textsc{(}}}\vee
    I'\textsc{\textsc{\textsc{)}}}$ $\leq$ $-Supp\textsc{\textsc{\textsc{(}}}\vee I\textsc{\textsc{\textsc{)}}}$ $+$ $\sum_{I'\subset I}\textsc{\textsc{(}}-1\textsc{\textsc{)}}^{\mid I'\mid - 1}Supp\textsc{\textsc{\textsc{(}}}\vee
    I'\textsc{\textsc{\textsc{)}}}$. Par conséquent, le support conjonctif calculé va être
    inférieur au support conjonctif correct. Étant donné que $I$ est infréquent par hypothèse,
    le support conjonctif calculé de $I$ sera strictement inférieur à
    \textit{minsup}. Ainsi, nous détectons correctement l'infréquence de $I$.
    \item \textbf{$I$ est de taille impaire}: En attribuant au support disjonctif de  $I$ celui de $I_{min}$, le statut déterminé de la fréquence de $I$  peut être erroné.
    En effet, puisque la taille de $I$ est impaire, nous obtenons $Supp\textsc{\textsc{\textsc{(}}}I\textsc{\textsc{\textsc{)}}}$
    $=$ $\sum_{I'\subseteq I}\textsc{\textsc{(}}-1\textsc{\textsc{)}}^{\mid I' - 1\mid}$$Supp\textsc{\textsc{\textsc{(}}}\vee
    I'\textsc{\textsc{\textsc{)}}}$ $=$ $Supp\textsc{\textsc{\textsc{(}}}\vee I\textsc{\textsc{\textsc{)}}}$ $+$ $\sum_{I'\subset I}\textsc{\textsc{(}}-1\textsc{\textsc{)}}^{\mid I'\mid - 1}Supp\textsc{\textsc{\textsc{(}}}\vee
    I'\textsc{\textsc{\textsc{)}}}$. Étant donné que $Supp\textsc{\textsc{\textsc{(}}}\vee I\textsc{\textsc{\textsc{)}}}
    \leq Supp\textsc{\textsc{\textsc{(}}}\vee I_{min}\textsc{\textsc{\textsc{)}}}$, nous obtenons $Supp\textsc{\textsc{\textsc{(}}}\vee I\textsc{\textsc{\textsc{)}}}$ $+$ $\sum_{I'\subset I}\textsc{\textsc{(}}-1\textsc{\textsc{)}}^{\mid I'\mid - 1}Supp\textsc{\textsc{\textsc{(}}}\vee
    I'\textsc{\textsc{\textsc{)}}}$ $\leq$ $Supp\textsc{\textsc{\textsc{(}}}\vee I_{min}\textsc{\textsc{\textsc{)}}}$ $+$ $\sum_{I'\subset I}\textsc{\textsc{(}}-1\textsc{\textsc{)}}^{\mid I'\mid - 1}Supp\textsc{\textsc{\textsc{(}}}\vee
    I'\textsc{\textsc{\textsc{)}}}$. D'où, le support conjonctif calculé de $I$ peut
    être strictement supérieur au support conjonctif exact voire dépasser le seuil
    \textit{minsup}. Ainsi, en attribuant au support disjonctif de $I$ celui de $I_{min}$, nous pouvons inférer la fréquence de $I$ alors qu'en réalité, cet itemset est infréquent.
    Afin d'éviter le calcul erroné du support conjonctif de $I$, la
    détermination \textit{exacte} de son support disjonctif s'impose. Pour ce faire, il suffit d'ajouter $h_d\textsc{\textsc{\textsc{(}}}I\textsc{\textsc{\textsc{)}}}$, munie de son support disjonctif, à l'ensemble
    $\mathcal{IFDE}$. Étant donné que $I$ est un itemset essentiel
    infréquent de taille impaire tel que tous ses sous-ensembles
    stricts sont fréquents \textsc{\textsc{(}}\textit{i.e.}, $I\in\mathcal{IEII}$\textsc{\textsc{)}}, nous déduisons que $h_d\textsc{\textsc{(}}I\textsc{\textsc{)}}$ appartient à
    $\mathcal{IFDA}$. Ainsi, l'ensemble des fermés disjonctifs de $\mathcal{IFDA}$, munis de leurs supports disjonctifs, permet la détection exacte de l'infréquence de $I$ en adoptant le même mécanisme que dans le sous-cas précédent.
\end{enumerate}
\end{enumerate}

Il est important de noter que la représentation sus-mentionnée est
:
\begin{enumerate}
    \item \textbf{Homog{\`e}ne:} En effet, tous ses {\'e}l{\'e}ments sont des ferm{\'e}s disjonctifs.
    \item \textbf{De taille r{\'e}duite}: En effet, en
se basant sur le fait que les éléments de l'ensemble
$\mathcal{IFDA}_\mathcal{K}$ doivent v{\'e}rifier une multitude de
conditions, nous pouvons affirmer que la cardinalité de l'ensemble
$\mathcal{IFDA}_\mathcal{K}$ sera, dans la plupart des cas,
largement inf{\'e}rieure à celle de l'ensemble
$\mathcal{IFDE}_\mathcal{K}$. En outre, puisque l'ensemble
$\mathcal{IFDE}_\mathcal{K}$ est de taille largement
inf{\'e}rieure à celle l'ensemble des itemsets fr{\'e}quents, nous
pouvons d{\'e}duire que l'ensemble $\mathcal{IFDE}_\mathcal{K}\cup
\mathcal{IFDA}_\mathcal{K}$ est de cardinalit{\'e} inf{\'e}rieure
à celle de l'ensemble des itemsets fr{\'e}quents.

Toutefois, la représentation concise exacte, que nous proposons,
ne constitue pas  une représentation parfaite, puisque sa
cardinalit{\'e} peut parfois d{\'e}passer celle de l'ensemble des
itemsets fr{\'e}quents. En effet, consid{\'e}rons le contexte
donn{\'e} par la Figure \ref{contexte_non_couverture_parfaite}.
Pour $minsup = 1$, l'ensemble des itemsets fr{\'e}quents est
{\'e}gal à $\mathcal{IF}_\mathcal{K}$ $=$ \{\mbox{A}, \mbox{B},
\mbox{C}, \mbox{D}, \mbox{AB}, \mbox{AC}, \mbox{AD}, \mbox{BC},
\mbox{BD}, \mbox{CD}, \mbox{ABC}\} alors que
$\mathcal{IFDE}_\mathcal{K}\cup \mathcal{IFDA}_\mathcal{K}$ $=$
\{\mbox{A}, \mbox{B}, \mbox{C}, \mbox{D}, \mbox{AB}, \mbox{AC},
\mbox{AD}, \mbox{BC}, \mbox{BD}, \mbox{CD}, \mbox{ABD},
\mbox{BCD}, \mbox{ABCD}\}. En comptabilisant la cardinalit{\'e} de
chaque ensemble, nous remarquons que
$\mid\mathcal{IF}_\mathcal{K}\mid = 11$ alors que $\mid
\mathcal{IFDE}_\mathcal{K}\cup \mathcal{IFDA}\mid = 13$.

     \item \textbf{Sans surcoût en temps extraction}: Afin de
     construire la représentation concise fondée sur les essentiels
     fréquents, l'algorithme \textsc{MEP}~\cite{casalidawak105} fait appel à un algorithme
     \emph{externe} d'extraction des itemsets maximaux fréquents ce qui engendre un surcoût
     en temps d'extraction. La représentation
     fondée sur les fermés disjonctifs a pu éviter un tel surcoût.
     En effet, si nous utilisons un algorithme du type "Générer-et-tester", adoptant la
     stratégie "en largeur d'abord" lors du parcours de l'espace de
     recherche, les fermés disjonctifs de $\mathcal{IFDA}_\mathcal{K}$ peuvent être extraits au fur et à mesure de l'extraction
     des fermés disjonctifs des essentiels
     fréquents.
\end{enumerate}

\begin{figure}[!ht]
  \centering
\begin{tabular}{|c|c|c|c|c|}
  \hline
   & \textsc{A} & \textsc{B} & \textsc{C} & \textsc{D} \\
  \hline
  1 & $\times$ &  &  &  \\
  \hline
  2 & $\times$ &  &  & $\times$ \\
  \hline
  3 & $\times$ &  & $\times$ &  \\
  \hline
  4 & $\times$ & $\times$ & $\times$ &  \\
  \hline
  5 & $\times$ & $\times$ &  &  \\
  \hline
  6 &  & $\times$ &  & $\times$ \\
  \hline
  7 &  & $\times$ & $\times$ &  \\
  \hline
  8 &  &  & $\times$ &  \\
  \hline
  9 &  &  & $\times$ & $\times$ \\
  \hline
\end{tabular}
  \caption{Contexte d'extraction}\label{contexte_non_couverture_parfaite}
\end{figure}

\subsection{L'algorithme \textsc{DCPR-Miner}}
\begin{table}[!ht]

\begin{center}
{\small
\begin{tabular}{|ll|}
\hline
$\mathcal{K}$ & Contexte d'extraction\\
$C_{i}$ & L'ensemble des itemsets essentiels fréquents
candidats de taille $i$.\\
$L_{i}$ & L'ensemble des itemsets essentiels fréquents
de taille $i$.\\
$\mathcal{IFDE}_{i}$ & L'ensemble des itemsets fermés disjonctifs,
relatifs aux essentiels fréquents de\\ &  taille $i$.\\
$\mathcal{IFDA}_{i}$ & L'ensemble des itemsets fermés disjonctifs,
relatifs aux essentiels infréquents de\\ &  taille impaire égale à  $i$.\\

$X_i$ & Itemset de taille $i$.\\
$X_{i}.h_d$ & Fermeture disjonctive de l'itemset $X_{i}$. \\
$X_{i}.h_d^-$ & L'ensemble des items qui ne doivent pas appartenir
à $X_{i}.h_d$. Autrement dit, c'est \\ & l'ensemble des items qui
sont apparus dans les transactions qui ne contiennent
\\ &aucun item appartenant à $X_i$.\\
$X_{i}.Supp\_Conj$ &  Support conjonctif de $X_{i}$.\\
$X_{i}.Supp\_Disj$ &  Support disjonctif de $X_{i}$.\\
\hline
\end{tabular}
}
\end{center}
\caption{Notations utilisées dans l'algorithme
\textsc{DCPR-Miner}.}\label{tab_notation}
\end{table}

Afin d'extraire l'ensemble $\mathcal{IFDE}_\mathcal{K}\cup
\mathcal{IFDA}_\mathcal{K}$, nous proposons un algorithme appelé
\textsc{DCPR-Miner} \footnote{\textsc{DCPR\_Miner} est l'acronyme
de \underline{D}isjucntive \underline{C}losed
\underline{P}attern-based \underline{R}epresentation
\underline{Miner}}, de type \textit{"Générer-et-tester"}. Afin
d'expliciter les stratégies d'élagage utilisées, nous introduisons
la proposition suivante.
\begin{proposition}\label{propositionsupportitemsetcontenudanslafermeturedisjonctivedelundesesensembles}
Soient $X$ et $Y$ deux itemsets tels que $X\subseteq Y$. Si
$Y\subseteq\mbox{ }h_{d}(X)$, alors $h_d(X)$ = $h_d(Y)$ et
$Supp(\vee X)$ = $ Supp(\vee Y)$.
\end{proposition}

\begin{table}[!ht] {\small
\begin{center}
\begin{tabular}{|l|}
  \hline
  Algorithme 1 : \textsc{DCPR-Miner}\\
  \hline
\textbf{Entrées :} Le contexte d'extraction $\mathcal{K}$ et
le support minimal \textit{minsup}.\\
\textbf{Sortie :} L'ensemble $\mathcal{IFDE}_\mathcal{K}\cup
\mathcal{IFDA}_\mathcal{K}$.\\
  1 :\hspace{20pt}$i$ = $1$ ;\\
  2 :\hspace{20pt}$C_i$= $\mathcal{I}$ ; $L_i$= $\emptyset$; $\mathcal{IFDE}_{i}$=$\emptyset$; $\mathcal{IFDA}_{i}$=$\emptyset$;\\
  3 :\hspace{20pt}Tant que ($C_{i}\neq\emptyset$) faire\\
  4 :\hspace{40pt}$\textsc{Calcul\_supports\_fermeture}(\mathcal{K}, \mbox{ }minsup, \mbox{ }C_{i}, \mbox{ }L_{i},
  \mbox{ }\mathcal{IFDE}_{i}
  \mbox{ }\mathcal{IFDA}_{i})$ ; /*\textit{L'élagage}\\
  \hspace{50pt}\textit{par rapport à $minsup$ s'effectue au sein} \textit{de cette procédure}*/\\
  5 :\hspace{40pt}$C_{i+1}$ = \textsc{Apriori\_Gen}($L_{i}$) ;\\
  6 :\hspace{40pt}$C_{i+1}$ = \{$X_{i+1}$ $\in$ $C_{i+1}\mid$ $\forall$ $X_{i}$ $\subset$ $X_{i+1}$, ($X_{i}$
  $\in$ $L_{i}$) $\wedge$ ($X_{i+1}$ $\not\subseteq$ $X_{i}.h_d)$\} ; /*\\
  \hspace{50pt}\textit{Élagage des itemsets essentiels candidats par rapport à l'idéal}\\
  \hspace{50pt}\textit{ d'ordre et par rapport à l'inclusion du candidat dans la fermeture}\\
  \hspace{50pt}\textit{disjonctive de l'un de ses sous-ensembles immédiats}*/\\
  7 :\hspace{40pt}$i$ = $i$ + $1$ ;\\
  8 :\hspace{20pt}fin Tant que\\
  9: \hspace{20pt}$\mathcal{IFDE}_\mathcal{K}$ =
  $\cup_{j=1,\ldots,i} \ \mathcal{IFDE}_j$ ;\\
  10 :\hspace{20pt}$\mathcal{IFDA}_\mathcal{K} = \{X\in \cup_{j=1,\ldots,i} \ \mathcal{IFDA}_{j}\mid \exists Y\in \mathcal{IFDE}_\mathcal{K}, X\subset Y\} $ ;\\
  11 :\hspace{15pt}retourner $\mathcal{IFDE}_\mathcal{K} \cup \mathcal{IFDA}_\mathcal{K}$ ;\\
  \hline
\end{tabular}
\end{center}
}
\end{table}

Les stratégies d'élagage utilisées par \textsc{DCPR-Miner} sont
les suivantes :
\begin{enumerate}
    \item Un élagage par rapport au support conjonctif des
itemsets essentiels candidats (\textit{i.e.}, par rapport au
support minimal \textit{minsup}) ;
    \item Un élagage par
rapport à la propriété d'idéal d'ordre que doit vérifier
l'ensemble des itemsets essentiels fréquents ;
    \item Un élagage par rapport à l'inclusion
dans la fermeture de l'un des sous-ensembles (\textit{cf.}
Proposition
\ref{propositionsupportitemsetcontenudanslafermeturedisjonctivedelundesesensembles}).
\end{enumerate}

Les notations utilisées dans notre algorithme sont données dans la
table \ref{tab_notation}. Le pseudo-code de l'algorithme
\textsc{DCPR-Miner} est décrit dans Algorithme 1.
\textsc{Apriori\_Gen} est la procédure utilisée dans \cite{Agra94}
pour générer les candidats de taille $(i$ + $1)$ à partir des
éléments retenus de taille $i$. Le pseudo-code de la procédure
\textsc{Calcul\_supports\_fermeture} est donné par Procédure 1.
Cette procédure permet de calculer, en effectuant un seul parcours
du contexte d'extraction, les supports conjonctifs et disjonctifs
ainsi que la fermeture disjonctive de chaque itemset $X_{i}$
appartenant à $C_{i}$.

\begin{table}[htbp]
{\small
\begin{center}
\begin{tabular}{|l|}
  \hline
  Procédure 1 : \textsc{Calcul\_supports\_fermeture} $(\mathcal{K}, \mbox{ }minsup, \mbox{ }C_{i}, \mbox{ }\mbox{ } L_{i},\mbox{ } \mathcal{IFDE}_{i},\mbox{ } \mathcal{IFDA}_{i})$\\
  \hline
  1 :\hspace{20pt}Pour chaque (transaction $T$ $\in$ $\mathcal{K}$) faire\\
  2 :\hspace{40pt}Pour chaque (itemset $X_{i}$ $\in$ $C_{i}$) faire\\
  3 :\hspace{60pt}$\Omega$ = $X_{i}$ $\cap$ $T$ ;\\
  4 :\hspace{60pt}Si ($\Omega$ = $\emptyset$) alors\\
  5 :\hspace{80pt}$X_{i}.h_d^-$ = $X_{i}.h_d^-$ $\cup$ $T$ ;\\
  6 :\hspace{70pt}Sinon\\
  7 :\hspace{80pt}$X_{i}.Supp\_Disj$ = $X_{i}.Supp\_Disj$ + $1$ ;\\
  8 :\hspace{80pt}Si ($\Omega$ = $X_{i}$) alors\\
  9 :\hspace{100pt}$X_{i}.Supp\_Conj$ = $X_{i}.Supp\_Conj$ + $1$ ;\\
  10 :\hspace{75pt}fin Si\\
  11 :\hspace{55pt}fin Si\\
  12 :\hspace{35pt}fin Pour\\
  13 :\hspace{15pt}fin Pour\\
  14 :\hspace{15pt}Si($i == 1$ ou $i \mod 2 == 0$) alors\\
   15:\hspace{35pt}Pour chaque (itemset $X_{i} \in C_{i}$)
  faire\\
   16:\hspace{55pt}Si ($X_{i}.Supp\_Conj$ $\geq$ $minsup$) alors\\
   17:\hspace{75pt}$L_{i}$ = $L_{i}$ $\cup$ $\{X_{i}\}$ ;\\
   18:\hspace{75pt}$X_{i}.h_d$ = $\mathcal{I}$ $\setminus$ $X_{i}.h_d^-$ ;\\
   19:\hspace{75pt}$\mathcal{IFDE}_{i}$ = $\mathcal{IFDE}_{i}$ $\cup$ \{($X_{i}.h_d$, $X_{i}.Supp\_Disj$)\} ;\\
   20:\hspace{55pt}fin Si\\
   21:\hspace{35pt}fin Pour\\
   22:\hspace{15pt}Sinon\\
   23:\hspace{35pt}Pour chaque (itemset $X_{i} \in C_{i}$) faire\\
   24:\hspace{55pt}Si ($X_{i}.Supp\_Conj$ $\geq$ $minsup$) alors\\
   25:\hspace{75pt}$L_{i}$ = $L_{i}$ $\cup$ $\{X_{i}\}$ ;\\
   26:\hspace{75pt}$X_{i}.h_d$ = $\mathcal{I}$ $\setminus$ $X_{i}.h_d^-$ ;\\
   27:\hspace{75pt}$\mathcal{IFDE}_{i}$ = $\mathcal{IFDE}_{i}$ $\cup$ \{($X_{i}.h_d$, $X_{i}.Supp\_Disj$)\} ;\\
   28:\hspace{55pt}Sinon \\
   29:\hspace{75pt}$\mathcal{IFDA}_{i}$ = $\mathcal{IFDA}_{i}$ $\cup$ \{($X_{i}.h_d$, $X_{i}.Supp\_Disj$)\} ;\\
   30:\hspace{55pt}fin Si\\
   31:\hspace{35pt}fin Pour\\
   32:\hspace{15pt}fin Si\\
  \hline
\end{tabular}
\end{center}
}
\end{table}

\subsection{Évaluation expérimentale}

Dans ce qui suit, nous présentons quelques résultats que nous
avons obtenus en comparant la taille de notre représentation avec
l'ensemble total des itemsets fréquents ainsi que les trois autres
types de représentations : (\emph{i}) celle utilisant l'ensemble
des itemsets non dérivables fréquents; (\emph{ii}) celle fondée
sur les itemsets fermés fréquents; et (\emph{iii}) celle utilisant
les itemsets essentiels fréquents.

Les résultats obtenus sont présentés dans les Tables
\ref{résultatsconnect}-\ref{résultatst10}. \`{A} la lecture de ces
tables, nous pouvons noter ce qui suit :
\begin{enumerate}
    \item Même pour des valeurs élevées de
\textit{minsup}, la cardinalité notre représentation concise
$\mathcal{IFD}_\mathcal{K}$ est réduite par rapport à celle des
itemsets fréquents $\mathcal{IF}_\mathcal{K}$;
    \item  Dans le cas des contextes \textsc{Connect} et \textsc{Chess},
la cardinalité de $\mathcal{IFD}_\mathcal{K}$ est largement
inférieure à celle des itemsets fermés fréquents
$\mathcal{IFF}_\mathcal{K}$;
    \item La cinquième colonne montre l'effet néfaste de
l'addition de l'ensemble $BD^{+}(\mathcal{IF}_\mathcal{K})$ par
rapport au taux de compacité réalisé par l'ensemble des itemsets
essentiels fréquents $\mathcal{IEF}_\mathcal{K}$. L'ensemble
$\mathcal{IFD}_\mathcal{K}$ a surmonté cet handicap en excluant
$BD^{+}(\mathcal{IF}_\mathcal{K})$ de la représentation concise.
    \item Pour le contexte épars \textsc{T10I4D100K}, nous
remarquons que nous arrivons quand même à obtenir un taux de
compacité acceptable, pour des valeurs faibles de \textit{minsup}.
Il est important de souligner que ce type de contextes est
considéré comme "difficile", vu que les approches fondées sur les
fermetures n'apportent pas de gains intéressants sur de tels
contextes. Il est aussi important de souligner que la taille de la
représentation concise fondée sur les itemsets essentiels
fréquents dépasse même celle de l'ensemble de tous les itemsets
fréquents. En effet, les éléments de $\mathcal{IEF}_\mathcal{K}$
sont caractérisés uniquement par leurs supports disjonctifs alors
que ceux de $BD^{+}(\mathcal{IF}_\mathcal{K})$ le sont uniquement
par leurs supports conjonctifs. Pour ce contexte, quelques
éléments font alors partie des deux ensembles et jouent à chaque
fois un rôle différent selon l'ensemble auquel ils appartiennent.
\item Finalement, il est nettement remarquable que la variation de
la cardinalité de $\mathcal{IFD}_\mathcal{K}$ est moins sensible à
la variation de \textit{minsup} que les autres représentations
concises.

\end{enumerate}

\begin{table}[htbp]
\begin{center}
{\small
\begin{longtable}{|p{35pt}||p{130pt}|r|r|p{140pt}|}
   \hline
    \textit{minsup} & $ \mid
\mathcal{IFD}_{\mathcal{K}}\mid=$& $\mid \mathcal{IFF}_\mathcal{K}\mid$ & $\mid \mathcal{IND}_\mathcal{K}\mid$ &$\mid\mathcal{IEF}_\mathcal{K}\mid + \mid BD^{+}(\mathcal{IF}_\mathcal{K})\mid $\\
&$\mid \mathcal{IFDE}_\mathcal{K}\mid +
\mid\mathcal{IFDA}_\mathcal{K}\mid $& & &\\\hline \hline
90\%&22+0=22&3 487&199&176+222=398\\
80\%&83+0=83&15 108&348&304+673=977\\
70\%&161+0=161&35 876&545&490+1 220=1 710\\
60\%&265+28=293&68 344&894&822+2 103=2 925\\
50\%&462+127=589&130 112&1 396& 1 315+3 748=5 063\\
40\%&819+243=1 062&239 373&2 066&1 948+6 213=8 161\\
30\%&1 625+361=1 986&460 357&3 220& 3 044+ 11 039=14 083 \\
20\%&4 725+789=5 514&1 483 199&7 573&6 620+32 583=39 203\\
10\%&19 797+2 603=22 400&8 035 412&29 167&22 942+130 986=153 928\\
5\%&70 798+11 940=82 738&28 384 574&91 050&75 345+413 053=488 398\\
\hline \caption{Taille des différentes représentations concises
pour le contexte \textsc{Connect}.} \label{résultatsconnect}
\end{longtable}
}\end{center}
\begin{center}
{\small
\begin{longtable}{|p{35pt}||p{140pt}|r|r|p{140pt}|}
   \hline
    \textit{minsup} & $ \mid
\mathcal{IFD}_{\mathcal{K}}\mid=$& $\mid \mathcal{IFF}_\mathcal{K}\mid$ & $\mid \mathcal{IND}_\mathcal{K}\mid$ &$\mid\mathcal{IEF}_\mathcal{K}\mid + \mid BD^{+}(\mathcal{IF}_\mathcal{K})\mid $\\
&$\mid \mathcal{IFDE}_\mathcal{K}\mid +
\mid\mathcal{IFDA}_\mathcal{K}\mid $& & &\\\hline \hline
 40\%&79+12=91&140&146&110+41=151\\

 30\%&182+31=213&427&329&247+63=310\\

 20\%&759+182=941&1 197&1 141&1 100+158=1 258\\

 10\%&4 432+1 025=5 457&4 885&4 340&5 983+547=6 530\\

 5\%&17 814+2 740=20 554&12 843&11 556&22 965+1 442=24 407\\

 1\%&186 273+10 782=197 055&51 640&486 47&230 474+6 768=237 242\\
\hline \caption{Taille des différentes représentations concises
pour le contexte \textsc{Mushroom}.} \label{résultatsmushroom}
\end{longtable} }
\end{center}
\begin{center}
{\small
\begin{longtable}{|p{35pt}||p{130pt}|r|r|p{140pt}|}
   \hline
    \textit{minsup} & $ \mid
\mathcal{IFD}_{\mathcal{K}}\mid=$& $\mid \mathcal{IFF}_\mathcal{K}\mid$ & $\mid \mathcal{IND}_\mathcal{K}\mid$ &$\mid\mathcal{IEF}_\mathcal{K}\mid + \mid BD^{+}(\mathcal{IF}_\mathcal{K})\mid $\\
&$\mid \mathcal{IFDE}_\mathcal{K}\mid +
\mid\mathcal{IFDA}_\mathcal{K}\mid $& & &\\\hline \hline
 90\%&40+3=43&499&95&84+34=118
\\

 80\%&129+21=150&5 084 & 279&241+226=467
\\

 70\%&349+71=420&23 893
&684&591+891=1 410
\\
 60\%&773+144=917&98 393&1 594&1314+3 323=4 637\\
 50\%&1 695+276=1 971&369 451&3 425&2 809+11 463=14 272
\\
 40\%&3 563+555=4 118&1 361 158&7 172&5 977+38 050=44 027
\\
 30\%&7 957+867=8 824&5 316 468&15 124&13 153+134 624=147 777
\\
 20\%&20 368+2 149=22 517&22 808 625&34 761&33 185+509 355=542 540
\\
 10\%&70 354+5 844=76 198&123 243 073&98 664&114 219+2 339 525=2 453 744
\\
\hline \caption{Taille des différentes représentations concises
pour le contexte \textsc{Chess}.} \label{résultatschess}
\end{longtable}
}
\end{center}
\end{table}

\begin{table}[!ht]
\begin{center}
{\small
\begin{longtable}{|p{35pt}||p{130pt}|r|r|p{140pt}|}
   \hline
    \textit{minsup} & $ \mid
\mathcal{IFD}_{\mathcal{K}}\mid=$& $\mid \mathcal{IFF}_\mathcal{K}\mid$ & $\mid \mathcal{IND}_\mathcal{K}\mid$ &$\mid\mathcal{IEF}_\mathcal{K}\mid + \mid BD^{+}(\mathcal{IF}_\mathcal{K})\mid $\\
&$\mid \mathcal{IFDE}_\mathcal{K}\mid +
\mid\mathcal{IFDA}_\mathcal{K}\mid $& & &\\\hline \hline

5 & 10+0=10 & 11 & 11& 10+10=20\\
4 & 26+0=26 & 27 & 27& 26+26=52\\
3 & 60+0=60&61 & 61& 60+60=120 \\
2 & 155+0=155 & 156 & 156& 155+155=310 \\
1& 385+0=385 & 386 &  386&385+370=755 \\
0,5 & 1 073+0=1 073 & 1074 & 1074 &1 073+585=1 685 \\

0,4 & 2 000+0=2 000 & 1 993 & 1 987 &2 000+761=2 761\\

0,3 & 4 474+2=4 476 &  4 510& 4 377 &4 474+1 293=5 767\\

0,2 & 12 944+6=12 950& 13 108 & 11 475 &12 949+1 938=14 887 \\

0,1 &26  677+5=26 682&  26 807&  24 120 &26 687+4 054=30 741\\

\hline \caption{Taille des différentes représentations concises
pour le contexte \textsc{T10I4D100K}.}  \label{résultatst10}
\end{longtable}
}\end{center}
\end{table}


\begin{table}[!ht]
\begin{center}
\begin{tabular}{|c||c||r||r|r|r|r|}
  \hline
Base &  \textit{minsup} (\%) & $|$ $\mathcal{IFD}_\mathcal{K}$ $|$
& $\frac{| \mathcal{IF}_\mathcal{K} |}{|
\mathcal{IFD}_\mathcal{K}|}$ &
$\frac{|\mathcal{IFF}_\mathcal{K}|}{| \mathcal{IFD}_\mathcal{K}|}$
& $\frac{|\mathcal{IND}_\mathcal{K}|}{|
\mathcal{IFD}_\mathcal{K}|}$& $\frac{| \mathcal{IEF}_\mathcal{K}
|+| BD^{+}(\mathcal{IF}_\mathcal{K}) |}{|
\mathcal{IFD}_\mathcal{K}|}$\\

\hline\hline \textsc{Connect}& 50\% &589 & 149 700,07  &220,90       & 2,37      & 8,60      \\
                & 40\% &1 062 & 320 070,86        &225,40       & 1,95      & 7,68      \\
                & 30\% &1986 & 670 530,40        & 231,80      &  1,62     &  7,09     \\
                & 20\% &5514 & 1 116 704,82        & 268,99      &  1,37     &  7,11     \\
\hline\hline

\textsc{Mushroom}& 20\% &941&    56,94     &      1,27 &     1,21  &     1,34  \\
                & 10\% &5457 &  105,27       &   0,90    &  0,80     &  1,20    \\
                & 5\% &20 554 & 182,71        &  0,62     & 0,56      & 1,19      \\
                & 1\% &197055 &460,54         & 0,26      &0,25       &1,20       \\
\hline\hline

\textsc{Chess}& 60\% &917 &    278,02     &    107,30   &    1,74   &    5,06   \\
                & 50\% &1971 &  645,83       &  187,44     &  1,74     &  7,24     \\
                & 40\% &4118 &  1 563,79       & 330,54      & 1,74      & 10,69       \\
                & 30\% &8824 &  4 225,18       & 602,50      & 1,71      & 16,75       \\
\hline\hline

\textsc{T10I4D100K}& 0,4\% &2000 & 1,00        & 1,00      &0,99       &1,38       \\
                & 0,3\% &4476 &    1,02     &    1,01   &   0,98    &   1,29    \\
                & 0,2\% &12950 &   1,02      &   1,01    &  0,89     &  1,15     \\
                & 0,1\% &26682 &   1,03      &   1,00    &  0,90     &  1,15     \\
\hline

\end{tabular}
\end{center}
\caption{Taux de compacité de la représentation
$\mathcal{IFD}_\mathcal{K}$ \emph{vs.} les représentations
concises connues de la littérature.} \label{table_rates}

\end{table}

La Table~\ref{table_rates} compare les cardinalités obtenues avec
la représentation concise que nous proposons, \emph{vs.} les
représentations concises les plus connues de la littérature.
Ainsi, une lecture transversale de ce tableau permet de mettre en
exergue la compacité de notre représentation concise même par
rapport à celle des itemsets non dérivables, notée
$\mathcal{IND}_\mathcal{K}$. Seul le nombre des itemsets fermés
suit la même évolution que le nombre de disjonctifs fermés. En
effet, le rapport de compacité augmente à fur et mesure que la
valeur de minsup \emph{diminue}.

\section{Conclusion}
Dans ce chapitre, nous avons introduit une nouvelle représentation
concise exacte des itemsets fréquents. Ceci a nécessité la
proposition d'un opérateur de fermeture disjonctive permettant le
calcul des éléments de cette nouvelle représentation. Les
expérimentations que nous avons menées ont permis de mettre en
valeur la compacité de notre représentation. La principale
caractéristique de cette représentation est qu'elle permet
d'inspirer une piste de recherche intéressante que nous
envisageons explorer à l'avenir. Cette piste a pour but de combler
le vide entre les travaux sur les représentations concises, d'une
part, et ceux portant sur la formalisation de bases génériques de
règles associatives, d'autre part. En effet, la représentation
concise que nous avons proposée, permet de dériver aisément les
supports disjonctifs et négatifs des itemsets fréquents. Ainsi, la
génération des règles (génériques) associatives généralisées -- ne
se limitant pas à l'utilisation de l'opérateur de conjonction
entre des items positifs -- deviendrait quasi-immédiate. Ainsi,
les différents types de règles qu'on pourrait extraire sont comme
suit :
\begin{enumerate}
    \item $\mathcal{R}_1$: $x_1\wedge x_2\wedge\ldots\wedge x_n\Rightarrow y_1\wedge y_2\wedge\ldots\wedge
    y_m$.
    \item $\mathcal{R}_2$: $x_1\wedge x_2\wedge\ldots\wedge x_n\Rightarrow y_1\vee
    y_2\vee\ldots\vee y_m$.
    \item $\mathcal{R}_3$: $x_1\wedge x_2\wedge\ldots\wedge x_n\Rightarrow \overline{y_1}\wedge \overline{y_2}\wedge\ldots\wedge
    \overline{y_m}$.
    \item $\mathcal{R}_4$: $x_1\vee x_2\vee\ldots\vee x_n\Rightarrow y_1\wedge y_2\wedge\ldots\wedge
    y_m$.
    \item $\mathcal{R}_5$: $x_1\vee x_2\vee\ldots\vee x_n\Rightarrow y_1\vee
    y_2\vee\ldots\vee y_m$.
    \item $\mathcal{R}_6$: $x_1\vee x_2\vee\ldots\vee x_n\Rightarrow \overline{y_1} \wedge \overline{y_2}\wedge\ldots\wedge
    \overline{y_m}$.
    \item $\mathcal{R}_7$: $\overline{x_1}\wedge \overline{x_2}\wedge\ldots\wedge \overline{x_n}\Rightarrow y_1\wedge y_2\wedge\ldots\wedge
    y_m$.
    \item $\mathcal{R}_8$: $\overline{x_1}\wedge \overline{x_2}\wedge\ldots\wedge \overline{x_n}\Rightarrow y_1\vee
    y_2\vee\ldots\vee y_m$.
    \item $\mathcal{R}_9$: $\overline{x_1}\wedge \overline{x_2}\wedge\ldots\wedge \overline{x_n}\Rightarrow \overline{y_1}\wedge \overline{y_2}\wedge\ldots\wedge
    \overline{y_m}$.
\end{enumerate}

Notons que dans ce chapitre, nous avons mis l'accent sur l'aspect
quantitatif, \emph{i.e.}, l'estimation de la taille de la
représentation concise. Néanmoins, l'aspect performance fait
partie des pistes que nous souhaiterions développer. En outre,
nous allons nous intéresser à l'aspect r{\'e}g{\'e}n{\'e}ration
des itemsets fr{\'e}quents. Ainsi, il est important de d{\'e}finir
des algorithmes, calculant les sommes d'inclusion-exclusion de
    mani{\`e}re efficace, afin de calculer les supports conjonctifs des
    itemsets candidats et de d{\'e}river ceux qui sont fr{\'e}quents de
    mani{\`e}re exacte.

 \addcontentsline{toc}{chapter}{Conclusion
et Perspectives}
\chapter*{Conclusion et Perspectives}
\markboth{Conclusion et Perspectives}{Conclusion et Perspectives}

Le besoin d'interpréter et d'analyser de grandes masses de données
constitue un véritable défi pour la communauté informatique au vu
de l'incapacité des outils existants à répondre aux nouveaux
besoins des utilisateurs, \emph{e.g.}, extraire une connaissance
cachée dans ces masses de données. Pour remédier à ce manque de
connaissances sur les données, de nouvelles méthodes d'extraction
de connaissances ont vu le jour, et qui sont regroupées sous le
terme générique de \emph{fouille de
données}~\cite{techniqdataminingag}. Les fondements mathématiques
de l'Analyse Formelle de Concepts ont été d'un grand apport dans
la proposition d'algorithmes efficaces d'extraction de
connaissances, dont la technique d'extraction des règles
associatives a été la principale bénéficiaire. En outre, ces
fondements ont été largement utilisés dans la définition d'un
cadre formel pour l'extraction d'un noyau irréductible et compact
de règles associatives. Ceci a ouvert la voie à la possibilité de
présenter un ensemble minimal de règles à l'utilisateur afin de
lui permettre de mieux exploiter les connaissances extraites.
Cependant, l'application "algorithmique", dans le seul souci
d'amélioration des performances, n'est pas arrivée à tenir les
promesses affichées par les approches recensées dans la
littérature. En effet, le principal inconvénient des approches
existantes réside dans le fait qu'elles se sont focalisées sur
l'extraction des itemsets fermés fréquents -- et au plus de leurs
générateurs minimaux associés -- sans se soucier de la relation
d'ordre sous-jacente.

\bigskip

Dans ce cadre, nous avons fait un certain nombre de propositions,
dont la plus importante est l'algorithme \textsc{Prince}, offrant
la possibilité à l'utilisateur de dériver toute base générique de
règles utilisant les générateurs minimaux et les itemsets fermés,
en évitant le surcoût du calcul des fermetures des itemsets
fréquents. Une deuxième contribution, entrant dans le cadre de la
quête d'informativité et plus de compacité de la connaissance
extraite, consiste en la proposition d'une base générique, appelée
$\mathcal{IGB}$, ainsi que le système axiomatique, associé.
Contrairement à la catégorisation classique des règles
associatives (\textit{i.e.,} exactes et approximatives), nous
avons proposé deux nouveaux types de règles, \emph{factuelles et
implicatives}, et nous avons montré leur intérêt dans un processus
de prise de décision. Enfin, nous avons proposé une nouvelle
représentation concise des itemsets fréquents fondée sur
l'exploration de l'espace de recherche disjonctif. Les
expérimentations que nous avons menées ont permis de mettre en
valeur la compacité de notre représentation concise par rapport à
celles existantes dans la littérature. La dérivation de
différentes formes de règles associatives, utilisant l'opérateur
de disjonction et de négation, est aussi à signaler au crédit de
cette dernière contribution.
\\
Les résultats obtenus ne font qu'ouvrir un certain nombre de
perspectives, qui s'articulent autour des deux thèmes
complémentaires suivants :
\begin{enumerate}
    \item \textbf{Distribution/Parallélisation du processus d'extraction des représentation concises :}
Suite à l'expansion des supports physiques de stockage et les
besoins incessants de sauvegarder de plus en plus des données, les
algorithmes séquentiels de recherche de règles associatives se
sont avérés inefficaces du point de vue performances. Ainsi,
l'utilisation du parallélisme est devenue une approche
incontournable dans le domaine de l'algorithmique associée à la
recherche de règles associatives. Dans ce cadre, l'étude que nous
avons menée concernant les itemsets "fermés" fréquents a permis de
mettre en relief l'absence d'algorithmes parallèles significatifs
permettant l'extraction de tels itemsets (ou d'une façon plus
générale des représentations concises).

Ainsi, la mise en oeuvre d'une algorithmique distribuée/parallèle
serait intéressante en vue de son application sur de grandes bases
réelles qui peuvent contenir plusieurs millions de transactions.
Par ailleurs, les algorithmes existants sont généralement
performants sur un type de bases (avec des dimensions et une
corrélation particulière entre les transactions, etc.) et beaucoup
moins sur d'autres. Il serait alors intéressant de mettre en place
un algorithme adaptatif suivant les caractéristiques des contextes
de fouille de données à traiter. En effet, il serait intéressant
de proposer, un environnement distribué muni d'une base de
connaissances, qui pourra affecter chaque "portion" de données à
l'algorithme le plus adapté pour l'extraction de la représentation
concise la plus adaptée. Ainsi, les pistes qui méritent d'être
explorées, dans le cadre d'une parallélisation par les données,
sont:

\begin{itemize}
    \item Exploration d'éventuelles relations entre les
opérateurs de fermeture disjonctive et conjonctive. Par exemple,
l'ensemble des fermés disjonctifs peut être retrouvé en calculant
les fermés conjonctifs sur le dual complémenté du contexte
d'extraction originel. Grâce à une telle relation, nous pouvons
adapter les algorithmes d'extraction des fermés
conjonctifs/disjonctifs proposés dans la littérature en optant
soit pour le contexte originel, ou soit pour le contexte dual.

    \item Caractérisation fine des contextes d'extraction. Dans ce cadre, nous avons amorcé
    une étude basée sur le Système Succinct des
Générateurs Minimaux (SSGM)~\cite{acke07}. Dans cette étude, nous
avons proposé une première définition formelle de la notion
d'éparsité/densité d'un contexte d'extraction. L'objectif visé par
cette étude était de déterminer l'algorithme d'extraction des
itemsets fermés le plus adéquat selon les estimations de densités
des contextes de fouille de données. Cependant, dans un souci
d'efficacité, cette détection devrait se faire à volée,
\emph{i.e.}, à partir d'un échantillon significatif dont le choix
reste encore à fixer.
\end{itemize}


    \item \textbf{Extraction des règles associatives (généralisées) multi-relationnelles:}

La prolifération de données riches ou complexes (génomiques, de
télécommunication, multi-relationnel, multimedia, etc.) a montré
que la structure particulière de ces données représente un défi
véritable pour les techniques de fouille de données classiques. En
outre, la majorité des algorithmes d'extraction des règles
associatives ont été développés pour des données centralisées.
Cependant, dans la pratique, les données se trouvent dispersées
dans plusieurs tables. Par conséquent, il est indispensable de
passer par une phase de transformation pour que les données soient
adaptées aux algorithmes existants d'extraction de règles
associatives. Toutefois, dans plusieurs cas, la transformation
s'avère désavantageuse menant ainsi à une perte d'information. Les
approches visant à extraire des Règles Associatives Relationnelles
($\mathcal{RAR}$) directement à partir des données décentralisées,
essaient de surmonter cette difficulté~\cite{jenson,eric,etoile}.
Cependant, les approches d'extraction des $\mathcal{RAR}$ à partir
d'un schéma en étoile~\cite{farmer} ainsi que les approches
d'extraction fondées sur la programmation logique inductive
(\textsc{pli})~\cite{toivonen} souffrent aussi d'un certain nombre
de limites. Ainsi, les approches qui permettent d'extraire des
règles associatives à partir de bases de données représentées sous
la forme d'un schéma en étoile, souffrent des problèmes de
redondance, de changement d'objet d'intérêt et de perte
d'informations. Par ailleurs, les approches basées sur la
\textsc{Pli}, soulèvent le problème de la taille des connaissances
présentées à l'utilisateur. En effet, bien que l'utilisation de la
\textsc{Pli} permettait d'enrichir l'expressivité des
connaissances extraites, l'espace de recherche exploré par les
algorithmes de la \textsc{Pli} est très large.

Dans ce cadre, il nous semble intéressant d'explorer les pistes
suivantes:
\begin{itemize}
\item Rechercher un noyau de règles associatives, à partir duquel
l'ensemble de règles restantes pourra être retrouvé via des
mécanismes d'inférence~\cite{garrigaijcai}.

\item La prise en considération des termes négatifs pour la
définition de règles génériques généralisées
(multi-relationnelles). En effet, les règles associatives ont été
introduites pour exprimer des corrélations entre les occurrences
des items. Toutefois, l'utilisateur a besoin aussi d'une
connaissance qui exprime la corrélation entre la non occurrence
des items. Le nombre de règles associatives présentées à
l'utilisateur sera évidemment très important, nécessitant ainsi
une sélection informative d'un sous-ensemble réduit de règles.
Pour cela, nous pensons étudier la possibilité d'appliquer des
approches d'extraction des bases génériques pour les règles
associatives positives et négatives~\cite{Kat05} serait une
perspective de recherche très pertinente.


\end{itemize}

Un cadre applicatif de cette perspective serait \textit{la
préservation de la vie privée et fouille de données}. Ce cadre est
strictement lié à l'utilisation croissante des systèmes
"multi-contextes", qui nécessite la maintenance d'un grand nombre
de contextes de fouille de données distribuées. Dans un objectif
d'aide à la décision, les grandes organisations souhaitent alors
pouvoir extraire de la connaissance à partir de l'ensemble de ces
contextes. Récemment, de nombreux travaux se sont intéressés à la
définition d'algorithmes de fouille de données préservant la vie
privée \cite{clifton03,Ev04,pinkas03}. Ainsi, préserver la vie
privée dans un contexte de fouille de données impose de n'offrir
des connaissances que si celles-ci garantissent la non divulgation
d'informations sensibles sur les personnes. Pour garantir que les
algorithmes de fouille de données ne violent pas la vie privée des
individus, certaines approches ont considéré qu'elles disposaient
d'une connaissance préalable sur ce qui était sensible ou non.
Cependant, ce type d'approches reste, malheureusement, très
subjectif et est très difficile à mettre en oeuvre~\cite{ponc07}.

%
%
%


\end{enumerate}


%

\nocite{ref1}
\nocite{ref2}
\nocite{ref3}
\nocite{ref4}
\nocite{ref5}
\nocite{ref6}
\nocite{ref7}
\nocite{ref8}
\nocite{ref9}
\nocite{ref10}
\nocite{ref11}
\nocite{ref12}
\nocite{ref13}
\nocite{ref14}
\nocite{ref15}
\nocite{ref16}
\nocite{ref17}
\nocite{ref18}
\nocite{ref19}
\nocite{ref20}
\nocite{ref21}
\nocite{ref22}
\nocite{ref23}
\nocite{ref24}
\nocite{ref25}
\nocite{ref26}
\nocite{ref27}
\nocite{ref28}
\nocite{ref29}
\nocite{ref30}
\nocite{ref31}
\nocite{ref32}
\nocite{ref33}
\nocite{ref34}
\nocite{ref35}
\nocite{ref36}
\nocite{ref37}

\addcontentsline{toc}{chapter}{Bibliographie} \small
\bibliography{Biblio/tarek_recherche,Biblio/THESE060904,Biblio/arXiv}
\bibliographystyle{plain}

\end{document}